\documentclass[aps,prb,superscriptaddress,twocolumn,floats]{revtex4-2}

\usepackage{amsmath}
\usepackage{amssymb}
\usepackage{bbold}
\usepackage{graphicx}
\usepackage{bbm}
\usepackage{braket}
\usepackage[section]{placeins}
\usepackage{color}
\usepackage[normalem]{ulem}
\newcommand{\mr}{\mathrm}

\def\expect#1{\langle #1\rangle}

\begin{document}

\title{Cross-correlation spectra  in interacting quantum dot  systems}

\author{Andreas Fischer}
\author{Iris Kleinjohann}
\affiliation{Department of Physics, Technische Universit\"at Dortmund,
Otto-Hahn-Stra{\ss}e 4, 44227 Dortmund, Germany}

\author{Nikolai A.\ Sinitsyn}
\affiliation{Los Alamos National Laboratory, Los Alamos, USA}
\author{Frithjof B.\ Anders}
\affiliation{Department of Physics, Technische Universit\"at Dortmund,
Otto-Hahn-Stra{\ss}e 4, 44227 Dortmund, Germany}

\date{\today}

\begin{abstract}
Two-color spin-noise spectroscopy of interacting electron spins in singly
charged semiconductor quantum dots provides information on the inter-quantum dot interactions.  We investigate the spin
cross-correlation function in a quantum dot ensemble employing
a modified
semiclassical approach.
Spin-correlation functions are calculated using a Hamilton quaternion
approach that maintains
local quantum mechanical properties of the spins.
This method takes into account the effects of the
nuclear-electric quadrupolar interactions, the randomness of the coupling
constants, and the variation of the electron $g$ factor on the spin-noise
power-spectra.  We demonstrate that the quantum dot ensemble can be
mapped on an effective two-quantum dot problem and discuss how the
characteristic length scale of the inter-dot interaction modifies the
low-frequency cross-correlation spectrum.  We argue that details
on the interaction strength distribution can be extracted from the
cross-correlation spectrum when applying a longitudinal or a
transversal external magnetic field.
\end{abstract}

\maketitle

\section{Introduction}

Trapped charge carriers in semiconductor quantum dots (QDs)
\cite{Hanson2007} are of interest for quantum functionality and the
implementation of spintronic devices \cite{Burkard2000}.  Major
progress has been made in the initialization, manipulation and read
out of the localized spin by using fast optical methods
\cite{Atatuere2006,GreilichEconomou2009,Press2008,Xu2007}.  The
confinement of electrons or holes in QDs removes the motion-related
spin relaxation, such that
the hyperfine interaction with surrounding nuclear
spins becomes the main source of decoherence.  With periodical optical
excitation exploiting nuclear-induced frequency-focusing, the
single spin coherence time can be increased to the order of
microseconds \cite{GreilichYakovlev2006,Greilich2007}.

Spin-noise spectroscopy \cite{Glazov2012} has served as a powerful
tool to identify the relevant energy and time scales in QD ensembles
due to their strong optical response.  Particularly insightful have
been
spin-noise experiments that explored spin correlations of QDs
in the thermal equilibrium
\cite{Crooker2010,Yan2012,Zapasskii2013,Hackmann2015,Sinitsyn_2016,Glasenapp2016}.
The spectrum also reveals the importance of the nuclear-electric
quadrupolar interaction for qubit decoherence
\cite{Sinitsyn2012,Bulutay2012,Hackmann2015}.
Recently, higher-order spin correlators attracted attention,
because they contain additional information that is absent in the
standard spin-noise power-spectra
\cite{LiSinitsyn2016,Froehling2017,Haegle2018,FroelingJaeschke2019,Norris2016,Szankowski2017}.

The investigation of the dynamics of the electron spin
subject to a hyperfine coupling \cite{Fermi1930}
with the surrounding nuclear spins has a long history.
Although the central spin model, the Gaudin model, is exactly solvable
by a Bethe ansatz \cite{Gaudin1976}, the complexity of the solutions 
does only allow to
access the exact dynamics for a small number of nuclear spins \cite{Bortz2007} or  requires the combination with a Monte Carlo approach \cite{FaribautSchuricht2013a,FaribautSchuricht2013b} for their evaluation.
Merkulov et al. \cite{Merkulov2002} addressed the electron spin relaxation by nuclei in a semiconductor QD using a frozen Overhauser field approximation. 
A linked-cluster expansion \cite{SaikinSham2007} and a cluster-correlation expansion \cite{Yang2008} in a finite size bath have been used to investigate
single-electron spin decoherence by the nuclear spin bath.
An exact quantum mechanical treatment of the problem 
using Chebyshev polynomials \cite{TalEzer-Kosloff-84}
was applied to calculate the spin-noise spectroscopy \cite{Smirnov_2021,Hackmann2014}
including nuclear-electric quadrupolar interactions \cite{Hackmann2015}.
An adaptation of the density matrix renormalization group approach \cite{Schollwoeck-2005}
was able to treat large numbers of nuclear bath spins \cite{StanekRaasUhrig2013} but was limited to the short time dynamics.
Coish and Loss took into account the temporal fluctuation of the nuclear spins within a generalized master equation \cite{CoishLoss2004}.
A rate equation approach was used 
to incorporate nuclear quadrupolar interactions in a double QD with spin and charge dynamics \cite{DengHu2005}.
Barnes et al.~\cite{BarnesDaSarma2011}
applied a Nakajima-Zwanzig  type master equation to the problem.
Recently a master equation was employed to address the nuclear polaron
formation at low temperatures \cite{fischer2020}.
Electric current noise in mesoscopic organic semiconductors induced by nuclear spin fluctuations
was studied by Smirnov et al.\ \cite{smirnov2021}.

Here, we study the spin cross-correlators in an ensemble of QDs or in
a QD molecule. We propose that the cross-correlation spectrum can be
used for extracting details on the inter-QD coupling
distribution and identifying the microscopic origin of the inter-QD
spin coupling in experiments.  However cross correlations have been
much less studied theoretically \cite{Roy2015} and have not beenGreilichYakovlev2006
investigated by spin-noise spectroscopy in QDs yet.

We extend a semiclassical approach (SCA)
\cite{Merkulov2002,Al-Hassanieh2006,ChenBalents2007,Glazov2012,StanekRaasUhrig2013,HackmannGlazov2014,Jaeschke2017} 
based on spin-coherent states to interactions and
characteristics that have not been investigated previously but are relevant
for the studies of cross-correlators in coupled QDs, as well as more
realistic interactions in the nuclear spin bath.  This is done by
mapping the quantum mechanical time evolution onto a quaternionic
representation \cite{Girard_1984} originally introduced by Hamilton
more than 150 years ago. It allows to simulate the system by an
effective classical dynamics and still maintains the non-commutativity of quantum mechanical Heisenberg operators at different times as required in spin correlation functions.

The basic idea of the quaternionic representation is the mapping of the 
unitary time-evolution operator that rotates an electron spin-coherent
state on the Bloch sphere onto the equation of motion of a
time-dependent classical 3d rotation matrix: The quantum mechanical
expectation values in the correlation functions can be evaluated
analytically in a fixed representation and the time-dependency is
shifted into the
time-dependent rotation matrix. This scheme is also generalizable 
to
higher-order spin correlation functions.
In the context of spin noise the importance of different third and fourth order spin correlation
functions \cite{LiSinitsyn2016,Froehling2017,FroelingJaeschke2019} has been
discussed. It was pointed out \cite{Froehling2017} that  certain types of fourth order
spin-correlation functions are useful for understanding the 
magnetic field dependency of the decoherence time 
in spin-echo experiments \cite{Bechtold2016,Press2008}.
Our approach might be useful to apply to 
higher order Carr-Purcell-Meiboom-Gill (CPMG)  pulse sequences as well
as optical pumping of the QD ensembles with periodic laser 
pulses \cite{GreilichYakovlev2006,Greilich2007,Evers2018}.

SCAs to the coupled electron-nuclear spin
dynamics in QDs in their various incarnations
\cite{Merkulov2002,Glazov2012,FausewehSchering2017,Jaeschke2017} have
been explored over the last 20 years.  The large number of nuclear
spins is used as a justification for replacing the hyperfine
interaction operator by a classical random variable, the so-called   Overhauser field that acts as an additional magnetic field onto the
electron spin \cite{Merkulov2002}.  Using the strong separation of
time scales, the fast electron spin precession and the several orders
of magnitude slower nuclear spin dynamics lead to the frozen
Overhauser field approximation where the nuclear spin dynamics is
replaced by a static field characterised by a Gaussian distribution
\cite{Merkulov2002} which includes the proper thermodynamic limit
\cite{FausewehSchering2017} of infinitely many nuclear spins.

Such a static approximation already yields  an excellent prediction of the
high-energy parts of the spin-noise spectrum but fails to predict the
low-frequency parts of the spectrum that are connected to the nuclear
spin dynamics.  Furthermore, large spectral weight is accumulated in a
zero-frequency peak since conservation laws protect the spin
correlations from
decay \cite{Merkulov2002,Uhrig2014}.  Violation of
the conservation laws by relevant interactions such as nuclear-electric
quadrupolar interactions \cite{Bulutay2012} leads to a broadening of
this zero-frequency peak \cite{Sinitsyn2012} that can be
experimentally resolved \cite{Hackmann2015}.  Therefore, we employ a
SCA that accounts for the long-time nuclear spin
dynamics as well as includes these additional symmetry breaking
terms in the Hamiltonian.

Our numerical data suggests that a genuine interacting QD ensemble can
be mapped onto a coupled two-QD model augmented by a distribution
function of the effective coupling constants when targeting two-color
spin-noise spectroscopy.  This is backed by the analytic structure of
the equations of motion where the total effect of all other electron
spins of the ensemble onto the dynamics of the electron spin in an
individual QD is included into a single additional effective noise
field in addition to the Overhauser field \cite{Fischer2018}. The
mapping is constructed such that the first two momenta of this
fluctuating field are reproduced.

Our article is structured as follows. We
introduce the model for the coupled QD ensemble in Sec.~\ref{sec:model}, provide an overview
of the modified SCA in Sec.~\ref{sec:sca}, and present the general
properties of cross-correlation functions in Sec.\
\ref{sec:correlation_functions}.  Section \ref{sec:results} is devoted
to the numerical results of our simulations.  We start with a
justification of the mapping of a genuine QD ensemble onto an effective
two-QD model in Sec.\ \ref{sec:JDist} by demonstrating that the
cross-correlation  spectra perfectly agree
independent on the number
of QDs in the ensemble.  In Section \ref{sec:QM}, we remove the
randomness of coupling constants by investigating the correlation
spectra for a fixed hyperfine coupling, a fixed QD-QD interaction and
fixed electron $g$ factors to discuss the elementary properties in
the single-color spin-noise as well as in the cross-correlation
function.  The frozen Overhauser field solution in Sec.\
\ref{sec:FOA} provides a better understanding of the additional
features observed in the spectra.  In Sec.\ \ref{sec:parameters}, the effect
of randomness of the coupling parameters, the
nuclear-electric quadrupolar interactions, and the electron $g$ factor variations
in the QD ensemble are discussed.  We examine the cross-correlation
spectrum at zero magnetic field, Sec.~\ref{sec:QI}, transversal
magnetic field, Sec.~\ref{sec:gfactors}, and longitudinal magnetic
field, Sec.~\ref{sec:longit}, where we demonstrate the relevance and
influence of the various interactions on the cross-correlation spectra
in the different regimes.  We make a connection to experiment in Sec.\
\ref{sec:exp} and summarize the results in the conclusion.

\section{Extended central spin model}
\label{sec:model}

We focus on the spin dynamics in an ensemble of singly-charged QDs.  
A two-color spectroscopic investigation has demonstrated that
the experimental data \cite{Spatzek2011} is consistent with an
interaction between the electron spins of a Heisenberg type. Its
microscopic origin is still unclear, but an optical RKKY interaction
\cite{Spatzek2011} can be ruled out due to its time dependency.
Potential candidates that are currently discussed are either a direct
exchange interaction for spatially adjacent QDs
\cite{Mizel04} or a long-range RKKY-like interaction mediated by a
weak carrier doping of the conduction bands as a side effect of
charging the QD with a single electron that was found for
self-assembled semiconductor QD
ensembles~\cite{Spatzek2011,Fischer2018}.

Here we assume an ensemble description of  $N_{QD}$ QDs by the Hamiltonian
\begin{eqnarray}
H &=& \sum_{i=1}^{N_{QD}}  H^{(i)} + \sum_{i<j}  J_{ij} \vec{S}^{(i)}\vec{S}^{(j)},
\label{eq:H}
\end{eqnarray}
where $H^{(i)}$ is the Hamiltonian of the i-th QD and $\vec{S}^{(i)}$
denotes its electron spin operator; $i$ runs over all QDs of
the ensemble.  Note, in this general form, $H$ also applies for
QD molecules \cite{Patanasemakul2012}, QD chains \cite{Wang2006} or QD
super-lattices \cite{Sugaya2013}, whereas the specific geometry of the
QDs is encoded in the interaction matrix $J_{ij}$.
Such a Hamiltonian has been proposed by Smirnov et al. \cite{Smirnov2014}
who focused on the calculation of spin autocorrelation functions.

For the description of each individual QD, we include the
hyperfine interaction between the resident electron spin and the
surrounding nuclear spins, an external magnetic
field and the static nuclear-electric quadrupolar interactions. We
neglect weaker and, therefore, less prominent effects such as the
nuclear dipole-dipole interactions.  As a first part of our model, we
introduce the Hamiltonian of the central spin model
(CSM)~\cite{Gaudin1976} with $\hbar = 1$
\begin{align}
H_{\mr{CSM}}^{(i)} &= \sum_{k=1}^{N_i} A_k^{(i)} \vec{S}^{(i)} \vec{I}^{(i)}_k + \mu_B g^{(i)} \vec{B}_\mr{ext} \vec{S}^{(i)} \notag \\ &+ \sum_{k=1}^{N_i} \mu_I g_k^{(i)} \vec{B}_\mr{ext}\vec{I}_k^{(i)} , \label{eq:H_CSM}
\end{align}
that accounts for the hyperfine interaction of the electron spin
$\vec{S}^{(i)}$ with the nuclear spins $\vec{I}^{(i)}_k$
and the external magnetic field $\vec{B}_\mr{ext}$.  The index
$i\in[1,N_{QD}]$ labels the QD in the ensemble and the index
$k\in[1,N_i]$ labels the nuclear spin in the respective QD.
The coupling constants $A_k^{(i)}$ of the hyperfine interaction
produce a characteristic time scale of the system
\begin{align}
T^*_i = \left( \sum_k (A_k^{(i)})^2 \braket{(\vec{I}_k^{(i)})^2}\right)^{-\frac{1}{2}}
.
 \label{eq:Tstar}
\end{align}
In real QD ensembles, the dot size variation generates a variation of
$T^*_i$. In this work, however, we assume $T^*_i$ to be equal for
all QDs and use it as a reference scale for measuring time
dependencies.  Furthermore, $1/T^*$ defines the intrinsic energy
scale of a QD.  In
real semiconductor QDs, $T^*$ is in the order of a few
nanoseconds \cite{Greilich2007}.  

Due to the strong coupling of the
electron spin to the external magnetic field via the Bohr magneton
$\mu_B$ and the g-factor $g^{(i)}$, that is approximately $0.5$ in 
electron-doped InGaAs QDs~\cite{Greilich2007}, the precession of the electron spin
provides the fastest dynamics in the spin system for a magnetic field
strength above $20\,\mr{mT}$~\cite{Schwan2011}.  In comparison the
precession frequency of the nuclear spins is typically three orders of
magnitude smaller as a result of the weaker nuclear magnetic moment
with a ratio $\mu_I g_k^{(i)} / (\mu_B g^{(i)}) \ll 1$
\cite{Beugeling2016,Beugeling2017}.

The CSM governs the short-time dynamics of a single QD. However, the nuclear-electric quadrupolar
interactions induce disorder in the nuclear spin bath and lead
to an additional electron spin dephasing on a time scale of $100
\,\mr{ns}$
\cite{Coish2009,Sinitsyn2012,Bulutay2012,Bechtold2015,Dyakonov2008}.
The additional contribution to the Hamiltonian reads
\begin{align}
H_{Q}^{(i)} =& \sum_k q^{(i)}_k \Big[ \left( \vec{I}_k^{(i)} \cdot \vec{n}_{Q,k}^{(i)} \right)^2 \notag\\
  &+ \frac{\eta}{3} \left( \left( \vec{I}_k^{(i)} \cdot \vec{n}_{Q,k}^{x,(i)} \right)^2 - \left( \vec{I}_k^{(i)} \cdot \vec{n}_{Q,k}^{y,(i)} \right)^2 \right) \Big], \label{eq:H_QI}
\end{align}
where $q^{(i)}_k$ is the quadrupolar coupling  of the $k$-th nuclear spin, $\eta$ the anisotropy factor \cite{Slichter1996}, which is set equal for all spins, and $\vec{n}_{Q,k}^{(i)}$ the quadrupolar easy axis of an individual nuclear spin \cite{Bulutay2012,Hackmann2015}.
The support vectors $\vec{n}_{Q,k}^{x,(i)}$ and $\vec{n}_{Q,k}^{y,(i)}$ are chosen in such a way, that $\vec{n}_{Q,k}^{(i)}$, $\vec{n}_{Q,k}^{x,(i)}$ and $\vec{n}_{Q,k}^{y,(i)}$ form an orthonormal basis.
Details of the theoretical description of static nuclear-electric quadrupolar interactions can be found in Ref.~\cite{Hackmann2015}.
Combining the CSM and the nuclear quadrupolar effects yields the complete Hamiltonian of the individual QD
\begin{align}
 H^{(i)} = H_{\mr{CSM}}^{(i)} + H_{Q}^{(i)} .
\end{align}

\section{Semiclassical approach}
\label{sec:sca}

For realistic system sizes of $N_i=10^4\ldots10^6$ nuclei in a QD \cite{Urbaszek2013}, the Hamiltonian, Eq.~\eqref{eq:H}, is not
solvable, analytically or numerically, in an exact manner due to the
Hilbert space dimension that grows exponentially with $N_i$ as the
Hilbert space incorporates $2^{N_{QD}} (2I+1)^{N_{QD} N_i}$ states for nuclear spins with length $I$.  A
pure quantum mechanical description is limited to a small number of
spins~\cite{FaribautSchuricht2013a,FaribautSchuricht2013b,Beugeling2016,Beugeling2017,Kleinjohann2018}.

In this paper, we use a SCA \cite{Glazov2012}
that retains quantum mechanical features on a single spin level.  This
approach is very similar to the spin-coherent-state representation by
Al-Hassanieh et al. \cite{Al-Hassanieh2006} but it is based on a path
integral formulation introduced by Chen et al. \cite{ChenBalents2007}.
While these two papers focused on the propagation of a well defined
central spin state in time, our approach extends the SCA to general
correlation functions.
Variations of this approach have already already been used in previous
publications \cite{Jaeschke2017,Fischer2018,FroelingJaeschke2019}.

\subsection{ Spin-coherent states and path integral}

To obtain the equations of motion (EOM) for spin-coherent states in
the SCA, we first decompose the mixed quantum mechanical density
operator $\rho$ of the system into a set of pure product states of
spin-coherent states.  Second, we solve the time evolution of the pure
product states using the semiclassical limit of a quantum mechanical
path integral.  This procedure
preserves many quantum effects, like vanishing quadrupolar
interactions for spin $1/2$ or complex-valued correlation functions, due to
the invertible map from classical vectors to spin-coherent states.

Let $\ket{s,m}$ be the spin basis with quantum numbers resulting from the eigenvalues of the quantum mechanical operators for the square of total spin $S^2$ and the spin $z$ component $S_z$
\begin{subequations}
\begin{align}
S^2 \ket{s,m} &= s(s+1) \ket{s,m} \\
S_z \ket{s,m} &= m \ket{s,m} .
\end{align}
\end{subequations}
We define the spin-coherent state parametrized by a classical vector $\vec{s} \in \mathbb{R}^3$ and length $|\vec{s}|=s$ as \cite{Fradkin1988,Fradkin2013}
\begin{align}
\ket{\vec{s}} = e^{ -i \vec{\theta} \cdot \vec{S} } \ket{s,s} .
\label{eq:scs}
\end{align}
The spin-coherent state describes a spin fully aligned along the quantization axis $\vec{n}_0$, here $\vec{e}_z$, that is rotated around the axis $\vec{\theta}/\theta = \vec{s} \times \vec{n_0} / |\vec{s} \times \vec{n_0}|$ and the angle $\cos(\theta) = \vec{s} \cdot \vec{n}_0$.

The set of spin-coherent states builds an over-complete basis of the
Hilbert space for a spin $\vec{S}$.  Their completeness relation is
given by
\begin{align}
	\mathbbm{1} = \int d\mu(\vec{s}) \ket{\vec{s}} \bra{\vec{s}} \label{eq:comp}
\end{align}
with the identity matrix $\mathbbm{1}$ and the integration measure
\begin{align}
	d\mu(\vec{s}) = \left(\frac{2s + 1 }{4 \pi} \right) \delta(\vec{s}^2 -s^2) \;d^3s   \; . 
\end{align}

In the following, we combine information on all electron and nuclear
spins in the coupled QD system into the state
$\ket{\{\vec{s}_j\}}$ where the index $j$ labels the individuals spins.  This state $\ket{\{\vec{s}_j\}}$ is
a product state of the states of individual spins $\ket{\vec{s}_j}$.
For the transition from quantum mechanical spin states to the limit of
classical spin vectors, we employ a path integral formulation.  The
starting point is the propagator,
\begin{align}
\label{eq:10}
	K(\{\vec{s}_{j,f}\},\{\vec{s}_{j,i}\},t) = \bra{\{\vec{s}_{j,f}\}} e^{-i H t} \ket{\{\vec{s}_{j,i}\}} ,
\end{align}
which provides the transition amplitude from the initial state
$\ket{\{\vec{s}_{j,i}\}}$ at time $0$ to the final state
$\ket{\{\vec{s}_{j,f}\}}$ at time~$t$.  Making use of $e^{- i H t} =
\lim\limits_{N \rightarrow \infty} \prod_{n=0}^{N-1} e^{-i H t / N}$,
where the unity $\eqref{eq:comp}$ is inserted between all
infinitesimal time evolution propagators, we rewrite the propagator in
the standard path integral form (see Appendix~\ref{sec:appPathInt} for
details,)
\begin{align}
K(\{\vec{s}_{j,f}\},\{\vec{s}_{j,i}\},t) &=  \int \mathcal{D}[\{\vec{s}_j\}]\; e^{i \mathcal{S}[\{\vec{s}_j\},\{\partial_t \vec{s}_j\}]}
,
\end{align}
with the action
\begin{eqnarray}
&&\mathcal{S}[\{\vec{s}_j\},\{\partial_t \vec{s}_j\}]  \\
&&=  \int \left(i \bra{\{\vec{s}_j\}}\frac{d}{dt}\ket{\{\vec{s}_j\}} - \bra{\{\vec{s}_j\}}H \ket{\{\vec{s}_j\}}\right) dt,
\nonumber
\end{eqnarray}
and the appropriate integral measure $\mathcal{D}[\{\vec{s}_j\}]$.
The action functional comprises of two contributions, (i) the
topological or kinetic action, which only depends on the topology of
the Hilbert space, and (ii) the action from the Hamiltonian.

\subsection{Semiclassical equations of motion}

Every possible spin-path contributes to the kernel in Eq.\
\eqref{eq:10} with equal probability but destructive
interference favors the so-called classical path where the variation
$\delta \mathcal{S}$ is zero and neighboring paths interfere
constructively.  This approximation is justified by the large number
of (nuclear) spins involved in the system \cite{ChenBalents2007}.  

Under the conditions of a stationary action $\delta \mathcal{S} = 0$ 
and a fixed spin length $(\vec{s}_j)^2 = s_j^2$, we obtain the general EOM for a classical spin $\vec{s}_j$,
\begin{align}
\partial_t \vec{s}_j &= \frac{\partial\tilde{H}(\{\vec{s}_j\})}{\partial \vec{s}_j} \times \vec{s}_j ,
\label{eq:eom}
\end{align}
with the classical Hamilton function $\tilde{H}(\{\vec{s}_j\}) = \bra{\{\vec{s}_j\}}H\ket{\{\vec{s}_j\}}$.
The EOM is form-invariant for different types of spins 
though their distinction is encoded in the Hamilton function $\tilde{H}(\{\vec{s}_j\})$.

For the full Hamiltonian~\eqref{eq:H} of the extended CSM, the EOMs 
are derived from
Eq.~\eqref{eq:eom} and read
\begin{subequations}
\label{eq:17}
\begin{align}
	\partial_t \vec{s}^{(i)} &= \vec{b}_{\mr{eff}}^{(i)} \times \vec{s}^{(i)} \label{eq:eom_s}\\
	\partial_t \vec{i}^{(i)}_k &= \vec{b}_{\mr{eff},k}^{(i)} \times \vec{i}_k^{(i)} \label{eq:eom_i}
\end{align}
\end{subequations}
for the classical electron spin vectors $\vec{s}^{(i)}$ as well as
nuclear spin vectors $\vec{i}^{(i)}_k$ with the respective effective
magnetic fields,
\begin{subequations}
\begin{align}
	\vec{b}_{\mr{eff}}^{(i)} &= \vec{b}_\mr{ext}^{(i)} + \vec{b}_N^{(i)} +  \vec{b}_J^{(i)}, \\
	\vec{b}_{\mr{eff},k}^{(i)} &= \vec{b}_{\mr{ext},k}^{(i)} + \vec{b}_{K,k}^{(i)} + \vec{b}_{Q,k}^{(i)} \; . \label{eq:beffk}
\end{align}
\end{subequations}
The external magnetic fields $\vec{b}_\mr{ext}^{(i)} = \mu_B g^{(i)}
\vec{B}_\mr{ext}$ and $\vec{b}_{\mr{ext},k}^{(i)} = \mu_I g_k^{(i)}
\vec{B}_\mr{ext}$ comprise $g$ factor and magneton of the spins.  The hyperfine
interaction manifests itself as the Overhauser field $\vec{b}_N^{(i)}
= \sum_k A_k^{(i)} \vec{i}_k^{(i)}$ acting on the central spin and the
much weaker Knight-field $\vec{b}_{K,k}^{(i)} = A_k^{(i)}
\vec{s}^{(i)}$ acting on the nuclear spins.  The inter-QD interactions
result in an effective magnetic field $\vec{b}_J^{(i)}= \sum_j J_{ij}
\vec{s}^{(j)}$ given as a weighted sum over the electron spins
$\vec{s}^{(j)}$ with the weights $J_{ij}$.  Less trivial are the
quadrupolar fields
\begin{align}
\vec{b}_{Q,k}^{(i)} &= 2 q_k^{(i)} \left( 1 - \frac{1}{2I} \right)  \Big\{ \left( \vec{i}^{(i)}_k \cdot \vec{n}_{Q,k}^{(i)} \right) \vec{n}_{Q,k}^{(i)} \notag \\&+ \frac{\eta}{3} \left[ \left( \vec{i}^{(i)}_k \cdot \vec{n}_{Q,k}^{x,(i)} \right) \vec{n}_{Q,k}^{x,(i)} - \left( \vec{i}^{(i)}_k \cdot \vec{n}_{Q,k}^{y,(i)} \right) \vec{n}_{Q,k}^{y,(i)} \right] \Big\} \; , \label{eq:b_Q}
\end{align}
which are derived in Appendix \ref{sec:appQI}.  Here, the prefactor
$(1-1/2I)$ ensures that quadrupolar interactions vanish for a spin $1/2$
as required by quantum mechanics. 
In contrast to $\vec{b}_{\mr{ext},k}$ the field $\vec{b}_{Q,k}^{(i)}$ depends on the nuclear spin itself and is time dependent like the other fields stemming from spin-spin interactions.

While the analytic form of Eq.\ \eqref{eq:eom} has been widely
used in the literature
\cite{Merkulov2002,Glazov2012,Al-Hassanieh2006,ChenBalents2007,StanekRaasUhrig2013,Jaeschke2017,FausewehSchering2017,ScheringSD-SCA2018}
the definition of the classical Hamilton function
$\tilde{H}(\{\vec{s}_j\})$ allows to access additional types of
interactions perviously not much explored with a SCA such as the
nuclear-electric quadrupolar interaction. This part of the Hamiltonian
is non-linear in spin operators, e.g. $H_{Q,k}^{(i)}$ in
Eq.~\eqref{eq:H_QI}, and the evaluation of $\tilde{H}(\{\vec{s}_j\})$
was performed rigorously and is beyond the simple replacement
\cite{FausewehSchering2017} of each spin operator $\vec{S}_j$ by its
classical counterpart $\vec{s}_j$.

\section{Correlation functions}
\label{sec:correlation_functions}

In the high-temperature limit, which is valid for nuclear spins at standard
cryostatic temperatures and moderate magnetic fields, we can summarize the SCA as follows:
First we decompose the density operator and replace the full integration over the Bloch-sphere 
by a Monte-Carlo integration 
\cite{Al-Hassanieh2006,ChenBalents2007},
\begin{align}
\rho &= \int d\mu(\{\vec{s}_j\}) \ket{\{\vec{s}_j\}} 
\bra{\{\vec{s}_j\} } 
\notag\\
&
\approx \frac{1}{N_C} \sum_\mu \ket{\{\vec{s}_j\}}_\mu \bra{\{\vec{s}_j\}}_\mu , \label{eq:sca_rho}
\end{align}
with $N_C$ samples (here we use $N_C\approx10^6$) whose states $\ket{\{\vec{s}_j\}}_\mu$ are picked
randomly from the Bloch-sphere. 
 In a second step, the time evolution
is calculated by solving the EOMs, Eqs.~\eqref{eq:eom_s} and
\eqref{eq:eom_i}, independently for each configuration $\mu$.  This
step is computationally the most costly, but can be massively
parallelized.  As a third step, the calculated trajectories
$\ket{\{\vec{s}_j\}(t)}_\mu$ can be used to evaluate the
time-dependent expectation value of an observable~$O$,
\begin{align}
	\braket{O(t)} &= \mr{Tr}[ \rho(t) O ] 
	\notag \\ 
&\approx 
	\frac{1}{N_C} \sum_\mu \bra{\{\vec{s}_j\}(t)}_\mu O \ket{\{\vec{s}_j\}(t)}_\mu \; ,
\end{align}
average over $N_C$ classical configurations $\mu$ by using
Eq.~\eqref{eq:sca_rho}.  For observables, that are linear in a spin
operator \cite{Al-Hassanieh2006,ChenBalents2007}, one can use
$\bra{\vec{s}}\vec{S}\ket{\vec{s}} = \vec{s}$ and simply replace the
spin operators by their corresponding classical vectors.  When an
observables $O$ is non-linear in spin operators, such as it is the case
for spin-spin correlators studied in the following, the expression
$\bra{\{\vec{s}_j\}} O \ket{\{\vec{s}_j\}}$ has to be evaluated
rigorously and reveals the quantum nature of the system.  We explore
this behavior using the second order auto-correlation function $C_2$
in the following.

\subsection{Auto-correlation function} 
  
While the power or noise spectrum in QDs is well established
\cite{Merkulov2002,Khaetskii2003,Hanson2007, Hackmann2014}, the inter-QD interactions enter just as an additional small noise source into
the auto-correlation function, which makes them hardly distinguishable
from the other central spin interactions \cite{Fischer2018}. However,
a non-vanishing cross-correlation function is explicitly generated by
the inter-QD interactions and can reveal the true origin of these
interactions.

For time-translational invariant systems, the second order auto-correlation function,
\begin{align}
C_2(t) =   \expect{ S_z(0) S_z(t) },\label{eq:c2}
\end{align}
can be written as a function of  the relative time~$t$.
As $C_2$ is non-linear in the spin operators, the configuration
average over the product of the classical variables $ s_{z,\mu}(0)
s_{z,\mu}(t)$ fails to evaluate $C_2$ in the SCA.  This becomes
particularly clear when looking at the time $t=0$, where the classical
configuration average $\ll (s_z(0))^2\gg \, \leq 1/4$ for classical
spins of a fixed length $s=1/2$, while quantum mechanics yield
$C_2(0) = 1/4$.  This fundamental problem of the SCA is usually
circumvented by exploiting the equivalence of $C_2(t)$ with
$\expect{S_z(t)}$ up to a factor of $1/2$ starting for $S_z(0)=1/2$
\cite{Al-Hassanieh2006,ChenBalents2007}.
Alternatively, a different SCA was
proposed \cite{FausewehSchering2017} where the fixed spin length is
replaced by a Gaussian distribution function that meets the condition
$C_2(0) = 1/4$.

Purely classical approaches \cite{Merkulov2002,Glazov2012} suffer one
fundamental drawback: the correlation function can only be a real
quantity whereas a correlation function as defined in Eq.\ \eqref{eq:c2}
must be a complex-valued function at low-temperatures in quantum
mechanics. Therefore, quantum mechanical correlators are usually
symmetrized to connect with measurable correlators. 

In this paper, we aim for an approach to arbitrary correlation
functions that (i) maintains the properties related to the quantum
mechanical spin algebra and (ii) uses only the dynamics of classical
vectors given by Eq.~\eqref{eq:17}.  The contribution of
a single spin-coherent state $\ket{\vec{s}_0}$ to Eq.~\eqref{eq:c2}
can be transformed into the form
\begin{align}
\bra{\vec{s}_0} S_z(0) S_z(t) \ket{\vec{s}_0}
&= \frac{1}{4} \vec{n}_z(0) \cdot \vec{n}_z(t)  \nonumber \\
& + \frac{i}{2} (\vec{n}_z(0) \times \vec{n}_z(t)) \cdot \vec{s}_0 , \label{eq:c2_sca}
\end{align}
using the basic spin $1/2$ algebra, which clearly differs from the product of
the classical variables $ s_{z,\mu}(0) s_{z,\mu}(t)$.
  Here, $\vec{n}_z(t) =
\tilde{R}(t) \vec{e}_z$ is the $z$-axis rotated by the effective field
$\vec{b}_\mr{eff}(t)$.  The derivation of Eq.\ \eqref{eq:c2_sca} and the EOM of the $SO(3)$ rotation matrix $\tilde{R}(t)$, using
the Hamilton quaternion representation, can be found in Appendix
\ref{sec:approt}.

Equation~\eqref{eq:c2_sca} can be easily evaluated within the SCA
and brings two relevant deviations from the naive product $ s_{z,\mu}(0) s_{z,\mu}(t)$.  
The demand $C_2(0)=1/4$ is fulfilled for any $\vec{s}_0$ and an
imaginary part is generally possible. In the high-temperature limit, the averaging over all $\vec{s}_0$ leads to a cancellation of the  
imaginary part. This does not hold at  temperatures well below the hyperfine energy scale.

The procedure to obtain $C_2(t)$ in the classical limit is not
restricted to the second-order autocorrelation function and can also
be applied systematically to higher-order correlation functions,
though their analysis is beyond the scope of this paper.

\subsection{Cross-correlation function}

The cross-correlation function of two spins in different QDs vanishes if
they are uncorrelated.  Therefore, it is an unambiguous tool for the
identification of inter-QD interactions.  However, we will see that
not only interactions between QDs appear, but also interactions within the individual QDs provide a contribution to the cross-correlation spectrum and reveal the rich
dynamics of the system.

For cross-correlation spin-noise spectroscopy, typical experimental
spin-noise setups have to be complemented by an additional probe laser
with a photon energy distinct from the probe laser of the original
setup \cite{Yang2014,Roy2015}.  Using only one of the probe lasers,
the individual auto-correlation functions
\begin{subequations}
\begin{align}
C_2^{(1)}(t) = \braket{S_z^{(1)}(0)S_z^{(1)}(t)} \\
C_2^{(2)}(t) =\braket{S_z^{(2)}(0)S_z^{(2)}(t)} 
\end{align}
\end{subequations}
of the electron spins are accessible.
A combined measurement of both probe lasers provides
\begin{align}
\label{eq:C2-tot-def}
C_2^{(1+2)}(t) &= \Braket{\left(S_z^{(1)}(0)+S_z^{(2)}(0)\right)\left(S_z^{(1)}(t)+S_z^{(2)}(t)\right)} \notag\\
               &= C_2^{(1)}(t) + C_2^{(2)}(t) + C_2^{(\times)}(t) ,
\end{align}
comprising of the sum of the single autocorrelation functions and
the cross-correlation function
\begin{align}
C_2^{(\times)}(t) &=  \braket{S_z^{(1)}(0)S_z^{(2)}(t)} + \braket{S_z^{(2)}(0)S_z^{(1)}(t)} .
\label{eq:defcross}
\end{align}
The cross-correlation function contains the actual information on the interaction between the QDs excited by the two photon energies.
Within the SCA the evaluation of the cross-correlation function yields
\begin{align}
C_2^{(\times)}(t,(\{\vec{s}_j\}) = \frac{1}{N_C}\sum_\mu s_{z,\mu}^{(1)}(0)s_{z,\mu}^{(2)}(t) + s_{z,\mu}^{(2)}(0)s_{z,\mu}^{(1)}(t) \; . \label{eq:sca_cross}
\end{align}
Since spin operators of different spins commute with each other, we can just replace the spin operators by their classical counterpart, from 
Eq.~\eqref{eq:defcross} to Eq.~\eqref{eq:sca_cross}.

\subsection{Correlation spectra}
\label{sec:spectra}

The spin dynamics in coupled QDs covers a wide range of time
scales as a result of the energy hierarchy of the relevant interactions
which leads to distinct features in frequency space.  We define
the correlation spectra as the
Fourier-transformed time-dependent correlation functions
\begin{align}
\tilde{C}_2^{(\alpha)}(\omega) = \frac{1}{2 \pi} \int_{-\infty}^{\infty} C_2^{(\alpha)}(t) e^{-i \omega t} \mathrm{d}t .
\end{align}
Accordingly, the inverse transformation is given by
\begin{align}
C_2^{(\alpha)}(t) = \int_{-\infty}^{\infty} \tilde{C}_2^{(\alpha)}(\omega) e^{i \omega t} \mathrm{d}\omega .
\end{align}
Note, numerically we are restricted to a finite measurement time $T_m$ which limits the frequency resolution of the Fourier transformation.
 However, we use a measurement time of $T_m=10^4 T^*$, where the relevant dynamics is resolved.

Since the relations $
C_2^{(1)}(0)=C_2^{(2)}(0)= 1 / 4$ and $
C_2^{(\times)}(0)=0$
hold in the high-temperature limit, we deduce the sum-rules \cite{Roy2015}
\begin{subequations}
\begin{align}
\int_{-\infty}^{\infty} \tilde{C}_2^{(1)}(\omega) d\omega &= \frac{1}{4} \label{eq:sumA1}\\
\int_{-\infty}^{\infty} \tilde{C}_2^{(2)}(\omega) d\omega &= \frac{1}{4} \label{eq:sumA2}\\
\int_{-\infty}^{\infty} \tilde{C}_2^{(\times)}(\omega) d\omega &= 0 \; . \label{eq:sumX}
\end{align}
\end{subequations}
While the autocorrelation spectra are always positive due to their
relation to the power spectrum and have a total spectral weight of
$1/4$, the cross-correlation spectrum will only have nonzero
contributions when the QDs interact with each other.  Correspondingly
the sum rule \eqref{eq:sumX} ensures that every positive component in
the cross-correlation spectrum is complemented by a negative
contribution at some other frequency, which will be useful for the
interpretation of the spectra in the upcoming section.

\section{Cross-correlation spin-noise results}
\label{sec:results}

We present results for the cross-correlation spectrum of an
interacting QD system obtained by means of the SCA.
For a better interpretation we discuss the autocorrelation spectrum
along-side the cross-correlation spectrum and highlight the
differences of both.  First we focus on simplified models and
introduce a two-QD reduction (Sec.~\ref{sec:JDist}), use the box model
(Sec.~\ref{sec:QM}) or assume frozen nuclear spins
(Sec.~\ref{sec:FOA}) to explore the generic behavior and extract
analytic expressions.

In Sec.~\ref{sec:parameters} we introduce randomness of
parameters capable to describe a typical semiconductor QD ensemble or
QD molecule.  We study the effects of the distribution of
hyperfine coupling constants, the nuclear quadrupolar interactions and
the electron $g$ factors on the cross-correlation spectrum at zero
magnetic field (Sec.~\ref{sec:QI}), transversal magnetic field
(Sec.~\ref{sec:gfactors}) and longitudinal magnetic field
(Sec.~\ref{sec:longit}), where we observe a rich interplay of the
different interactions on a broad range of time-scales.  Finally, in
Sec.~\ref{sec:exp} we relate our results to previous experiments.

\subsection{Reduction of the QD ensemble to an effective two-QD system}
\label{sec:JDist}

To reduce the numerical effort for the investigation of two-color spin-noise
spectroscopy, we introduce an effective mapping of a QD ensemble to a
system of two representative QDs. 

First we address the original problem of a large QD ensemble to
define the appropriate reference.  We assume, that $N_{QD}$ QDs in the
ensemble are distributed randomly on a 2D square of length $L = \sqrt{N_{QD}} \times 100\,\mr{nm}$
corresponding to a QD density of $n = (100\,\mr{nm})^{-2}$
\cite{Spatzek2011} and implement periodic
boundary conditions.  Due to the QD size, we assume a minimum distance
of $20\, \mr{nm}$ between the centers of the QDs.  The randomly drawn
positions $\vec{R}_i$ of the QD centers are used to calculate the
coupling constants $J_{ij} = J(r_{ij})$ with
$r_{ij}=|\vec{R}_i-\vec{R}_j|$, where we assume an exponentially
decreasing coupling strength \cite{Spatzek2011},
\begin{align}
J(r_{ij}) = \alpha \exp\left(- \frac{r_{ij}}{\rho_0}\right),
\label{eq:expdist}
\end{align}
with the characteristic length scale $\rho_0$ of the interaction and the prefactor $\alpha$, which parameterizes the interaction strength.
Note that since an electron spin does not interact with itself, we set $J_{ii} =0$.

In order to enable the simulation of a large number of QDs in the array, we resort to the so-called
box model approximation \cite{FroehlingGlazov2018}, i.\ e.,
$A_k^{(i)}=A_0$, restrict
ourselves to $N_i=100$ nuclear spins of length $I=1/2$ and, hence, neglect quadrupolar interactions.
This way the complexity of the SCA reduces by the
order $\mathcal{O}(N_i)$ since all nuclear spins of QD~$i$ experience the
same field $\vec{b}_{\mr{eff},k}^{(i)}$ (see Eq.~\eqref{eq:beffk}) and
can be summarized to $\vec{i}^{(i)}_\mr{tot} = \sum_k
\vec{i}^{(i)}_k$.

\begin{figure}[t]
\includegraphics[scale=1]{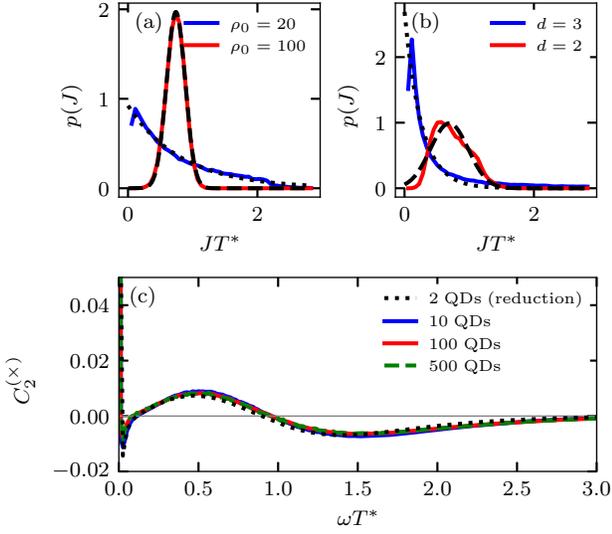}

\caption{
(a) Distribution $p(\overline{J}_i)$ for an exponential distance dependency $J(r_{ij}) \propto \exp( - r_{ij} / \rho_0)$ for two different values of $\rho_0$ measured in $\mr{nm}$.
(b) Power law $J(r_{ij}) \propto r_{ij}^{-d}$ with $d=2$ or $d=3$.
The curves can be approximated by an exponential function (dotted lines)  or a Gaussian function (dashed lines).
(c) The cross-correlation spectra for ensembles of different sizes are compared to a reduced two-QD model. 
For the QD ensembles the distance-dependent coupling constant, Eq.~\eqref{eq:expdist} with $\rho_0=20\,\mr{nm}$, and for the reduced two-QD model the exponential distribution $p(J)$ is used. 
} 
\label{fig:JDistapp}

\end{figure}

By inspecting the EOM, the effect of all other electron spins onto the electron spin of the i-th QD,
\begin{eqnarray}
\label{eq:bj-i}
\vec{b}_J^{(i)}&= &\sum_j J_{ij} \vec{s}^{(j)},
\end{eqnarray}
is identified as an additional noise source. Its expectation value vanishes in the high-temperature limit but its variance
\begin{eqnarray}
\langle (\vec{b}_J^{(i)})^2 \rangle &=& \sum_j J^2_{ij} \langle (\vec{s}^{(j)})^2 \rangle 
= (\bar J_i)^2 \langle S^2 \rangle 
\label{eq:Bj-variance}
\end{eqnarray}
is finite for coupling constants $J_{ij}$ that decay fast enough with the distance 
$r_{ij}$  between the QDs.
In the last step, we defined the quadratically averaged coupling
constant for each QD $i$:
\begin{eqnarray}
\overline{J}_i&=& \sqrt{\sum_j J^2_{ij}} , \label{eq:Jeff}
\end{eqnarray}
which is possible, since all spins have the same length $(\vec{s}^{(j)})^2=S^2$.
In our simulations, we choose $\alpha$ in Eq.\ \eqref{eq:expdist} such
that the averaged fluctuation strength 
\begin{eqnarray}
\bar{J} &=& \frac{1}{N_{\rm QD}} \sum_{i=1}^{N_{\rm QD}} \overline{J}_i = \frac{1}{T^*}
\end{eqnarray}
is fulfilled. This is a reasonable value since experimental data on samples with interacting QDs \cite{Fischer2018} suggests that the dephasing via the QD interaction is of the same order of magnitude as the dephasing by the local Overhauser field.

In the ensemble, $\overline{J}_i$ differs for the individual QDs $i$. Therefore, we not only calculate the average $\bar{J}$ but also 
compile the histogram of all values $\overline{J}_i$ and average over many
different realizations of the randomly generated QD ensembles to 
obtain a probability distribution $p(J)$ for finding the coupling strength $J=\overline{J}_i$
in the i-th QD.

Assuming that we have a microscopic
model that provides the coupling constant $J_{ij}=
J(\vec{R}_i,\vec{R}_j,\rho_0)$ for each QD pair of the original
ensemble, we calculate the effective coupling constant $\bar J_i$
according to Eq.~\eqref{eq:Jeff} for each QD of a random configuration
$\{\vec{R}_i\}$ to obtain the distribution $p(\bar
J_i)$ of the effective coupling constants for a given characteristic
length scale $\rho_0$.  
This distribution $p(\bar J_i)$ is presented for two different
values of $\rho_0$ in Fig.~\ref{fig:JDistapp}(a) using the exponential
form of Eq.\ \eqref{eq:expdist}.  While $p(\bar
J_i)$ can be well approximated by an exponential distribution for a
short-range interaction $\rho_0 = 20\,\mr{nm}$, $p(\bar J_i)$ approaches a Gaussian for a long-range interaction, $\rho_0 = 100\,\mr{nm}$, due to the
central-limit theorem which is displayed as dashed line. 

We added the distribution $p(J)$
for a power law dependency
\begin{align} J(r_{ij}) = \alpha r_{ij}^{-d},
\end{align} with the prefactor $\alpha$ and exponent $d$
as Fig.~\ref{fig:JDistapp}(b).  We observe the
same generic behavior with a short-range interaction for $d=3$ and a
long-range interaction for $d=2$.  Such a power law distance dependency could occur for an RKKY mediated effective coupling when some of the donator charges populate the conduction band of the wetting layer.

To measure the cross-correlation function in the ensemble, we divide the QDs into two classes and replace $S_z^{(i)}$ in Eq.~\eqref{eq:defcross} by the averaged spin of the respective class.
We compare the cross-correlation spectra for ensembles of
different sizes in the absence of an external magnetic field.
In Fig.~\ref{fig:JDistapp}(c) the cross-correlation spectra for different numbers of QDs from $N_{QD}=10$ to $N_{QD}=500$ are presented as colored lines.
We postpone the physical interpretation of the curves to latter
and only notice that the spectra for different ensemble sizes coincide, which is attributed to the short-range nature of the interaction between the QDs.
In the high-temperature limit long-range correlations between the QDs are absent and the consideration of a small cluster is sufficient \cite{Smirnov2014}.

For a further reduction of the system size we note that the influence of the other QDs onto the electron spin dynamics is encoded in a single
effective magnetic field, $\vec{b}_J^{(i)}$ introduced in Eq.\ \eqref{eq:bj-i}.
Since the electron spin dephasing in a QD is governed by the overall fluctuations of
$\vec{b}_J^{(i)}$, we can efficiently simulate the effect of a large
number of QDs onto the spin dynamics by replacing the sum over all other QDs by a single additional QD employing an effective coupling constant
$J=\overline{J}_i$.
The system reduces to an effective two-QD problem where the two QDs represent their respective class, however, we ignore higher-order inter-QD
interactions.
In order to mimic the randomness of the coupling constants $p(\overline{J}_i)$
in the ensemble of QDs, we simulate many such coupled QD pairs by drawing the effective coupling constant $J$ from a distribution $p(J)$ for each classical configuration, such that the distribution of $\vec{b}_J^{(i)}$ remains unchanged.

We added the cross-correlation spectrum for the reduced two-QD model
 as a dotted black line to Fig.~\ref{fig:JDistapp}(c), 
where we use the exponential distribution $p(J)$.  
The agreement between the full simulation of a random array of
QDs and the averaging over a distribution of many effective two-QD systems is remarkable. 
The reduced two-QD model reproduces the
ensemble behavior with only very small deviations even for an average
interaction strength $\bar{J} = 1/T^*$ that is comparable to the
Overhauser field fluctuation scale. In the absence of
an external magnetic field, we find increasing deviations in the
low-frequency part of the spectra once $\overline{J}_i$ significantly
exceeds the Overhauser field strength (not shown here): when a single effective
inter-QD interaction starts to dominate the electron spin correlations,
a larger amount of QDs must be considered in order to capture the
correct physics.

\subsection{Auto- and cross-correlation spectrum in box-model limit}
\label{sec:QM}

To explore the generic behavior of the system we study the auto
and cross-correlation spectrum of two interacting QDs, which either
represent a reduced ensemble, see Sec.~\ref{sec:JDist}, or a QD
molecule.  We stick to the box-model approximation, i.\ e.\ $A_k^{(i)}=A_0$,
restrict ourselves to $N_i=100$ nuclear spins of length $I=1/2$
which enables the comparison to an exact quantum
mechanical approach (QMA).
$\vec{I}_\mr{tot}^{(i)} = \sum_k
\vec{I}_k^{(i)}$ can be replaced in Eq.~\eqref{eq:H_CSM} and the
commutator relation $[ (I_\mr{tot}^{(i)})^2, H]=0$ holds.  The
Hamiltonian becomes block diagonal in the subspaces of fixed quantum
numbers $\{I_\mr{tot}^{(i)}\}$, which reduces the complexity of the
problem considerably \cite{Kozlov2007}
and larger values of $N_i$ become
accessable. In order to relate the SCA
and QMA the rescaled coupling constant $ J' = \sqrt{3/(4
\braket{S^2})} J$ is used where $\braket{S^2}=1/4$ in the SCA and
$\braket{S^2}=3/4$ in the QMA.

\begin{figure}
\includegraphics[scale=1]{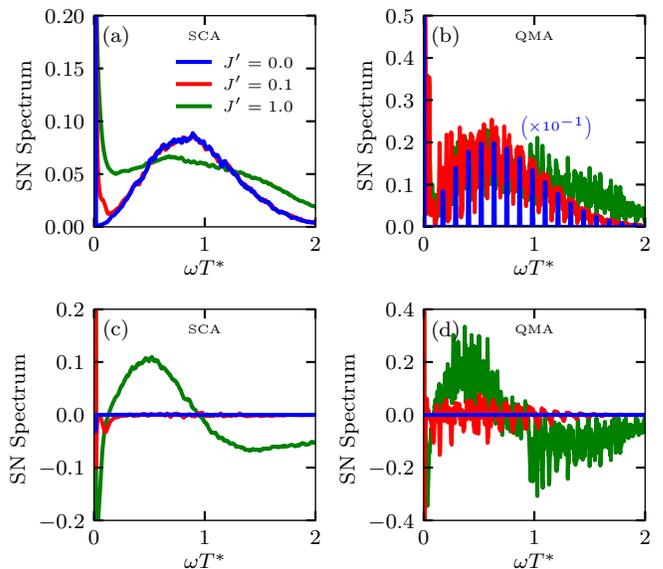}
\caption{Comparison of the SCA (left panels) and QMA (right panels) for fixed $J'$ measured in units of $1/T^*$. The upper panels, (a) and (b), show the autocorrelation spectrum, the lower panels, (c) and (d), depict the related cross-correlation spectrum. For a better presentation, the amplitude of the blue curve in panel (b) is reduced by the factor $10^{-1}$.}
\label{fig:qm}
\end{figure} 

\begin{figure}
\includegraphics[scale=1]{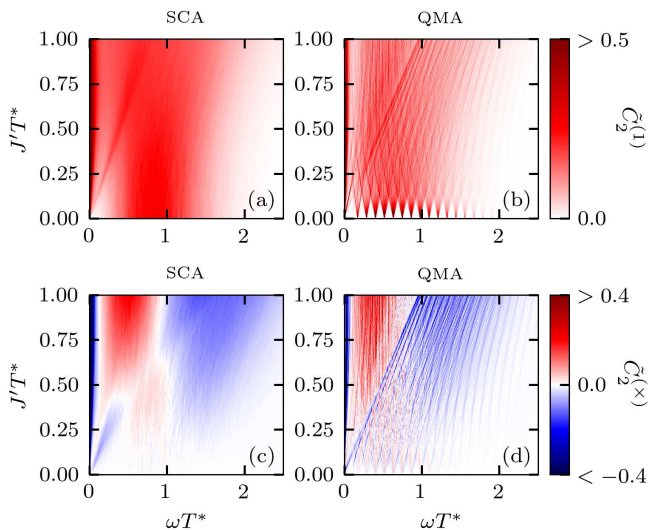}
\caption{Comparison of the SCA and QMA. The upper panels, (a) and (b), show the autocorrelation spectrum, and the lower panels, (c) and (d), depict the cross-correlation spectrum as a function of the frequency $\omega$ and interaction strength $J'$. The left panels, (a) and (c), correspond to the SCA and the right panels, (b) and (d), belong to the QMA.
The magnitude of the correlation function is color coded according to the legends on the right-hand side.}
\label{fig:qm_color}
\end{figure}

In a first step, we focus on the comparison of the autocorrelation
spectra obtained by the SCA and the QMA which are presented in the
upper panels of Figs.~\ref{fig:qm} and~\ref{fig:qm_color}
respectively.  In Fig.~\ref{fig:qm}(a) the autocorrelation spectrum
resulting from the SCA is depicted for various values of the inter-QD
interaction strength $J'$.  Without interactions between the QDs
($J'=0$), the total angular momentum $\vec{F}^{(i)} = \sum_k
\vec{I}_k^{(i)}+\vec{S}^{(i)}$ is conserved for each QD and the
autocorrelation spectrum coincides with the Frozen-Overhauser field
solution of a single QD \cite{Merkulov2002}.  In this case, the
correlation spectrum consists of a delta-peak at $\omega=0$, which
corresponds to the non-decaying part of $C_2(t\rightarrow \infty)$,
and a broader contribution from the Gaussian Overhauser field
fluctuations with location and width governed by $1/T^*$.  While the
Overhauser field contributions are continuous in the SCA, they have a
Dirac-comb substructure in the QMA, see Fig.~\ref{fig:qm}(b), that
reflects the transitions between the equidistant energy levels for a
finite number of quantum mechanical states.  In the box-model
approximation, the quantum mechanical energy levels have a separation
of $A_0 \propto 1/\sqrt{N_i}$, which recovers the continuous spectrum
for $N_i \rightarrow \infty$.

At finite inter-QD interaction strength ($J\neq 0$), the conservation
of the total spin $\vec{F}^{(i)}$ in each individual QD is lifted and
only $\vec{F}_\mr{tot} = \sum_i \vec{F}^{(i)}$ remains conserved.  As
a consequence the sharp lines in the quantum mechanical
autocorrelation spectrum fan out and, except for some noise, the
discrete quantum mechanical spectrum resembles the continuous
semiclassical spectrum for $N_i=100$ spins already.

In order to illustrate the evolution with increasing inter-QD coupling
strength $J'$, we present the autocorrelation spectra as function of
$\omega$ and $J'$ in two-dimensional color plots in
Fig.~\ref{fig:qm_color}(a) and (b).  A strong red color encodes a
large positive value of correlation while a dark blue color
corresponds to negative values of the correlation function, i.e. a
strong anticorrelation.  In the quantum mechanical results in
Fig.~\ref{fig:qm_color}(b), a broadening of the equidistant
$\delta$-peaks for increasing $J'$ is clearly visible which progresses
towards the classical continuum.  As common features of SCA and QMA we
observe three relevant aspects: First, the zero-frequency
$\delta$-peak is broadened in both approaches
(Fig.~\ref{fig:qm_color}(a) and (b)) since $\vec{F}^{(i)}$ and hence
the Overhauser field is not conserved anymore and a long-term decay of
$C_2(t)$ sets in.  Secondly, the zero-frequency peak partially splits
into an additional peak at $\omega=J'$ due to the Heisenberg
interaction between the QDs: There is a splitoff line starting at the
origin and increasing linearly with $J'$ in the $(\omega,J')$ plane.
Thirdly, as soon as $J'$ becomes comparable to $1/T^*$, the
peak related to the Overhauser-field fluctuations broadens, which
implies that the inter-QD interaction enter as an additional dephasing
source and supplement the Overhauser field fluctuations.  In this
simplified picture, the electron spins of adjacent QDs provide an
additional source for the spin noise and act like additional spins in
the nuclear spin bath.  However, in contrast to the nuclear spins,
which provide a nearly static contribution, the adjacent electron spin
is fluctuating which results in a non Gaussian shape for $J'=1/T^*$.

In a next step we examine the cross-correlation spectra obtained by
SCA and QMA, that have been plotted below the autocorrelation spectra
in Figs.~\ref{fig:qm} and \ref{fig:qm_color}.  Without interaction,
$J'=0$, the cross-correlation spectrum is zero, since the individual
QDs are decoupled.  For $J'>0$, a frequency dependent spectrum emerges
whose positive and negative components have the same weights due to
the sum rule, Eq.~\eqref{eq:sumX}.  For $J'T^*\ll 1$, a sharp
zero-frequency peak is found, indicating a synchronization between the
electron spins of different QDs at very long time scales.  With
growing $J'$ this zero-frequency peak splits into three contributions:
a $\delta$-peak with positive spectral weight at exactly $\omega=0$, a
second peak close to $\omega=0$ with negative spectral weight and
another peak with negative spectral weight at $\omega=J'$.  This
behavior is observed in the SCA data in Fig.~\ref{fig:qm_color}(c), as
well as the corresponding QMA results in Fig.~\ref{fig:qm_color}(d).
As an additional feature in the cross-correlation spectrum the
Overhauser field fluctuations, which are responsible for the broad
peak around $1/T^*$ of the autocorrelation spectrum, are reflected by
a positive cross-correlation for the smaller frequencies and a
negative cross-correlation for larger frequencies.  Note, when $J'$
becomes comparable to $1/T^*$, some of the mentioned features might
interfere destructively and are not solely visible anymore.  The
cross-correlation spectrum for the QMA maintains its discrete nature
for $J'>0$, as shown in Fig.~\ref{fig:qm}(d) and
Fig.~\ref{fig:qm_color}(d).  The envelope function, however, is
similar to the semiclassical spectrum supporting the application of
the SCA for large spin systems.

\subsection{Frozen Overhauser field solution}
\label{sec:FOA}

To understand the origin of the different positive and negative
components in the cross-correlation spectrum, shown in
Fig.~\ref{fig:qm_color}, we study a simplified and analytically
solvable system.  We exploit the separation of the time scales between
the electron and nuclear spin dynamics and study the problem on a
short time scale (up to 100 ns) where the nuclear spins can be
considered as frozen.  Thus, approximation errors for small
frequencies, where the assumption is not valid, are expected.  To that
end we start with a Hamiltonian in the electronic subspace for two-QDs
\begin{align}
H_e=\vec{S}^{(1)} \vec{b}^{(1)}+\vec{S}^{(2)} \vec{b}^{(2)} + J \vec{S}^{(1)} \cdot \vec{S}^{(2)}
\end{align}
that are subject to individual time-constant fields
$\vec{b}^{(i)}=\vec{b}^{(i)}_\mr{ext}+\vec{b}_N^{(i)}$.  Without loss
of generality we choose the $z$ axis along the averaged field
$\vec{\overline{b}}= (\vec{b}^{(1)}+\vec{b}^{(2)})/2$ and define the
field deviating from the average $\Delta \vec{b} =
(\vec{b}^{(1)}-\vec{b}^{(2)})/2$.  Then the Hamiltonian of the
electronic subspace reads
\begin{align}
H_e=\overline{b}(S_z^{(1)}+S_z^{(2)}) + \Delta \vec{b} (\vec{S}^{(1)}-\vec{S}^{(2)})+ J \vec{S}^{(1)} \cdot \vec{S}^{(2)} \notag \\
\approx \overline{b} (S_z^{(1)}+S_z^{(2)}) + \Delta b_z (S_z^{(1)}-S_z^{(2)})+ J \vec{S}^{(1)} \cdot \vec{S}^{(2)}  \; , \label{eq:toyHfinal}
\end{align}
where in the last step we neglect the transversal components of the
deviating field.  This approximation is justified for large magnetic
fields.  Details as well as a solution of the Hamiltonian
\eqref{eq:toyHfinal} can be found in Appendix~\ref{sec:apptoy}.  This
model is exactly solvable and can be used to extract the
autocorrelation spectra and cross-correlation spectra.  For the
longitudinal functions we obtain the result
\begin{subequations}
\begin{align}
\braket{S_z^{(1)}(t) S_z^{(1)}} =& \frac{1}{4}+\frac{\alpha^2(1-\alpha^2)}{2} (\cos(\omega_{\parallel} t) -1) \label{eq:FOAlongA}\\
\braket{S_z^{(1)}(t) S_z^{(2)}} =& 
\frac{\alpha^2(1-\alpha^2)}{2} 
(1-\cos(\omega_{\parallel} t))
\label{eq:FOAlongB}
\end{align}
\end{subequations}
with the abbreviations
\begin{align}
\alpha = \frac{1}{\sqrt{2}}\frac{x}{\sqrt{1 + x^2 - \sqrt{1+x^2}}} \; \mathrm{and} \; x = \frac{J}{2 \Delta b_z}. \label{eq:alpha}
\end{align}

The Eqs.~\eqref{eq:FOAlongA} and \eqref{eq:FOAlongB} comprise an oscillating term of frequency 
\begin{align}
\omega_{\parallel} = \sqrt{J^2+ 4 \Delta b_z^2} \label{eq:omega_par}
\end{align} 
as well as a constant part.  Correspondingly, the Fourier
transformation contains a $\delta$-peak at $\omega =
\omega_{\parallel}$ from the oscillatory part and a $\delta(\omega)$
term from the constant part.  In accordance to the sum-rules
\eqref{eq:sumA1}-\eqref{eq:sumX}, both peaks carry positive spectral
weights in the auto-correlation function whereas the cross-correlation
spectrum contains a positive and a negative peak of equal weight.
Note the limit $\alpha \stackrel{J\ll\Delta b_z}{\rightarrow} 1$ ,
which guarantees vanishing cross-correlations for $J=0$.

The transversal correlation functions in $x$ direction yield
\begin{subequations}
\begin{align}
\braket{S_x^{(1)}(t) S_x^{(1)}} =& \frac{\alpha^2}{8} \left(\cos(\omega_{\perp}^{++}t)+\cos(\omega_{\perp}^{-+}t)\right) \notag\\ 
&+ \frac{1- \alpha^2}{8} \left(\cos(\omega_{\perp}^{+-}t)+\cos(\omega_{\perp}^{--}t)\right)\\
\braket{S_x^{(1)}(t) S_x^{(2)}} =& \frac{\sqrt{\alpha^2(1-\alpha^2)}}{8} (-\cos(\omega_{\perp}^{++}t)-\cos(\omega_{\perp}^{-+}t) \notag\\ 
&+ \cos(\omega_{\perp}^{+-}t)+\cos(\omega_{\perp}^{--}t)), 
\end{align}
\end{subequations}
and, due to the spin rotational symmetry around the $z$-axis, they are the same
as those in $y$ direction.  In the transversal case the four
frequencies
\begin{subequations}
\begin{align}
\omega_{\perp}^{\pm +} = \overline{b} \pm \left(\frac{J}{2}+\sqrt{\frac{J^2}{4}+(\Delta b_z)^2}\right) \label{eq:omega_perpA}\\
\omega_{\perp}^{\pm -} = \overline{b} \pm \left(\frac{J}{2}-\sqrt{\frac{J^2}{4}+(\Delta b_z)^2}\right) , \label{eq:omega_perpB}
\end{align}
\end{subequations}
centered around $\overline{b}$, occur.

The terms "longitudinal" and "transversal" are usually defined with
respect to $\vec{b}_\mr{ext}$ and not with respect to
$\vec{\overline{b}}$.  For large $b_\mr{ext} \gg b_N$ this distinction
becomes obsolete, however for small $b_\mr{ext}$ an averaging over an
isotropic distribution of random field $\vec{b}_N^{(i)}$ is required
leading to a mixing of correlation functions in $x$ and $z$ direction.
In the case $b_\mr{ext}=0$, the field $\vec{\overline{b}}$ is
isotropically distributed and the correlation functions of all spatial
directions $x$, $y$ and $z$ mix with equal proportion.  Therefore,
longitudinal and transversal components are comprised in the spectra
in Fig.~\ref{fig:qm_color}.

The frequencies $\omega_{\perp}^{\pm \pm}$ build up the positive and
negative cross-correlations around $1/T^*$.  The longitudinal spectrum
produces the negative cross-correlation at $\omega_\parallel \approx
J'$ and positive cross-correlations at $\omega=0$.  The
positive correlations at $\omega=0$ follow from the fact that the
frozen Overhauser field approximation does not allow the full
relaxation of the electronic spins, even in the presence of their
interactions with each other.  Only the strong anti-correlation at
very small frequencies, visible in Fig~\ref{fig:qm_color}, are absent
within this simplification: The frozen Overhauser field approximation
disregards the slow dynamics of the nuclear spins on long time-scales
and small frequencies. 

Indeed, the sharp $\delta$-peaks in the correlation spectra of the
simplified model are broadened by the randomness of the hyperfine
couplings leading the a continuous spectrum after the averaging as in
Figs.~\ref{fig:qm} and \ref{fig:qm_color}.

\subsection{Randomness of the coupling parameters}
\label{sec:parameters}

In Secs.~\ref{sec:JDist} and \ref{sec:QM} we established that
the simulation of full QD ensemble with a larger number of $N_{QD}$ is
equivalent to a coupled two-QD problem in the SCA supplemented with
appropriate distribution function $p(J)$ with respect to the
calculation of two-color cross-correlation functions.

The SCA allows  studies of a large spin system beyond the simplified box
model and facilitates the introduction of additional interactions
without significantly increasing the computational effort.  In the
following we study a system of $N_{QD}=2$ QDs with $N_i=100$ nuclear
spins each.  It either represents a large QD ensemble or a QD
molecule.  For the nuclear spins we use a spin length $I=3/2$, which
matches the spin length of Ga or As isotopes.  The hyperfine couplings
are proportional to the probability density of the electron at the
location of the nucleus $A_k \propto |\psi(\vec{r}_k)|^2$.  Under the
assumption of a flat 2-dimensional QD and a Gaussian wave function
$\psi(\vec{r}) \propto \exp( -r^2/ (2 L^2) )$ with the characteristic
QD length scale $L$, we obtain the probability distribution
\begin{align} p(A_k) = \frac{L^2}{R^2} \frac{1}{A_k} ,\label{eq:pak}
\end{align} where $R >L$ is the cutoff radius defining the smallest
hyperfine constant $A_\mr{min}$.  Here, we choose the cut-off radius
$R = 2 L$ and the maximum hyperfine constant $A_\mr{max}$ in such a
way that Eq.~\eqref{eq:Tstar} is fulfilled \cite{Hackmann2014}.

We use the exponential distribution for the inter-QD
interaction, see Sec.~\ref{sec:JDist}, with a mean value of
$\overline{J} = 1/T^*$ in accordance to
Refs.~\cite{Spatzek2011,Fischer2018}.  Therefore, we perform
configuration averaging over individual realizations of $A_k^{(i)}$ as
well as the effective coupling constant $J$ between two
QDs~\cite{Fischer2018}.

In addition we include the effect of nuclear-quadrupolar
interactions on the auto- and cross-correlation spectrum at low
frequencies.  We use uniformly distributed coupling constants
$q\in[0,2\overline{q}]$ with the mean value $\overline{q}=0.015
(T^*)^{-1}$ and the easy axis $\vec{n}_Q$ uniformly distributed in a
cone around $\vec{e}_z$ with apex angle $\theta_\mr{max}=1.19$ and a
biaxiality $\eta=0.5$ as a set of nuclear-quadrupolar interaction
parameters introduced in Refs.~\cite{Hackmann2015}.

\subsection{Effect of quduadrupolar interactions}
\label{sec:QI}

\begin{figure}
\includegraphics{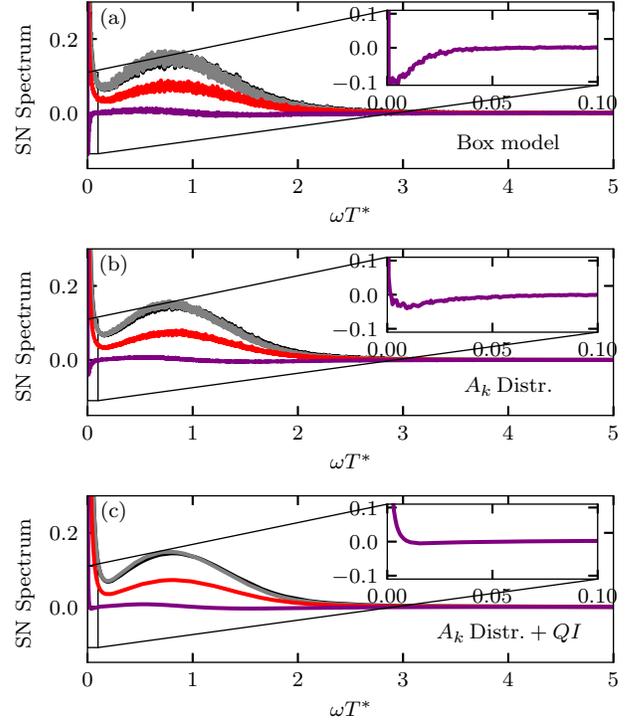} 
\caption{Comparison of the correlation spectra $\tilde{C}_2^{(1)}$ (red), $\tilde{C}_2^{(1+2)}$ (gray), $\tilde{C}_2^{(1)}+C_2^{(2)}$ (black) and $\tilde{C}_2^{(\times)}$ (violet) at ${B_\mr{ext} = 0}$.
Panel (a) uses the box-model approximation while in panels (b) and (c) the distribution in Eq.~\eqref{eq:pak} is included.
The quadrupolar interactions are omitted in panels (a) and (b), and are included in panel (c).
The cross-correlation spectra at low-frequencies are enlarged in the insets.}
\label{fig:b0}
\end{figure} 

In Fig.~\ref{fig:b0} we show various correlation spectra in the
absence of an external magnetic field: the individual
autocorrelation spectrum $\tilde{C}_2^{(1)}$ (red), the total
autocorrelation spectrum $\tilde{C}_2^{(1+2)}$ (gray) and the
cross-correlation spectrum $\tilde{C}_2^{(\times)}$ (violet).
Fig.~\ref{fig:b0}(a) depicts the results for the box-model
approximation with homogeneous hyperfine coupling constants
$A_k^{(i)}=A_0$ as a reference, in Fig.~\ref{fig:b0}(b) we plot the
data with an average over hyperfine coupling constants following the
distribution, Eq.~\eqref{eq:pak}, and in Fig.~\ref{fig:b0}(c) we add
the nuclear-quadrupolar interaction term as well.

As both QDs are modeled with the same $T^*_i$, the
individual autocorrelation functions of the electron spins are the
same, i.e.  $\tilde{C}_2^{(1)}=\tilde{C}_2^{(2)}$.  The
autocorrelation function of an individual electron spin displays two
peaks at $\omega=0$ and $\omega=1/T^*$.  Note, the inter-QD
interaction only manifests in an additional broadening of both peaks,
while the clear indications for inter-QD interactions, such as the
peak at $\omega=J$, are smoothed out due to the randomness of
$J$.

Figure~\ref{fig:b0}(b) illustrates the effect of the distribution of
hyperfine coupling constants.  The Overhauser field is no longer
proportional to the total angular momentum of the nuclear spins as in
the box model and as a consequence a long term decay of the Overhauser
field as well as the auto-correlation function $C_2^{(i)}(t)$ is
possible.  In the auto-correlation spectrum this reflects as a
broadening of the zero-frequency peak, while higher frequency
components are unchanged.  The effect is intensified, when quadrupolar
interactions are taken into account as well, see Fig.~\ref{fig:b0}(c).

The summed $\tilde{C}_2^{(1)}+\tilde{C}_2^{(2)}$ and combined
$\tilde{C}_2^{(1+2)}$ auto-correlation spectra are depicted in black
and gray, respectively.  While their general shape is rather similar
to the auto-correlation spectrum of a single QD, the difference
between the two defining the cross correlations contains additional
information on the interaction between the QDs since it vanishes for
QDs without inter-QD interaction.

In the box-model approximation, see Fig.~\ref{fig:b0}(a), the
cross-correlation spectrum contains a zero-frequency $\delta$-peak,
just like the analytical solution for the FOA presented above.
However, the anti-correlations at $\omega\propto J$ are not visible
due to the $J$ averaging.  The strong anti-correlations visible in the
inset of Fig.~\ref{fig:b0}(a), can be attributed to the slow dynamics
of the nuclear spins.

In a system with an $A_k$ distribution, the broadening of the
zero-frequency peak modifies the cross-correlation spectrum.  The
$\delta$-peak at $\omega=0$ is broadened and partially cancels out the
adjacent anti-correlations at low frequencies, which as a result are
strongly reduced as shown in the inset of Fig.~\ref{fig:b0}(b).

This effect is even more pronounced when quadrupolar interactions are
taken into account.  Nuclear-electric quadrupolar-interactions
originate from the strain induced by the self-assembled growth
process.  In the SCA they are reflected by time dependent fields
acting on each nuclear spin independently.  This introduces an
additional disorder on the Overhauser field and, therefore, a
broadening of the zero frequency peak.  In Fig.~\ref{fig:b0}(c) the
effect of quadrupolar interactions for a realistic set of parameters
\cite{Bulutay2012,Hackmann2015} is depicted.  The zero-frequency peak
is even more broadened.  As a consequence the strong anti
cross-correlations at low frequency vanish completely.

\subsection{Effect of electron g-factor variation}
\label{sec:gfactors}

\begin{figure} 

  \includegraphics[scale=1]{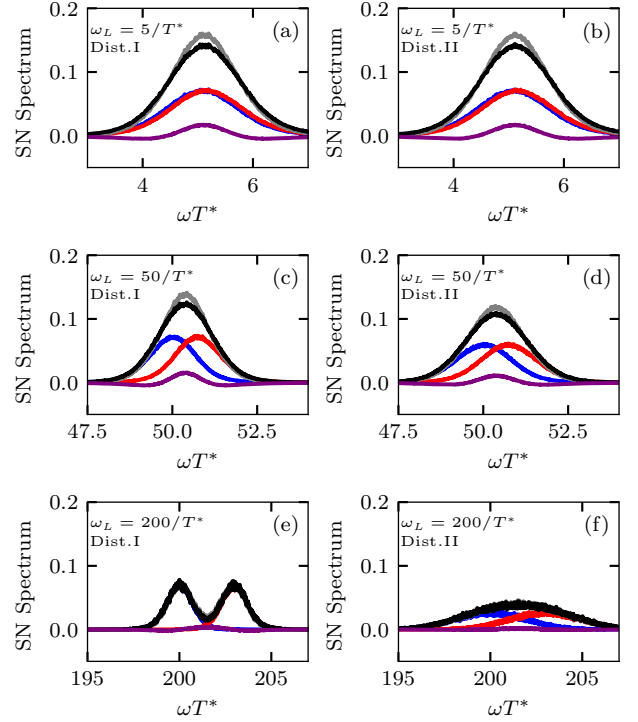}
  
\caption{Comparison of the transversal correlation spectra
$\tilde{C}_2^{(1)}$ (red), $\tilde{C}_2^{(2)}$ (blue),
$\tilde{C}_2^{(1+2)}$ (gray), $\tilde{C}_2^{(1)}+\tilde{C}_2^{(2)}$
(black) and $\tilde{C}_2^{(\times)}$ (violet) for various external
magnetic fields as well as g-factor distributions. The left
panels correspond to distribution I, Eq.~\eqref{eq:gdistI}, the
right panels relate to distribution II, Eq.~\eqref{eq:gdistII}. The
external magnetic fields are $\omega_L = 5 / T^*$
($B_\mr{ext}\approx100\mr{mT}$) in panel (a) \& (b), $\omega_L = 100 /
T^*$ ($B_\mr{ext}=2\mr{T}$) in panel (c) \& (d) and $\omega_L = 200 /
T^*$ ($B_\mr{ext}=4\mr{T}$) in panel (e) \& (f).}
\label{fig:gdist}
\end{figure}

In this section we introduce a transverse external magnetic field
acting on electron spins as well as nuclear spins.  The electron spins
precess in the external field with a Larmor frequency ${\omega_L =
\mu_B g_0 {B}_\mr{ext}}$, where $g_0=0.5$ is the typical effective
$g$-factor of the electron in a InGaAs QD. 
For the nuclear spins we use the value $g_k^{(i)} \mu_I / g_0 \mu_B \approx 1 / 800$ averaged over the different isotopes \cite{Beugeling2017}.
 At large external magnetic
fields, the spectral weight of the Gaussian shaped peak at
$\omega\approx 1/T^*$ in the auto-correlation spectrum is shifted by
this Larmor frequency.  In the FOA this can be directly seen in the
frequencies stated in Eq.~\eqref{eq:omega_perpA} and in Eq.~\eqref{eq:omega_perpB}:
They  are shifted by $\overline{b}$.  Consequently, effects relevant on
long time-scales such as the distribution of hyperfine coupling
constant or the nuclear quadrupolar interactions are suppressed.  Only
in spin-echo experiments \cite{Bechtold2016,Press2010} described by
fourth-order spin-spin correlation functions \cite{Froehling2017} a
second dephasing time associated with these interactions occurs. 
Their effect is not observable in the second order correlations .

In a transversal field, however, the distribution of electron $g$
factors becomes relevant.  Since the $g$ factor of an electron spin in
a semiconductor is correlated with the excitation energy of the QD due
to the Roth-Lax-Zwerdling relation, we assume that both QD
subensembles have a different average $g$ factor
\cite{Yugova2009,Fischer2018}. The optical selection of QDs used for
the measurement of cross-correlation functions \cite{Roy2015} is
facilitated by the usage of probe lasers with different photon
energies.

An additional variation of the $g$ factor for a fixed excitation
energy is caused by local differences in a self-assembled QD ensemble.
In order to study (i) the effect of a different  
g-factors in each subensemble as well as (ii) the $g$-factor variation
within each subensemble
we define the $g$-factor
distribution I
\begin{align}
g^{(1)} &= g_0 \notag \\
g^{(2)} &= 1.015 g_0 \label{eq:gdistI}
\end{align}
and $g$-factor distribution II
\begin{align}
g^{(1)} &\sim \mathcal{N}(g_0, 0.005 ) \notag\\
g^{(2)} &\sim \mathcal{N}(1.015 g_0, 0.005 )\label{eq:gdistII}
\end{align}
where we use parameters extracted from literature
\cite{Fischer2018,Schwan2011}.  While for the  distribution~I 
the electrons 1 and
2 have different but fixed $g$ factors, the $g$
factor is drawn from a Gaussian distribution with a very small width
\cite{Fischer2018}  for the distribution II.

The impact of both distributions on the spectral functions is
presented in Fig.~\ref{fig:gdist} for different transverse external
magnetic fields.  The left panels belong to distribution~I whereas the
right panels use distribution~II.  Various correlation spectra at
small magnetic fields of $\omega_L = 5 / T^*$ ($B_\mr{ext}\approx
100\mr{mT}$) applied in the $x$-direction are depicted in
Figs.~\ref{fig:gdist}(a) and \ref{fig:gdist}(b) using the color coding
of Fig.\ \ref{fig:b0} with an additional blue line for the individual
auto-correlation function $\tilde{C}_2^{(2)}$ (that now differs from
$\tilde{C}_2^{(1)}$).  For this small field the $g$-factors
distribution is rather irrelevant, since the energy scale
$\Delta\omega=B_\mr{ext} \mu_B (g^{(1)}-g^{(2)})$ is small in
comparison to the other energy scales of the system.  Due to the
transversal magnetic field the auto-correlation spectrum (red  and blue
line) has a Gaussian shape \cite{Merkulov2002} and is centered around
the Larmor frequency of the respective QD.
Without inter-QD interaction, the width of the Gaussian is given by
the Overhauser field fluctuations as well as the variation of
$g$-factors.  However, the latter effect is not relevant here which
leaves $1/T^*$ as the only influence.

As the auto-correlation spectrum, the cross-correlation spectrum
(purple line) is centered around the Larmor frequency $\omega_L$.  It
is positive in the center and is flanked by negative contributions at
the wings.  This behavior can be understood within the FOA.  Relevant
for the cross-correlation spectrum are the four frequencies in the
Eqs.~\eqref{eq:omega_perpA} and \eqref{eq:omega_perpB}.  While the two
outer frequencies $\omega_\perp^{\pm +}$ have negative prefactors, the
two inner frequencies $\omega_\perp^{\pm -}$ have a positive
contribution to the cross-correlation spectrum.  The sharp lines of
the single configuration FOA analysis presented above are broadened
due to the Overhauser-field fluctuations.

In a larger magnetic field, $\omega_L = 50 / T^*$ ($B_\mr{ext}\approx
1\mr{T}$), the splitting of the peak location due to the different
$g$ factors is clearly visible in the auto-correlation functions of
both QDs in Fig.~\ref{fig:gdist}(c).  The cross-correlation function,
however, retains its shape.  Its amplitude is significantly reduced by
the $g$ factor distribution II as can be seen in
Fig.~\ref{fig:gdist}(d).  The four lines obtained from the FOA, smear
out due to the $g$-factor spreading and cancel each other.

For an even larger field, $\omega_L = 200 / T^*$ ($B_\mr{ext}\approx
4\mr{T}$), the auto-correlation spectra of the different QDs shown in
Fig.~\ref{fig:gdist}(e) separate since the difference in the Larmor
frequencies $\Delta\omega$ becomes significant.  The cross-correlation
strongly reduces for distribution~I, since the prefactor $\alpha$
vanishes for $J/\Delta b \rightarrow 0$.  By taking into account
distribution II, the cross-correlation spectrum is almost completely
washed out as depicted in Fig.~\ref{fig:gdist}(f).

In a transversal magnetic field, dephasing effects on long time scales
such as caused by the quadrupolar interactions are suppressed since
the cross-correlation spectrum is shifted to higher frequencies.  The
spectral features in the vicinity of the Larmor frequency, however,
are suppressed in strong magnetic fields by a $g$-factor disorder
leading to an attenuation of the cross-correlation spectrum.  However,
at small magnetic fields neither long time dephasing nor the
$g$-factor dispersion attenuates the cross-correlation spectrum.

\subsection{Longitudinal magnetic field}
\label{sec:longit}

\begin{figure} 
\includegraphics{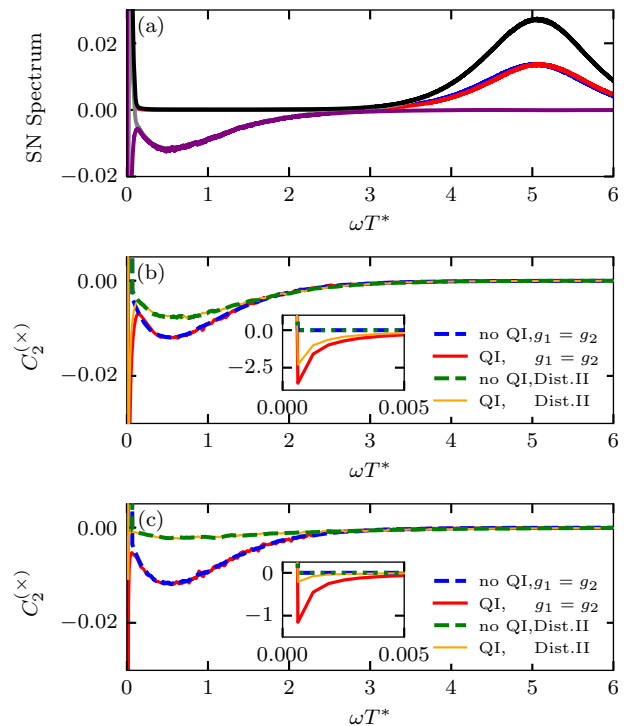}
\caption{In panel (a) the correlation spectra $C_2^{(1)}$ (blue),
$C_2^{(2)}$ (red), $C_2^{(1+2)}$ (gray), $C_2^{(1)}+C_2^{(2)}$ (black)
and $C_2^{(\times)}$ (violet) are depicted for a longitudinal magnetic
field of $\omega_L=5 / T^*$. In the lower panels the cross-correlation
spectrum $C_2^{(\times)}$ is depicted for the four combinations of:
(i) quadrupolar interactions switched on and off and (ii) identical
electron $g$-factors or distribution II, Eq.~\eqref{eq:gdistII}. Panel
(b) presents $\omega_L=50 / T^*$ and panel (c) $\omega_L=200 / T^*$.}
\label{fig:longit}
\end{figure}

In this section, we present the results for the cross-correlation
spectrum in a longitudinal external magnetic field, where both long-
and short-time effects are apparently relevant.  In
Fig.~\ref{fig:longit}(a) the correlation spectra in a weak
longitudinal field of $\omega_L=5 / T^*$ are depicted using the same
color code as in the previous figures. We included the distribution of
hyperfine coupling constants, the quadrupolar interactions and the
$g$-factor distribution II in the dynamics.

For the auto-correlation spectra (red and blue), we observe a
zero-frequency peak that is associated with the conserved spin
projection along the external magnetic field axis in FOA, broadened
due to the dephasing induced by quadrupolar interactions.  In
addition, some intermixing of the transversal spin components is
visible in the peak centered around the Larmor frequency.

To understand the cross-correlation spectrum (violet) we again refer
to the FOA results in Eq.~\eqref{eq:FOAlongB}.  It predicts a positive
cross-correlation at $\omega=0$ and an anti-correlation at
$\omega=\omega_\parallel$.  For large $J$ we can approximate
$\omega_\parallel\approx J$, accordingly the cross-correlations are
sensitive to the distribution of coupling constants $p(J)$.
Therefore, cross-correlation spin-noise spectroscopy in a longitudinal
field provides the possibility to study the distribution of inter-QD
couplings in an ensemble, an information that is not otherwise
accessible.

We present the
cross-correlation spectra for longitudinal fields $\omega_L=50 / T^*$
in Fig.~\ref{fig:longit}(b) and $\omega_L=200 / T^*$ in
Fig.~\ref{fig:longit}(c) for the four combinations of (i) quadrupolar
interactions included or not, and (ii) identical electron $g$-factors
or distribution II.  The low frequency behavior is enlarged in the
insets.  The positive cross-correlations predicted by the FOA, can be
found as the $\delta$-peak at $\omega=0$.  In addition, we find strong
anti-correlations at small frequencies (orange) that we assign to the
nuclear-quadrupolar interactions, that are absent in the FOA results,
since the nuclear spins are considered frozen.  The quadrupolar
interactions, however, only plays a role at very low frequencies.

In contrast the distribution of $g$-factors does not change the shape
of the cross-correlation spectrum, but lowers the amplitude.  This
attenuation effect is not as strong as in the transversal magnetic
field, since no destructive interference between positive and negative
lines occur and only enters via the prefactor $\alpha$
Eq.~\eqref{eq:alpha}.  At larger field of $\omega_L=200 / T^*$ the
effects of quadrupolar interactions and $g$-factor distribution are
modified.  The nuclear Zeeman term suppresses the effect of the
quadrupolar interactions
\cite{Glasenapp2016} which is indicated by a weakening of the peak
caused by the quadrupolar interactions.  In addition the attenuation
effect of the $g$-factor distribution is enhanced strongly reducing
the signal.

\subsection{Connection to experiment}
\label{sec:exp}

The auto-correlation spectrum $C_2^{(i)}$ corresponds to the noise
power-spectrum that is experimentally \cite{Glasenapp2016} extensively
studied and already well understood in terms of a single QD picture,
where inter-QD interactions are neglected
\cite{Merkulov2002,Hackmann2015}.  In our framework we demonstrated
that the auto-correlation spectrum qualitatively remains unchanged by
the inter-QD interaction only the broadening of the peaks is modified.
These modifications can be absorbed in a renormalization of $T^*$
translating into the parameters in a single-QD picture, so that all
previous results can be embedded into the extension to interacting
QDs.

Recently, the auto-correlation spectra $C_2^{(1)}$, $C_2^{(2)}$ and
$C_2^{(1+2)}$ are measured for a QD ensemble at zero magnetic field to
determine the homogeneous line widths \cite{Yang2014}.  Since the
cross-correlation function $C_2^{(\times)}(t)$ is determined by the
difference between the total two-color signal and the sum of the
individual contributions according to Eq.\ \eqref{eq:C2-tot-def},
$C_2^{(\times)}(t)$ could be extracted.

The experimental data in Ref. \cite{Yang2014} qualitatively
agrees with our results, as the cross-correlation spectrum is
strongly attenuated at low frequencies by quadrupolar interactions and
the hyperfine distribution -- see Fig.~\ref{fig:b0}(c).  One has to
bear in mind, however, that only 10\% of the QDs were charged in the
sample studied in Ref. \cite{Yang2014}, which strongly reduces
spin-spin interactions between the QDs.

\begin{figure}[t]
\includegraphics[scale=1]{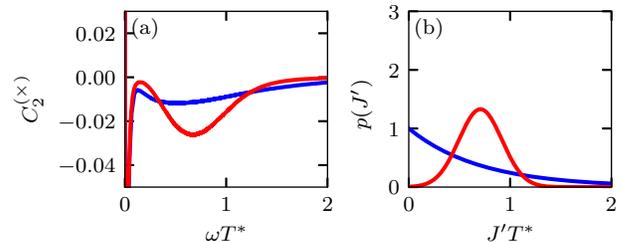}

\caption{Effect of the distribution $p(J')$ on the cross-correlation
spectrum in a longitudinal magnetic field of $\omega_L = 5 /
T^*$. Panel (a) depicts the cross-correlation spectrum while panel (b)
shows the associated distribution function $p(J')$. The red curves
belong to a Gaussian distribution with mean value $\overline{J}=1/T^*$
and standard deviation $\sigma=0.3 /T^*$. The blue curves belong to an
exponential distribution with $\overline{J}=1/T^*$.}
\label{fig:JDist}
\end{figure}

As a consequence of our investigation we propose to measure 
the cross-correlation spectrum at a finite but low external magnetic
field, where additional dephasing by the $g$-factor distribution is
negligible.  The strongest cross-correlation signal is expected for a
transversal magnetic field, for which the quadrupolar interaction is
suppressed.

>From the longitudinal cross-correlation spectrum, additional
information about the distribution of the interaction strength $J$ can
be obtained, even though the signal strength is expected to be lower
than in the transverse case.  We present the cross-correlation
spectrum for two different distribution function $p(J)$ in
Fig.~\ref{fig:JDist}.  The cross-correlation function obtained from a
Gaussian distribution $p(J)$ and depicted in red significantly
deviated from the one calculated with a exponential distribution
plotted in blue.  This way, one can extract additional details
about the microscopical origin  and the characteristic length scale
of the inter-QD interactions.

\section{Conclusion}

We studied the spin-noise spectrum for an ensemble of singly charged
QDs whose resident electron spins interact via a Heisenberg type
interaction.  To this end, we employed a SCA that
allows for the description of coupled spin systems comprising a larger
number of spins than e.g. would be manageable in straight forward
quantum mechanical calculations.  The approach is based on the
introduction of the classical Hamilton function derived from an
arbitrary quantum mechanical Hamiltonian via spin-coherent states.
This procedure accounts for a correct description of terms in the
Hamiltonian that are quadratic in the spin operators as it occurs in
the case of nuclear quadrupolar interactions.

For two-color spectroscopy we were able to map the ensemble problem
onto a pair of coupled QDs augmented with a coupling constant
distribution.  Studying the mapped system, we find features in the
spin-noise spectrum related to the inter-QD interaction.  In the
absence of an external magnetic field,  the
hyperfine interaction and the nuclear quadrupolar interactions
conceal a potential effect of an inter-QD interaction on the
correlation spectra.

The application of an external magnetic field either
transverse or longitudinal to the measurement axis allows for the
separation of the inter-QD interaction from other contributions in the
cross-correlation spectra.  In case of a transversal  external magnetic field, the field strength has to be adjusted in
an intermediate region in order to achieve a significant shift of the
Larmor frequencies for the two electron spins due to their slightly
differing $g$ factors, but not to suppress the effect of inter-QD
interaction completely.  However, the  longitudinal
field may provide easier access.  Here, the contribution of the
inter-QD interaction in the cross-correlation spectrum is strongly
pronounced over a broad range of the magnetic field strengths.  The
shape of the cross-correlation spectrum can even provide information
on the distribution of the coupling strength and thereby  hint on the
unclear physical origin of the interaction.  Experimental 
cross correlation data could be an interesting step towards further understanding to the
interaction between electron spins in QD ensembles.

\begin{acknowledgments} The authors would like to thank M. Bayer, K. Deltenre, A. Greilich and N. J\"aschke for fruitful discussions.  We
acknowledge financial support by the Deutsche Forschungsgemeinschaft
and the Russian Foundation of Basic Research through the transregio
TRR 160 within the Projects No. A4, and No. A7.  The authors
gratefully acknowledge the computing time granted by the John von
Neumann Institute for Computing (NIC) under Project HDO09 at the
J\"ulich Supercomputer center. N. Sinitsyn acknowledges support from 
LDRD program at LANL.

\end{acknowledgments}

\appendix

\section{Derivation of the SCA from a path integral formulation}
\label{sec:appPathInt}

We start with the propagator
\begin{align}
K(\{\vec{s}_{j,f}\},\{\vec{s}_{j,i}\},t) &=   \bra{\{\vec{s}_{j,f}\}} e^{-i H t} \ket{\{\vec{s}_{j,i}\}} 
\end{align}
which provides the transition amplitude from the initial state
$\ket{\{\vec{s}_{j,i}\}}$ at time $0$ to the final state
$\ket{\{\vec{s}_{j,f}\}}$ at time~$t$. $\{\vec{s}_{j,f}\}$ denotes the set of all electron and nuclear spins of
the ensemble.

We decompose the time evolution operator into small time intervals $\Delta t = t / N$
\begin{align}
K(\{\vec{s}_{j,f}\},\{\vec{s}_{j,i}\},t)
 = \lim\limits_{N \rightarrow \infty}\bra{\{\vec{s}_{j,f}\}} \prod_{n=0}^{N-1} e^{-i H \Delta t}\ket{\{\vec{s}_{j,i}\}}  \; .
\end{align}
Now we can insert the completeness relation \eqref{eq:comp} between all infinitesimal time evolution operators to get infinitesimal transition amplitudes
\begin{align}
K(\{\vec{s}_{j,f}\},\{\vec{s}_{j,i}\},t)
= \lim\limits_{N \rightarrow \infty} \int \prod_{n=1}^{N-1} d\mu(\{\vec{s}_{j,n})  
\nonumber
\\
\times
\prod_{n=0}^{N-1} \bra{\{\vec{s}_{j,n+1}\} } e^{-i H \Delta t} \ket{\{\vec{s}_{j,n}\}} 
\label{eq:B3}
\end{align}
with $\ket{\{\vec{s}_{j,0}\}}=\ket{\{\vec{s}_{j,i}\}}  $ and $\ket{\{\vec{s}_{j,N}\}}  =  \ket{\{\vec{s}_{j,f}\}}$.

Next, we estimate the transition amplitude
for very small time steps, $\bra{\{\vec{s}_{j,n+1}\} } e^{-i H \Delta t} \ket{\{\vec{s}_{j,n}\}}$. Using the Taylor expansion of
the exponential up to linear order, and including the difference between $\langle \{\vec{s}_{j,n+1}\}  \ket{\{\vec{s}_{j,n}\} }$
and   $1=\langle \{\vec{s}_{j,n}\}  \ket{\{\vec{s}_{j,n}\} }$, we obtain
\begin{widetext}
\begin{eqnarray}
 \bra{\{\vec{s}_{j,n+1}\} } e^{-i H \Delta t} \ket{\{\vec{s}_{j,n}\}}
 &=&
\langle \{\vec{s}_{j,n+1}\}  \ket{\{\vec{s}_{j,n}\} }
- i\Delta t  \bra{ \{\vec{s}_{j,n+1}\}}  H  \ket{ \{\vec{s}_{j,n}\}}  + O(\Delta t^2)
 \nonumber \\
 &=&
 e^{-i \left( i \bra{ \{\vec{s}_{j,n}\}}\frac{d}{dt}\ket{ \{\vec{s}_{j,n}\}} + \bra{ \{\vec{s}_{j,n}\}}H \ket{ \{\vec{s}_{j,n}\}} \right) \Delta t}
 + O(\Delta t^2).
\end{eqnarray}

Making use of the fact that $\Delta t$ is arbitrary small and inserting the exponential into the Eq.~\eqref{eq:B3}
yields
\begin{eqnarray}
K(\{\vec{s}_{j,f}\},\{\vec{s}_{j,i}\},t)
&=& \lim\limits_{N \rightarrow \infty} \int \prod_{n=1}^{N-1} d\mu(\{\vec{s}_{j,n}\})
e^{-i \sum_{n=0}^{N-1}\left(i \bra{ \{\vec{s}_{j,n}\}}\frac{d}{dt}\ket{ \{\vec{s}_{j,n}\}} + \bra{ \{\vec{s}_{j,n}\}}H \ket{ \{\vec{s}_{j,n}\}} \right) \Delta t}
\nonumber
\end{eqnarray}
\end{widetext}
At this point the transition amplitude can be identified as a path integral that can be comprehensively written as
\begin{align}
K(\{\vec{s}_{j,f}\},\{\vec{s}_{j,i}\},t)&=  \int \mathcal{D}[\{\vec{s}_j\}]\; e^{i S[\{\vec{s}_j\}, \{\partial_t \vec{s}_j\}]}
\end{align}
with the action
\begin{align}
\label{eq:action}
S[\{\vec{s}_{j}\}, \{\partial_t \vec{s}_{j}\}] =  
 \int \left(i \bra{\{\vec{s}_j\}}\frac{d}{dt}\ket{\{\vec{s}_j\}} - \bra{\{\vec{s}_j\}}H \ket{\{\vec{s}_j\}}\right) dt,
\end{align}
and the integration measure
\begin{align}
\mathcal{D}[\{\vec{s}_j\}]= \lim\limits_{N \rightarrow \infty}  \prod_{n=1}^{N-1} d\mu(\{\vec{s}_{j,n}\}) \; .
\end{align}
This path integral formulation of the transition amplitude was used to derived the SCA using a saddle-point approximation leading to the Euler-Lagrange Eq.~\eqref{eq:eom} for the coherent state
using the action $S$.

\section{Quadrupolar Hamiltonian}
\label{sec:appQI}

The quadrupolar interaction acting on a single nuclear spin is quantum mechanically described by the Hamiltonian
\begin{align}
H_Q = q \left[ \left( \vec{I} \cdot \vec{n} \right)^2 + \frac{\eta}{3} \left( \left( \vec{I} \cdot \vec{n}_x \right)^2 - \left( \vec{I} \cdot \vec{n}_y \right)^2 \right) \right].
\label{eq:hq}
\end{align}
In the following, our aim is to derive the semiclassical EOM for the classical nuclear spin from this Hamiltonian.

First, we have to calculate the classical Hamilton function $\tilde{H}(\vec{i})$ with the classical nuclear vector $\vec{i}$
entering the action S defined in Eq.~\eqref{eq:action}.
Since the quadrupolar Hamiltonian consists of three similar terms, we evaluate a term of the general form
\begin{align}
	\bra{\vec{i}} ( \vec{I} \cdot \vec{m} )^2 \ket{\vec{i}} &=  \bra{I,I} e^{ i (\vec{\theta} \times\vec{n}_0) \cdot \vec{I} } ( \vec{I} \cdot \vec{m} )^2  e^{ -i (\vec{\theta} \times \vec{n}_0) \cdot \vec{I} } \ket{I,I} \notag\\
	&= \bra{I,I} ( \vec{I} \cdot \tilde{\vec{m}} )^2  \ket{I,I} \; .
\end{align}
>From the first to the second line, we insert the rotation that relates 
the spin-coherent state $\ket{\vec{i}}$ to the spin state with maximum quantum spin number in the $z$ component $\ket{I,I}$.
These two unitary operators are used to rotated the orientation vector $\vec{m}$ to  $\tilde{\vec{m}}$, so that the total
matrix element remains invariant.

In a next step, we insert the spin operator
\begin{align}
\vec{I} = \left( I^x, I^y, I^z \right)^T = \left( \frac12 \left( I^+ + I^- \right), \; \frac{1}{2i} \left( I^+ - I^- \right), \; I^z \right)^T
\end{align}
leading to
\begin{align}
	\bra{\vec{i}} ( \hat{\vec{I}} \cdot \vec{m} )^2 \ket{\vec{i}} &= \bra{I,I} \frac14 \left[ (\tilde{m}^x)^2 + (\tilde{m}^y )^2 \right] I^+ I^-  \notag \\ 
	&+ (\tilde{m}^z)^2 (I^z)^2 \ket{I,I} \; .
\end{align}
We insert $\hat{I}^+ \hat{I}^- \ket{I,I} = 2 I \ket{I,I}$ and $(\hat{I}^z)^2 \ket{I,I} = I^2 \ket{I,I}$ and obtain
\begin{align}
	\bra{\vec{i}} ( \hat{\vec{I}} \cdot \vec{m} )^2 \ket{\vec{i}} 
	&= \frac{I}{2} + \left( 1 - \frac{1}{2I} \right) (\tilde{m}^z)^2 I^2 \notag\\
	&= \frac{I}{2} + \left( 1 - \frac{1}{2I} \right) \left( \tilde{\vec{m}} \cdot \vec{e}^z I \right)^2.
\end{align}
After rotating back into the original coordinate space, the final result reads
\begin{align}
	\bra{\vec{i}} ( \hat{\vec{I}} \cdot \vec{m} )^2 \ket{\vec{i}} &= \frac{I}{2} + \left( 1 - \frac{1}{2I} \right) \left( \vec{i} \cdot \vec{m} \right)^2 \; .	
\end{align}

When we apply this relation to the three terms of the quadrupolar Hamiltonian in Eq.~\eqref{eq:hq}, the classical Hamilton function entering the action $S$, derived from the quantum mechanical operator $\hat{H}_Q$, is given by
\begin{align}
	\tilde{H}_Q =& q \left\{ \frac{I}{2} + \left( 1 - \frac{1}{2I} \right) \left( \vec{i} \cdot \vec{n} \right)^2  \right. \notag\\
	&\left. + \frac{\eta}{3} \left( 1 - \frac{1}{2I} \right) \left[ \left( \vec{i} \cdot \vec{n}_x \right)^2 - \left( \vec{i} \cdot \vec{n}_y \right)^2 \right] \right\} \; .
\end{align}
$\tilde{H}_Q$ has been used to obtain the quadrupolar field in Eq.~\eqref{eq:b_Q} by applying
the general EOM stated in Eq.~\eqref{eq:eom}.

\section{Auto-correlation function in semiclassical form}
\label{sec:approt}

Let us transform the auto-correlation function
\begin{align}
C_2(t) = \left< S^z(0) S^z(t) \right> .
\end{align}
into a suitable form  for the SCA.
As a demonstration, we start by the calculation of $C_2$ for a single isolated spin $1/2$ and only then generalize the results to  more complex systems.
The most general time-dependent Hamiltonian for a single spin $1/2$ 
has four degrees of freedom and can be parameterized by $H(t) = \vec{B}(t) \cdot \vec{S} + E_0(t)$, whereby $E_0(t)$ solely creates a global phase factor that does not have any impact in our case.

The time evolution is governed by the  operator
\begin{align}
U(t) &= \mathcal{T} \exp\left( -i\int_0^t H(t) dt \right) \notag \\
&= e^{i \varphi(t)} \exp\left( - i \, \vec{\alpha}(t)\cdot \vec{S} \right) = e^{i \varphi(t)} R_{\vec{\alpha}(t)} ,
\label{eq:ut}
\end{align}
with the time-ordering operator $\mathcal{T}$ and the global phase $e^{i \varphi(t)}$.
Since $U(t)$ is an element of the rotation group SU(2), we have replaced the time-ordered integral in the second line of Eq.~\eqref{eq:ut} by a rotation along a specific axis $\vec{\alpha}\in \mathbb{R}^3$, whose explicit calculation shall be postponed for the moment.
We use the latter representation of $U(t)$ to calculate the correlator $C_2$ with respect to an initial spin-coherent state $\ket{\vec{s}_0}$
\begin{align}
C_2 &= \bra{\vec{s}_0} S^z(0) S^z(t) \ket{\vec{s}_0}\\
&= \bra{\vec{s}_0} (\vec{e}_z \cdot \vec{S}) R_{-\vec{\alpha}(t)} (\vec{e}_z \cdot \vec{S}) R_{\vec{\alpha}(t)} \ket{\vec{s}_0} .
\end{align}
In a next step we make use of the relation
\begin{align}
R_{-\vec{\alpha}} (\vec{e}_z \cdot \vec{S}) R_{\vec{\alpha}} &= (\tilde{R}_{-\vec{\alpha}} \vec{e}_z) \cdot \vec{S} \;,
\end{align}
that translates a rotation $R_{\vec{\alpha}}$ of the SU(2) to a $3\times3$ rotation matrix $\tilde{R}_{-\vec{\alpha}}= \exp( -\left[\vec{\alpha}\right]_x )$ of the SO(3), using the cross-product matrix $\left[\vec{\alpha}\right]_x$.
This yields
\begin{align}
C_2
&= \bra{\vec{s}_0} (\vec{n}_z(0) \cdot \vec{S}) (\vec{n}_z(t) \cdot \vec{S}) \ket{\vec{s}_0} ,
\label{eq:c2s0}
\end{align}
with the rotated $z$ axis $\vec{n}_z(t) = \tilde{R}_{-\vec{\alpha}(t)} \vec{e}_z$.

Finally, the auto-correlation function, Eq.~\eqref{eq:c2s0}, is converted to a fully classical form.
To this end, standard spin $1/2$ algebra, i.e. the relation $S^\alpha S^\beta = (1/4) \delta_{\alpha \beta} + (i/2) \sum_\gamma \epsilon_{ \alpha \beta \gamma} S^\gamma$, and the relation $\bra{\vec{s}_0} \vec{S} \ket{\vec{s}_0}=\vec{s}_0$ are exploited, such that the auto-correlation function for the single spin 1/2 becomes
\begin{align}
C_2 &= \frac{1}{4} \vec{n}_z(0) \cdot \vec{n}_z(t)  + \frac{i}{2} (\vec{n}_z(0) \times \vec{n}_z(t)) \cdot \vec{s}_0 \label{eq:appc2}
\end{align}
for a fixed initial coherent state $\ket{\vec{s}_0}$ and the fixed time $t$.

This leaves us with the calculation of the rotation matrix as function of time.
By identifying $\vec{B}(t)$ in the Hamiltonian of the single spin 1/2 with the effective field $\vec{b}_\mr{eff}(t) = \frac{\partial\tilde{H}(\vec{s})}{\partial \vec{s}}$, we can interpret Eq.~\eqref{eq:appc2} as a recipe for evaluation of $C_2$ in a more complex system described by the Hamilton function $\tilde{H}(\vec{s})$ in the semiclassical limit.

In a last step, we present an efficient way to compute the rotation matrix $\tilde{R}_{\vec{\alpha}}$ within the SCA.
Instead of calculating the rotation matrix directly, we use its quaternionic representation \cite{Girard_1984}, which is more compact and numerically stable.
Unit quaterions are isomorphic with the group SU(2), which becomes particular clear, considering that the basic quaternions $\{1,i,k,l\}$ and the Pauli matrices $\{\mathbb{1},i \sigma_1,i \sigma_2,i \sigma_3\}$ share the same algebra.
However, unit quaternions can also be used to represent the SO(3) rotation group due to the two-to-one homomorphism of the SU(2) onto the SO(3).
A SO(3) rotation $\tilde{R}_{\vec{\alpha}}$ can be represented by the unit quaternion
\begin{align}
	Q(\vec{\alpha}) = \cos( \alpha/2 ) + \frac{\vec{\alpha}}{\alpha} \cdot \vec{j} \sin( \alpha / 2 ) \; ,
\end{align}
with the basic quaternions $\vec{j} = (i,k,l)^T$.
As the SO(3) is covered twice in the SU(2), the quaterions $Q$ and $-Q$ represent the same SO(3) rotation.
It can be shown that the desired rotation can be applied to a vector  with the quaternion multiplication and its inverse
\begin{align}
\vec{j} \vec{n}(t) = Q (\vec{j}\vec{n}(0)) Q^{-1} \; .
\end{align}
Consequently two consecutive rotations are combined by the quaternion multiplication $Q=Q_1 Q_2$.

We can derive a simple EOM for $Q$ by applying an infinitesimal rotation $Q' \approx 1 + \vec{b}_\mr{eff} \vec{j} dt$ to $Q$ with the quaternion multiplication
\begin{align}
\frac{Q'Q - Q}{dt} &= -\vec{b}_\mr{eff} \cdot \vec{v} + ( q \vec{b}_\mr{eff} + \vec{b}_\mr{eff} \times \vec{v} )\cdot \vec{j},
\end{align}
where we divide the rotation $Q=q+\vec{v} \vec{j}$ into scalar part $q \in \mathbb{R}$ and a vector part $v \in \mathbb{R}^3$.
In conclusion we get the EOM for the  quaternionic representation of the rotation matrix
\begin{align}
	\frac{dq}{dt} &= - \vec{b}_\mr{eff} \cdot \vec{v} \; ,\\
	\frac{d \vec{v}}{dt} &= q \vec{b}_\mr{eff} + \vec{b}_\mr{eff} \times \vec{v} \; ,
\end{align}
which can be solved along-side the Eqs.~\eqref{eq:eom_s} and \eqref{eq:eom_i}.

\section{FOA - approximation and solution}
\label{sec:apptoy}
We start with the Hamiltonian
\begin{align}
H_e=\overline{b}(S_z^{(1)}+S_z^{(2)}) + \Delta \vec{b} (\vec{S}^{(1)}-\vec{S}^{(2)})+ J \vec{S}^{(1)} \cdot \vec{S}^{(2)} \;.
\end{align}
Using the operator $U=\exp(-i b (S_z^{(1)}+S_z^{(2)}))$, we transform into the rotating frame of $\vec{\overline{b}} $ 
\begin{align}
H' &= \Delta \vec{b}(t) (S_z^{(1)} - S_z^{(2)}) + J \vec{S}^{(1)}\cdot \vec{S}^{(2)} \\
&\approx \Delta b_z (S_z^{(1)} - S_z^{(2)}) + J \vec{S}^{(1)}\cdot \vec{S}^{(2)} \; ,
\end{align}
where $\overline{b}$ vanishes and $\Delta \vec{b}(t)$ obtains a dependence on time.
The time dependent deviational field $\Delta \vec{b}(t) = \Delta b_z \vec{e}_z + \Delta \vec{b}_\perp(t)$ can be split into a time-constant part along the $z$ axis and oscillating terms perpendicular to the $z$ direction.
For $b_{ext} \gg b_n$, it is well justified to omit $\Delta \vec{b}_\perp(t)$ due to fast oscillating contributions. 
After omitting $\Delta \vec{b}_\perp(t)$ we rotate back out of the rotating frame and obtain 
\begin{align}
H_e \approx b (S_z^{(1)}+S_z^{(2)}) + \Delta b_z (S_z^{(1)}-S_z^{(2)})+ J \vec{S}^{(1)} \cdot \vec{S}^{(2)} .
\end{align}
This Hamiltonian commutes with $S_z=S_z^{(1)}+S_z^{(2)}$ and has the four eigenenergies
\begin{align}
	\epsilon_{1,2} = \frac{J}{4} \pm b; \hspace{1cm} \epsilon_{3,4} = -\frac{J}{4} \pm \sqrt{\frac{J^2}{4}+(\Delta b_z)^2} .
\end{align}
The related eigenstates read
\begin{subequations}
\begin{align}
\ket{\epsilon_{1,2}} &= \Ket{\pm \frac{1}{2},\pm \frac{1}{2}} \\
	 \ket{\epsilon_{3,4}} &= \alpha \Ket{\pm \frac{1}{2},\mp \frac{1}{2}} \pm \sqrt{1-\alpha^2} \Ket{\mp \frac{1}{2},\pm \frac{1}{2}}
\end{align}
\end{subequations} 
with the abbreviation
\begin{align}
\alpha = \frac{1}{\sqrt{2}}\frac{x}{\sqrt{1 + x^2 - \sqrt{1+x^2}}} \; \mathrm{and} \; x = \frac{J}{2 \Delta b_z},
\end{align}
(Limits: $\lim_{x\rightarrow 0} \alpha = 1$ and $\lim_{x\rightarrow \infty} \alpha = \frac{1}{\sqrt{2}}$.)
which can be used to calculate arbitrary spin-spin correlators within this approximation.

\bibliography{references}

\begin{thebibliography}{76}%
\makeatletter
\providecommand \@ifxundefined [1]{%
 \@ifx{#1\undefined}
}%
\providecommand \@ifnum [1]{%
 \ifnum #1\expandafter \@firstoftwo
 \else \expandafter \@secondoftwo
 \fi
}%
\providecommand \@ifx [1]{%
 \ifx #1\expandafter \@firstoftwo
 \else \expandafter \@secondoftwo
 \fi
}%
\providecommand \natexlab [1]{#1}%
\providecommand \enquote  [1]{``#1''}%
\providecommand \bibnamefont  [1]{#1}%
\providecommand \bibfnamefont [1]{#1}%
\providecommand \citenamefont [1]{#1}%
\providecommand \href@noop [0]{\@secondoftwo}%
\providecommand \href [0]{\begingroup \@sanitize@url \@href}%
\providecommand \@href[1]{\@@startlink{#1}\@@href}%
\providecommand \@@href[1]{\endgroup#1\@@endlink}%
\providecommand \@sanitize@url [0]{\catcode `\\12\catcode `\$12\catcode
  `\&12\catcode `\#12\catcode `\^12\catcode `\_12\catcode `\%12\relax}%
\providecommand \@@startlink[1]{}%
\providecommand \@@endlink[0]{}%
\providecommand \url  [0]{\begingroup\@sanitize@url \@url }%
\providecommand \@url [1]{\endgroup\@href {#1}{\urlprefix }}%
\providecommand \urlprefix  [0]{URL }%
\providecommand \Eprint [0]{\href }%
\providecommand \doibase [0]{https://doi.org/}%
\providecommand \selectlanguage [0]{\@gobble}%
\providecommand \bibinfo  [0]{\@secondoftwo}%
\providecommand \bibfield  [0]{\@secondoftwo}%
\providecommand \translation [1]{[#1]}%
\providecommand \BibitemOpen [0]{}%
\providecommand \bibitemStop [0]{}%
\providecommand \bibitemNoStop [0]{.\EOS\space}%
\providecommand \EOS [0]{\spacefactor3000\relax}%
\providecommand \BibitemShut  [1]{\csname bibitem#1\endcsname}%
\let\auto@bib@innerbib\@empty
\bibitem [{\citenamefont {Hanson}\ \emph {et~al.}(2007)\citenamefont {Hanson},
  \citenamefont {Kouwenhoven}, \citenamefont {Petta}, \citenamefont {Tarucha},\
  and\ \citenamefont {Vandersypen}}]{Hanson2007}%
  \BibitemOpen
  \bibfield  {author} {\bibinfo {author} {\bibfnamefont {R.}~\bibnamefont
  {Hanson}}, \bibinfo {author} {\bibfnamefont {L.~P.}\ \bibnamefont
  {Kouwenhoven}}, \bibinfo {author} {\bibfnamefont {J.~R.}\ \bibnamefont
  {Petta}}, \bibinfo {author} {\bibfnamefont {S.}~\bibnamefont {Tarucha}},\
  and\ \bibinfo {author} {\bibfnamefont {L.~M.~K.}\ \bibnamefont
  {Vandersypen}},\ }\bibfield  {title} {\bibinfo {title} {Spins in few-electron
  quantum dots},\ }\href {https://doi.org/10.1103/RevModPhys.79.1217}
  {\bibfield  {journal} {\bibinfo  {journal} {Rev. Mod. Phys.}\ }\textbf
  {\bibinfo {volume} {79}},\ \bibinfo {pages} {1217} (\bibinfo {year}
  {2007})}\BibitemShut {NoStop}%
\bibitem [{\citenamefont {Burkard}\ \emph {et~al.}(2000)\citenamefont
  {Burkard}, \citenamefont {Engel},\ and\ \citenamefont {Loss}}]{Burkard2000}%
  \BibitemOpen
  \bibfield  {author} {\bibinfo {author} {\bibfnamefont {G.}~\bibnamefont
  {Burkard}}, \bibinfo {author} {\bibfnamefont {H.-A.}\ \bibnamefont {Engel}},\
  and\ \bibinfo {author} {\bibfnamefont {D.}~\bibnamefont {Loss}},\ }\bibfield
  {title} {\bibinfo {title} {Spintronics and quantum dots for quantum computing
  and quantum communication},\ }\href@noop {} {\bibfield  {journal} {\bibinfo
  {journal} {Fortschritte der Physik}\ }\textbf {\bibinfo {volume} {48}},\
  \bibinfo {pages} {965} (\bibinfo {year} {2000})}\BibitemShut {NoStop}%
\bibitem [{\citenamefont {Atat{\"u}re}\ \emph {et~al.}(2006)\citenamefont
  {Atat{\"u}re}, \citenamefont {Dreiser}, \citenamefont {Badolato},
  \citenamefont {H{\"o}gele}, \citenamefont {Karrai},\ and\ \citenamefont
  {Imamoglu}}]{Atatuere2006}%
  \BibitemOpen
  \bibfield  {author} {\bibinfo {author} {\bibfnamefont {M.}~\bibnamefont
  {Atat{\"u}re}}, \bibinfo {author} {\bibfnamefont {J.}~\bibnamefont
  {Dreiser}}, \bibinfo {author} {\bibfnamefont {A.}~\bibnamefont {Badolato}},
  \bibinfo {author} {\bibfnamefont {A.}~\bibnamefont {H{\"o}gele}}, \bibinfo
  {author} {\bibfnamefont {K.}~\bibnamefont {Karrai}},\ and\ \bibinfo {author}
  {\bibfnamefont {A.}~\bibnamefont {Imamoglu}},\ }\bibfield  {title} {\bibinfo
  {title} {Quantum-dot spin-state preparation with near-unity fidelity},\
  }\href {https://doi.org/10.1126/science.1126074} {\bibfield  {journal}
  {\bibinfo  {journal} {Science}\ }\textbf {\bibinfo {volume} {312}},\ \bibinfo
  {pages} {551} (\bibinfo {year} {2006})}\BibitemShut {NoStop}%
\bibitem [{\citenamefont {Greilich}\ \emph {et~al.}(2009)\citenamefont
  {Greilich}, \citenamefont {Economou}, \citenamefont {Spatzek}, \citenamefont
  {Yakovlev}, \citenamefont {Reuter}, \citenamefont {Wieck}, \citenamefont
  {Reinecke},\ and\ \citenamefont {Bayer}}]{GreilichEconomou2009}%
  \BibitemOpen
  \bibfield  {author} {\bibinfo {author} {\bibfnamefont {A.}~\bibnamefont
  {Greilich}}, \bibinfo {author} {\bibfnamefont {S.~E.}\ \bibnamefont
  {Economou}}, \bibinfo {author} {\bibfnamefont {S.}~\bibnamefont {Spatzek}},
  \bibinfo {author} {\bibfnamefont {D.~R.}\ \bibnamefont {Yakovlev}}, \bibinfo
  {author} {\bibfnamefont {D.}~\bibnamefont {Reuter}}, \bibinfo {author}
  {\bibfnamefont {A.~D.}\ \bibnamefont {Wieck}}, \bibinfo {author}
  {\bibfnamefont {T.~L.}\ \bibnamefont {Reinecke}},\ and\ \bibinfo {author}
  {\bibfnamefont {M.}~\bibnamefont {Bayer}},\ }\bibfield  {title} {\bibinfo
  {title} {Ultrafast optical rotations of electron spins in quantum dots},\
  }\href {http://dx.doi.org/10.1038/nphys1226} {\bibfield  {journal} {\bibinfo
  {journal} {Nature Physics}\ }\textbf {\bibinfo {volume} {5}},\ \bibinfo
  {pages} {262 EP } (\bibinfo {year} {2009})}\BibitemShut {NoStop}%
\bibitem [{\citenamefont {Press}\ \emph {et~al.}(2008)\citenamefont {Press},
  \citenamefont {Ladd}, \citenamefont {Zhang},\ and\ \citenamefont
  {Yamamoto}}]{Press2008}%
  \BibitemOpen
  \bibfield  {author} {\bibinfo {author} {\bibfnamefont {D.}~\bibnamefont
  {Press}}, \bibinfo {author} {\bibfnamefont {T.~D.}\ \bibnamefont {Ladd}},
  \bibinfo {author} {\bibfnamefont {B.}~\bibnamefont {Zhang}},\ and\ \bibinfo
  {author} {\bibfnamefont {Y.}~\bibnamefont {Yamamoto}},\ }\bibfield  {title}
  {\bibinfo {title} {Complete quantum control of a single quantum dot spin
  using ultrafast optical pulses},\ }\href
  {https://doi.org/10.1038/nature07530} {\bibfield  {journal} {\bibinfo
  {journal} {Nature}\ }\textbf {\bibinfo {volume} {456}},\ \bibinfo {pages}
  {218} (\bibinfo {year} {2008})}\BibitemShut {NoStop}%
\bibitem [{\citenamefont {Xu}\ \emph {et~al.}(2007)\citenamefont {Xu},
  \citenamefont {Wu}, \citenamefont {Sun}, \citenamefont {Huang}, \citenamefont
  {Cheng}, \citenamefont {Steel}, \citenamefont {Bracker}, \citenamefont
  {Gammon}, \citenamefont {Emary},\ and\ \citenamefont {Sham}}]{Xu2007}%
  \BibitemOpen
  \bibfield  {author} {\bibinfo {author} {\bibfnamefont {X.}~\bibnamefont
  {Xu}}, \bibinfo {author} {\bibfnamefont {Y.}~\bibnamefont {Wu}}, \bibinfo
  {author} {\bibfnamefont {B.}~\bibnamefont {Sun}}, \bibinfo {author}
  {\bibfnamefont {Q.}~\bibnamefont {Huang}}, \bibinfo {author} {\bibfnamefont
  {J.}~\bibnamefont {Cheng}}, \bibinfo {author} {\bibfnamefont {D.~G.}\
  \bibnamefont {Steel}}, \bibinfo {author} {\bibfnamefont {A.~S.}\ \bibnamefont
  {Bracker}}, \bibinfo {author} {\bibfnamefont {D.}~\bibnamefont {Gammon}},
  \bibinfo {author} {\bibfnamefont {C.}~\bibnamefont {Emary}},\ and\ \bibinfo
  {author} {\bibfnamefont {L.~J.}\ \bibnamefont {Sham}},\ }\bibfield  {title}
  {\bibinfo {title} {{Fast Spin State Initialization in a Singly Charged
  InAs-GaAs Quantum Dot by Optical Cooling}},\ }\href
  {https://doi.org/10.1103/PhysRevLett.99.097401} {\bibfield  {journal}
  {\bibinfo  {journal} {Phys. Rev. Lett.}\ }\textbf {\bibinfo {volume} {99}},\
  \bibinfo {pages} {097401} (\bibinfo {year} {2007})}\BibitemShut {NoStop}%
\bibitem [{\citenamefont {Greilich}\ \emph {et~al.}(2006)\citenamefont
  {Greilich}, \citenamefont {Yakovlev}, \citenamefont {Shabaev}, \citenamefont
  {Efros}, \citenamefont {Yugova}, \citenamefont {Oulton}, \citenamefont
  {Stavarache}, \citenamefont {Reuter}, \citenamefont {Wieck},\ and\
  \citenamefont {Bayer}}]{GreilichYakovlev2006}%
  \BibitemOpen
  \bibfield  {author} {\bibinfo {author} {\bibfnamefont {A.}~\bibnamefont
  {Greilich}}, \bibinfo {author} {\bibfnamefont {D.~R.}\ \bibnamefont
  {Yakovlev}}, \bibinfo {author} {\bibfnamefont {A.}~\bibnamefont {Shabaev}},
  \bibinfo {author} {\bibfnamefont {{\relax Al}.~L.}\ \bibnamefont {Efros}},
  \bibinfo {author} {\bibfnamefont {I.~A.}\ \bibnamefont {Yugova}}, \bibinfo
  {author} {\bibfnamefont {R.}~\bibnamefont {Oulton}}, \bibinfo {author}
  {\bibfnamefont {V.}~\bibnamefont {Stavarache}}, \bibinfo {author}
  {\bibfnamefont {D.}~\bibnamefont {Reuter}}, \bibinfo {author} {\bibfnamefont
  {A.}~\bibnamefont {Wieck}},\ and\ \bibinfo {author} {\bibfnamefont
  {M.}~\bibnamefont {Bayer}},\ }\bibfield  {title} {\bibinfo {title} {Mode
  locking of electron spin coherences in singly charged quantum dots},\ }\href
  {https://doi.org/10.1126/science.1128215} {\bibfield  {journal} {\bibinfo
  {journal} {Science}\ }\textbf {\bibinfo {volume} {313}},\ \bibinfo {pages}
  {341} (\bibinfo {year} {2006})}\BibitemShut {NoStop}%
\bibitem [{\citenamefont {Greilich}\ \emph {et~al.}(2007)\citenamefont
  {Greilich}, \citenamefont {Shabaev}, \citenamefont {Yakovlev}, \citenamefont
  {~}, \citenamefont {Yugova}, \citenamefont {Reuter}, \citenamefont {Wieck},\
  and\ \citenamefont {Bayer}}]{Greilich2007}%
  \BibitemOpen
  \bibfield  {author} {\bibinfo {author} {\bibfnamefont {A.}~\bibnamefont
  {Greilich}}, \bibinfo {author} {\bibfnamefont {A.}~\bibnamefont {Shabaev}},
  \bibinfo {author} {\bibfnamefont {D.~R.}\ \bibnamefont {Yakovlev}}, \bibinfo
  {author} {\bibfnamefont {{\relax Al}.~L.}\ \bibnamefont {~}}, \bibinfo
  {author} {\bibfnamefont {I.~A.}\ \bibnamefont {Yugova}}, \bibinfo {author}
  {\bibfnamefont {D.}~\bibnamefont {Reuter}}, \bibinfo {author} {\bibfnamefont
  {A.~D.}\ \bibnamefont {Wieck}},\ and\ \bibinfo {author} {\bibfnamefont
  {M.}~\bibnamefont {Bayer}},\ }\bibfield  {title} {\bibinfo {title}
  {Nuclei-induced frequency focusing of electron spin coherence},\ }\href@noop
  {} {\bibfield  {journal} {\bibinfo  {journal} {Science}\ }\textbf {\bibinfo
  {volume} {317}},\ \bibinfo {pages} {1896} (\bibinfo {year}
  {2007})}\BibitemShut {NoStop}%
\bibitem [{\citenamefont {Glazov}\ \emph {et~al.}(2012)\citenamefont {Glazov},
  \citenamefont {Yugova},\ and\ \citenamefont {Efros}}]{Glazov2012}%
  \BibitemOpen
  \bibfield  {author} {\bibinfo {author} {\bibfnamefont {M.~M.}\ \bibnamefont
  {Glazov}}, \bibinfo {author} {\bibfnamefont {I.~A.}\ \bibnamefont {Yugova}},\
  and\ \bibinfo {author} {\bibfnamefont {{\relax Al}.~L.}\ \bibnamefont
  {Efros}},\ }\bibfield  {title} {\bibinfo {title} {Electron spin
  synchronization induced by optical nuclear magnetic resonance feedback},\
  }\href {https://doi.org/10.1103/PhysRevB.85.041303} {\bibfield  {journal}
  {\bibinfo  {journal} {Phys. Rev. B}\ }\textbf {\bibinfo {volume} {85}},\
  \bibinfo {pages} {041303} (\bibinfo {year} {2012})}\BibitemShut {NoStop}%
\bibitem [{\citenamefont {Crooker}\ \emph {et~al.}(2010)\citenamefont
  {Crooker}, \citenamefont {Brandt}, \citenamefont {Sandfort}, \citenamefont
  {Greilich}, \citenamefont {Yakovlev}, \citenamefont {Reuter}, \citenamefont
  {Wieck},\ and\ \citenamefont {Bayer}}]{Crooker2010}%
  \BibitemOpen
  \bibfield  {author} {\bibinfo {author} {\bibfnamefont {S.~A.}\ \bibnamefont
  {Crooker}}, \bibinfo {author} {\bibfnamefont {J.}~\bibnamefont {Brandt}},
  \bibinfo {author} {\bibfnamefont {C.}~\bibnamefont {Sandfort}}, \bibinfo
  {author} {\bibfnamefont {A.}~\bibnamefont {Greilich}}, \bibinfo {author}
  {\bibfnamefont {D.~R.}\ \bibnamefont {Yakovlev}}, \bibinfo {author}
  {\bibfnamefont {D.}~\bibnamefont {Reuter}}, \bibinfo {author} {\bibfnamefont
  {A.~D.}\ \bibnamefont {Wieck}},\ and\ \bibinfo {author} {\bibfnamefont
  {M.}~\bibnamefont {Bayer}},\ }\bibfield  {title} {\bibinfo {title} {Spin
  noise of electrons and holes in self-assembled quantum dots},\ }\href
  {https://doi.org/10.1103/PhysRevLett.104.036601} {\bibfield  {journal}
  {\bibinfo  {journal} {Phys. Rev. Lett.}\ }\textbf {\bibinfo {volume} {104}},\
  \bibinfo {pages} {036601} (\bibinfo {year} {2010})}\BibitemShut {NoStop}%
\bibitem [{\citenamefont {Li}\ \emph {et~al.}(2012)\citenamefont {Li},
  \citenamefont {Sinitsyn}, \citenamefont {Smith}, \citenamefont {Reuter},
  \citenamefont {Wieck}, \citenamefont {Yakovlev}, \citenamefont {Bayer},\ and\
  \citenamefont {Crooker}}]{Yan2012}%
  \BibitemOpen
  \bibfield  {author} {\bibinfo {author} {\bibfnamefont {Y.}~\bibnamefont
  {Li}}, \bibinfo {author} {\bibfnamefont {N.}~\bibnamefont {Sinitsyn}},
  \bibinfo {author} {\bibfnamefont {D.~L.}\ \bibnamefont {Smith}}, \bibinfo
  {author} {\bibfnamefont {D.}~\bibnamefont {Reuter}}, \bibinfo {author}
  {\bibfnamefont {A.~D.}\ \bibnamefont {Wieck}}, \bibinfo {author}
  {\bibfnamefont {D.~R.}\ \bibnamefont {Yakovlev}}, \bibinfo {author}
  {\bibfnamefont {M.}~\bibnamefont {Bayer}},\ and\ \bibinfo {author}
  {\bibfnamefont {S.~A.}\ \bibnamefont {Crooker}},\ }\bibfield  {title}
  {\bibinfo {title} {{Intrinsic Spin Fluctuations Reveal the Dynamical Response
  Function of Holes Coupled to Nuclear Spin Baths in (In,Ga)As Quantum Dots}},\
  }\href {https://doi.org/10.1103/PhysRevLett.108.186603} {\bibfield  {journal}
  {\bibinfo  {journal} {Phys. Rev. Lett.}\ }\textbf {\bibinfo {volume} {108}},\
  \bibinfo {pages} {186603} (\bibinfo {year} {2012})}\BibitemShut {NoStop}%
\bibitem [{\citenamefont {Zapasskii}(2013)}]{Zapasskii2013}%
  \BibitemOpen
  \bibfield  {author} {\bibinfo {author} {\bibfnamefont {V.~S.}\ \bibnamefont
  {Zapasskii}},\ }\bibfield  {title} {\bibinfo {title} {Spin-noise
  spectroscopy: from proof of principle to applications},\ }\href
  {https://doi.org/10.1364/AOP.5.000131} {\bibfield  {journal} {\bibinfo
  {journal} {Adv. Opt. Photon.}\ }\textbf {\bibinfo {volume} {5}},\ \bibinfo
  {pages} {131} (\bibinfo {year} {2013})}\BibitemShut {NoStop}%
\bibitem [{\citenamefont {Hackmann}\ \emph {et~al.}(2015)\citenamefont
  {Hackmann}, \citenamefont {Glasenapp}, \citenamefont {Greilich},
  \citenamefont {Bayer},\ and\ \citenamefont {Anders}}]{Hackmann2015}%
  \BibitemOpen
  \bibfield  {author} {\bibinfo {author} {\bibfnamefont {J.}~\bibnamefont
  {Hackmann}}, \bibinfo {author} {\bibfnamefont {P.}~\bibnamefont {Glasenapp}},
  \bibinfo {author} {\bibfnamefont {A.}~\bibnamefont {Greilich}}, \bibinfo
  {author} {\bibfnamefont {M.}~\bibnamefont {Bayer}},\ and\ \bibinfo {author}
  {\bibfnamefont {F.~B.}\ \bibnamefont {Anders}},\ }\bibfield  {title}
  {\bibinfo {title} {{Influence of the Nuclear Electric Quadrupolar Interaction
  on the Coherence Time of Hole and Electron Spins Confined in Semiconductor
  Quantum Dots}},\ }\href@noop {} {\bibfield  {journal} {\bibinfo  {journal}
  {Phys. Rev. Lett.}\ }\textbf {\bibinfo {volume} {115}},\ \bibinfo {pages}
  {207401} (\bibinfo {year} {2015})}\BibitemShut {NoStop}%
\bibitem [{\citenamefont {Sinitsyn}\ and\ \citenamefont
  {Pershin}(2016)}]{Sinitsyn_2016}%
  \BibitemOpen
  \bibfield  {author} {\bibinfo {author} {\bibfnamefont {N.~A.}\ \bibnamefont
  {Sinitsyn}}\ and\ \bibinfo {author} {\bibfnamefont {Y.~V.}\ \bibnamefont
  {Pershin}},\ }\bibfield  {title} {\bibinfo {title} {The theory of spin noise
  spectroscopy: a review},\ }\href
  {https://doi.org/10.1088/0034-4885/79/10/106501} {\bibfield  {journal}
  {\bibinfo  {journal} {Reports on Progress in Physics}\ }\textbf {\bibinfo
  {volume} {79}},\ \bibinfo {pages} {106501} (\bibinfo {year}
  {2016})}\BibitemShut {NoStop}%
\bibitem [{\citenamefont {Glasenapp}\ \emph {et~al.}(2016)\citenamefont
  {Glasenapp}, \citenamefont {Smirnov}, \citenamefont {Greilich}, \citenamefont
  {Hackmann}, \citenamefont {Glazov}, \citenamefont {Anders},\ and\
  \citenamefont {Bayer}}]{Glasenapp2016}%
  \BibitemOpen
  \bibfield  {author} {\bibinfo {author} {\bibfnamefont {P.}~\bibnamefont
  {Glasenapp}}, \bibinfo {author} {\bibfnamefont {D.~S.}\ \bibnamefont
  {Smirnov}}, \bibinfo {author} {\bibfnamefont {A.}~\bibnamefont {Greilich}},
  \bibinfo {author} {\bibfnamefont {J.}~\bibnamefont {Hackmann}}, \bibinfo
  {author} {\bibfnamefont {M.~M.}\ \bibnamefont {Glazov}}, \bibinfo {author}
  {\bibfnamefont {F.~B.}\ \bibnamefont {Anders}},\ and\ \bibinfo {author}
  {\bibfnamefont {M.}~\bibnamefont {Bayer}},\ }\bibfield  {title} {\bibinfo
  {title} {{Spin noise of electrons and holes in (In,Ga)As quantum dots:
  Experiment and theory}},\ }\href {https://doi.org/10.1103/PhysRevB.93.205429}
  {\bibfield  {journal} {\bibinfo  {journal} {Phys. Rev. B}\ }\textbf {\bibinfo
  {volume} {93}},\ \bibinfo {pages} {205429} (\bibinfo {year}
  {2016})}\BibitemShut {NoStop}%
\bibitem [{\citenamefont {Sinitsyn}\ \emph {et~al.}(2012)\citenamefont
  {Sinitsyn}, \citenamefont {Li}, \citenamefont {Crooker}, \citenamefont
  {Saxena},\ and\ \citenamefont {Smith}}]{Sinitsyn2012}%
  \BibitemOpen
  \bibfield  {author} {\bibinfo {author} {\bibfnamefont {N.~A.}\ \bibnamefont
  {Sinitsyn}}, \bibinfo {author} {\bibfnamefont {Y.}~\bibnamefont {Li}},
  \bibinfo {author} {\bibfnamefont {S.~A.}\ \bibnamefont {Crooker}}, \bibinfo
  {author} {\bibfnamefont {A.}~\bibnamefont {Saxena}},\ and\ \bibinfo {author}
  {\bibfnamefont {D.~L.}\ \bibnamefont {Smith}},\ }\bibfield  {title} {\bibinfo
  {title} {{Role of Nuclear Quadrupole Coupling on Decoherence and Relaxation
  of Central Spins in Quantum Dots}},\ }\href
  {https://doi.org/10.1103/PhysRevLett.109.166605} {\bibfield  {journal}
  {\bibinfo  {journal} {Phys. Rev. Lett.}\ }\textbf {\bibinfo {volume} {109}},\
  \bibinfo {pages} {166605} (\bibinfo {year} {2012})}\BibitemShut {NoStop}%
\bibitem [{\citenamefont {Bulutay}(2012)}]{Bulutay2012}%
  \BibitemOpen
  \bibfield  {author} {\bibinfo {author} {\bibfnamefont {C.}~\bibnamefont
  {Bulutay}},\ }\bibfield  {title} {\bibinfo {title} {{Quadrupolar spectra of
  nuclear spins in strained InGaAs quantum dots}},\ }\href
  {https://doi.org/10.1103/PhysRevB.85.115313} {\bibfield  {journal} {\bibinfo
  {journal} {Phys. Rev. B}\ }\textbf {\bibinfo {volume} {85}},\ \bibinfo
  {pages} {115313} (\bibinfo {year} {2012})}\BibitemShut {NoStop}%
\bibitem [{\citenamefont {Li}\ and\ \citenamefont
  {Sinitsyn}(2016)}]{LiSinitsyn2016}%
  \BibitemOpen
  \bibfield  {author} {\bibinfo {author} {\bibfnamefont {F.}~\bibnamefont
  {Li}}\ and\ \bibinfo {author} {\bibfnamefont {N.}~\bibnamefont {Sinitsyn}},\
  }\bibfield  {title} {\bibinfo {title} {Universality in higher order spin
  noise spectroscopy},\ }\href@noop {} {\bibfield  {journal} {\bibinfo
  {journal} {Phys. Rev. Lett.}\ }\textbf {\bibinfo {volume} {116}},\ \bibinfo
  {pages} {026601} (\bibinfo {year} {2016})}\BibitemShut {NoStop}%
\bibitem [{\citenamefont {Fr\"ohling}\ and\ \citenamefont
  {Anders}(2017)}]{Froehling2017}%
  \BibitemOpen
  \bibfield  {author} {\bibinfo {author} {\bibfnamefont {N.}~\bibnamefont
  {Fr\"ohling}}\ and\ \bibinfo {author} {\bibfnamefont {F.~B.}\ \bibnamefont
  {Anders}},\ }\bibfield  {title} {\bibinfo {title} {Long-time coherence in
  fourth-order spin correlation functions},\ }\href
  {https://doi.org/10.1103/PhysRevB.96.045441} {\bibfield  {journal} {\bibinfo
  {journal} {Phys. Rev. B}\ }\textbf {\bibinfo {volume} {96}},\ \bibinfo
  {pages} {045441} (\bibinfo {year} {2017})}\BibitemShut {NoStop}%
\bibitem [{\citenamefont {H\"agele}\ and\ \citenamefont
  {Schefczik}(2018)}]{Haegle2018}%
  \BibitemOpen
  \bibfield  {author} {\bibinfo {author} {\bibfnamefont {D.}~\bibnamefont
  {H\"agele}}\ and\ \bibinfo {author} {\bibfnamefont {F.}~\bibnamefont
  {Schefczik}},\ }\bibfield  {title} {\bibinfo {title} {Higher-order moments,
  cumulants, and spectra of continuous quantum noise measurements},\ }\href
  {https://doi.org/10.1103/PhysRevB.98.205143} {\bibfield  {journal} {\bibinfo
  {journal} {Phys. Rev. B}\ }\textbf {\bibinfo {volume} {98}},\ \bibinfo
  {pages} {205143} (\bibinfo {year} {2018})}\BibitemShut {NoStop}%
\bibitem [{\citenamefont {Fr\"ohling}\ \emph {et~al.}(2019)\citenamefont
  {Fr\"ohling}, \citenamefont {J\"aschke},\ and\ \citenamefont
  {Anders}}]{FroelingJaeschke2019}%
  \BibitemOpen
  \bibfield  {author} {\bibinfo {author} {\bibfnamefont {N.}~\bibnamefont
  {Fr\"ohling}}, \bibinfo {author} {\bibfnamefont {N.}~\bibnamefont
  {J\"aschke}},\ and\ \bibinfo {author} {\bibfnamefont {F.~B.}\ \bibnamefont
  {Anders}},\ }\bibfield  {title} {\bibinfo {title} {Fourth-order spin
  correlation function in the extended central spin model},\ }\href
  {https://doi.org/10.1103/PhysRevB.99.155305} {\bibfield  {journal} {\bibinfo
  {journal} {Phys. Rev. B}\ }\textbf {\bibinfo {volume} {99}},\ \bibinfo
  {pages} {155305} (\bibinfo {year} {2019})}\BibitemShut {NoStop}%
\bibitem [{\citenamefont {Norris}\ \emph {et~al.}(2016)\citenamefont {Norris},
  \citenamefont {Paz-Silva},\ and\ \citenamefont {Viola}}]{Norris2016}%
  \BibitemOpen
  \bibfield  {author} {\bibinfo {author} {\bibfnamefont {L.~M.}\ \bibnamefont
  {Norris}}, \bibinfo {author} {\bibfnamefont {G.~A.}\ \bibnamefont
  {Paz-Silva}},\ and\ \bibinfo {author} {\bibfnamefont {L.}~\bibnamefont
  {Viola}},\ }\bibfield  {title} {\bibinfo {title} {Qubit noise spectroscopy
  for non-gaussian dephasing environments},\ }\href
  {https://doi.org/10.1103/PhysRevLett.116.150503} {\bibfield  {journal}
  {\bibinfo  {journal} {Phys. Rev. Lett.}\ }\textbf {\bibinfo {volume} {116}},\
  \bibinfo {pages} {150503} (\bibinfo {year} {2016})}\BibitemShut {NoStop}%
\bibitem [{\citenamefont {Sza{\'{n}}kowski}\ \emph {et~al.}(2017)\citenamefont
  {Sza{\'{n}}kowski}, \citenamefont {Ramon}, \citenamefont {Krzywda},
  \citenamefont {Kwiatkowski},\ and\ \citenamefont
  {Cywi{\'{n}}ski}}]{Szankowski2017}%
  \BibitemOpen
  \bibfield  {author} {\bibinfo {author} {\bibfnamefont {P.}~\bibnamefont
  {Sza{\'{n}}kowski}}, \bibinfo {author} {\bibfnamefont {G.}~\bibnamefont
  {Ramon}}, \bibinfo {author} {\bibfnamefont {J.}~\bibnamefont {Krzywda}},
  \bibinfo {author} {\bibfnamefont {D.}~\bibnamefont {Kwiatkowski}},\ and\
  \bibinfo {author} {\bibfnamefont {{\L}.}~\bibnamefont {Cywi{\'{n}}ski}},\
  }\bibfield  {title} {\bibinfo {title} {Environmental noise spectroscopy with
  qubits subjected to dynamical decoupling},\ }\href
  {https://doi.org/10.1088/1361-648x/aa7648} {\bibfield  {journal} {\bibinfo
  {journal} {Journal of Physics: Condensed Matter}\ }\textbf {\bibinfo {volume}
  {29}},\ \bibinfo {pages} {333001} (\bibinfo {year} {2017})}\BibitemShut
  {NoStop}%
\bibitem [{\citenamefont {Fermi}(1930)}]{Fermi1930}%
  \BibitemOpen
  \bibfield  {author} {\bibinfo {author} {\bibfnamefont {E.}~\bibnamefont
  {Fermi}},\ }\bibfield  {title} {\bibinfo {title} {{\"U}ber die magnetischen
  momente der atomkerne.},\ }\href@noop {} {\bibfield  {journal} {\bibinfo
  {journal} {Z. Phys.}\ }\textbf {\bibinfo {volume} {60}},\ \bibinfo {pages}
  {320} (\bibinfo {year} {1930})}\BibitemShut {NoStop}%
\bibitem [{\citenamefont {Gaudin}(1976)}]{Gaudin1976}%
  \BibitemOpen
  \bibfield  {author} {\bibinfo {author} {\bibfnamefont {M.}~\bibnamefont
  {Gaudin}},\ }\bibfield  {title} {\bibinfo {title} {Diagonalisation d'une
  classe d'hamiltoniens de spin},\ }\href@noop {} {\bibfield  {journal}
  {\bibinfo  {journal} {J. Phys. France}\ }\textbf {\bibinfo {volume} {37}},\
  \bibinfo {pages} {1087} (\bibinfo {year} {1976})}\BibitemShut {NoStop}%
\bibitem [{\citenamefont {Bortz}\ and\ \citenamefont
  {Stolze}(2007)}]{Bortz2007}%
  \BibitemOpen
  \bibfield  {author} {\bibinfo {author} {\bibfnamefont {M.}~\bibnamefont
  {Bortz}}\ and\ \bibinfo {author} {\bibfnamefont {J.}~\bibnamefont {Stolze}},\
  }\bibfield  {title} {\bibinfo {title} {Exact dynamics in the inhomogeneous
  central-spin model},\ }\href@noop {} {\bibfield  {journal} {\bibinfo
  {journal} {Phys. Rev. B}\ }\textbf {\bibinfo {volume} {76}},\ \bibinfo
  {pages} {014304} (\bibinfo {year} {2007})}\BibitemShut {NoStop}%
\bibitem [{\citenamefont {Faribault}\ and\ \citenamefont
  {Schuricht}(2013{\natexlab{a}})}]{FaribautSchuricht2013a}%
  \BibitemOpen
  \bibfield  {author} {\bibinfo {author} {\bibfnamefont {A.}~\bibnamefont
  {Faribault}}\ and\ \bibinfo {author} {\bibfnamefont {D.}~\bibnamefont
  {Schuricht}},\ }\bibfield  {title} {\bibinfo {title} {Integrability-based
  analysis of the hyperfine-interaction-induced decoherence in quantum dots},\
  }\href {https://doi.org/10.1103/PhysRevLett.110.040405} {\bibfield  {journal}
  {\bibinfo  {journal} {Phys. Rev. Lett.}\ }\textbf {\bibinfo {volume} {110}},\
  \bibinfo {pages} {040405} (\bibinfo {year} {2013}{\natexlab{a}})}\BibitemShut
  {NoStop}%
\bibitem [{\citenamefont {Faribault}\ and\ \citenamefont
  {Schuricht}(2013{\natexlab{b}})}]{FaribautSchuricht2013b}%
  \BibitemOpen
  \bibfield  {author} {\bibinfo {author} {\bibfnamefont {A.}~\bibnamefont
  {Faribault}}\ and\ \bibinfo {author} {\bibfnamefont {D.}~\bibnamefont
  {Schuricht}},\ }\bibfield  {title} {\bibinfo {title} {Spin decoherence due to
  a randomly fluctuating spin bath},\ }\href
  {https://doi.org/10.1103/PhysRevB.88.085323} {\bibfield  {journal} {\bibinfo
  {journal} {Phys. Rev. B}\ }\textbf {\bibinfo {volume} {88}},\ \bibinfo
  {pages} {085323} (\bibinfo {year} {2013}{\natexlab{b}})}\BibitemShut
  {NoStop}%
\bibitem [{\citenamefont {Merkulov}\ \emph {et~al.}(2002)\citenamefont
  {Merkulov}, \citenamefont {Efros},\ and\ \citenamefont
  {Rosen}}]{Merkulov2002}%
  \BibitemOpen
  \bibfield  {author} {\bibinfo {author} {\bibfnamefont {I.~A.}\ \bibnamefont
  {Merkulov}}, \bibinfo {author} {\bibfnamefont {{\relax Al}.~L.}\ \bibnamefont
  {Efros}},\ and\ \bibinfo {author} {\bibfnamefont {M.}~\bibnamefont {Rosen}},\
  }\bibfield  {title} {\bibinfo {title} {Electron spin relaxation by nuclei in
  semiconductor quantum dots},\ }\href
  {https://doi.org/10.1103/PhysRevB.65.205309} {\bibfield  {journal} {\bibinfo
  {journal} {Phys. Rev. B}\ }\textbf {\bibinfo {volume} {65}},\ \bibinfo
  {pages} {205309} (\bibinfo {year} {2002})}\BibitemShut {NoStop}%
\bibitem [{\citenamefont {Saikin}\ \emph {et~al.}(2007)\citenamefont {Saikin},
  \citenamefont {Yao},\ and\ \citenamefont {Sham}}]{SaikinSham2007}%
  \BibitemOpen
  \bibfield  {author} {\bibinfo {author} {\bibfnamefont {S.~K.}\ \bibnamefont
  {Saikin}}, \bibinfo {author} {\bibfnamefont {W.}~\bibnamefont {Yao}},\ and\
  \bibinfo {author} {\bibfnamefont {L.~J.}\ \bibnamefont {Sham}},\ }\bibfield
  {title} {\bibinfo {title} {Single-electron spin decoherence by nuclear spin
  bath: Linked-cluster expansion approach},\ }\href
  {https://doi.org/10.1103/PhysRevB.75.125314} {\bibfield  {journal} {\bibinfo
  {journal} {Phys. Rev. B}\ }\textbf {\bibinfo {volume} {75}},\ \bibinfo
  {pages} {125314} (\bibinfo {year} {2007})}\BibitemShut {NoStop}%
\bibitem [{\citenamefont {Yang}\ and\ \citenamefont {Liu}(2008)}]{Yang2008}%
  \BibitemOpen
  \bibfield  {author} {\bibinfo {author} {\bibfnamefont {W.}~\bibnamefont
  {Yang}}\ and\ \bibinfo {author} {\bibfnamefont {R.-B.}\ \bibnamefont {Liu}},\
  }\bibfield  {title} {\bibinfo {title} {Quantum many-body theory of qubit
  decoherence in a finite-size spin bath},\ }\href
  {https://doi.org/10.1103/PhysRevB.78.085315} {\bibfield  {journal} {\bibinfo
  {journal} {Phys. Rev. B}\ }\textbf {\bibinfo {volume} {78}},\ \bibinfo
  {pages} {085315} (\bibinfo {year} {2008})}\BibitemShut {NoStop}%
\bibitem [{\citenamefont {Tal-Ezer}\ and\ \citenamefont
  {Kosloff}(1984)}]{TalEzer-Kosloff-84}%
  \BibitemOpen
  \bibfield  {author} {\bibinfo {author} {\bibfnamefont {H.}~\bibnamefont
  {Tal-Ezer}}\ and\ \bibinfo {author} {\bibfnamefont {R.}~\bibnamefont
  {Kosloff}},\ }\bibfield  {title} {\bibinfo {title} {{An accurate and
  efficient scheme for propagating the time dependent Schr{\"o}dinger
  equation}},\ }\href@noop {} {\bibfield  {journal} {\bibinfo  {journal} {J.
  Chem. Phys}\ }\textbf {\bibinfo {volume} {81}},\ \bibinfo {pages} {3967}
  (\bibinfo {year} {1984})}\BibitemShut {NoStop}%
\bibitem [{\citenamefont {Smirnov}\ \emph {et~al.}(2021)\citenamefont
  {Smirnov}, \citenamefont {Mantsevich},\ and\ \citenamefont
  {Glazov}}]{Smirnov_2021}%
  \BibitemOpen
  \bibfield  {author} {\bibinfo {author} {\bibfnamefont {D.~S.}\ \bibnamefont
  {Smirnov}}, \bibinfo {author} {\bibfnamefont {V.~N.}\ \bibnamefont
  {Mantsevich}},\ and\ \bibinfo {author} {\bibfnamefont {M.~M.}\ \bibnamefont
  {Glazov}},\ }\bibfield  {title} {\bibinfo {title} {Theory of optically
  detected spin noise in nanosystems},\ }\href
  {https://doi.org/10.3367/ufne.2020.10.038861} {\bibfield  {journal} {\bibinfo
   {journal} {Physics-Uspekhi}\ }\textbf {\bibinfo {volume} {64}},\ \bibinfo
  {pages} {923} (\bibinfo {year} {2021})}\BibitemShut {NoStop}%
\bibitem [{\citenamefont {Hackmann}\ and\ \citenamefont
  {Anders}(2014)}]{Hackmann2014}%
  \BibitemOpen
  \bibfield  {author} {\bibinfo {author} {\bibfnamefont {J.}~\bibnamefont
  {Hackmann}}\ and\ \bibinfo {author} {\bibfnamefont {F.~B.}\ \bibnamefont
  {Anders}},\ }\bibfield  {title} {\bibinfo {title} {Spin noise in the
  anisotropic central spin model},\ }\href@noop {} {\bibfield  {journal}
  {\bibinfo  {journal} {Phys. Rev. B}\ }\textbf {\bibinfo {volume} {89}},\
  \bibinfo {pages} {045317} (\bibinfo {year} {2014})}\BibitemShut {NoStop}%
\bibitem [{\citenamefont {Schollw\"ock}(2005)}]{Schollwoeck-2005}%
  \BibitemOpen
  \bibfield  {author} {\bibinfo {author} {\bibfnamefont {U.}~\bibnamefont
  {Schollw\"ock}},\ }\bibfield  {title} {\bibinfo {title} {The density-matrix
  renormalization group},\ }\href@noop {} {\bibfield  {journal} {\bibinfo
  {journal} {Rev. Mod. Phys.}\ }\textbf {\bibinfo {volume} {77}},\ \bibinfo
  {pages} {259} (\bibinfo {year} {2005})}\BibitemShut {NoStop}%
\bibitem [{\citenamefont {Stanek}\ \emph {et~al.}(2013)\citenamefont {Stanek},
  \citenamefont {Raas},\ and\ \citenamefont {Uhrig}}]{StanekRaasUhrig2013}%
  \BibitemOpen
  \bibfield  {author} {\bibinfo {author} {\bibfnamefont {D.}~\bibnamefont
  {Stanek}}, \bibinfo {author} {\bibfnamefont {C.}~\bibnamefont {Raas}},\ and\
  \bibinfo {author} {\bibfnamefont {G.~S.}\ \bibnamefont {Uhrig}},\ }\bibfield
  {title} {\bibinfo {title} {Dynamics and decoherence in the central spin model
  in the low-field limit},\ }\href {https://doi.org/10.1103/PhysRevB.88.155305}
  {\bibfield  {journal} {\bibinfo  {journal} {Phys. Rev. B}\ }\textbf {\bibinfo
  {volume} {88}},\ \bibinfo {pages} {155305} (\bibinfo {year}
  {2013})}\BibitemShut {NoStop}%
\bibitem [{\citenamefont {Coish}\ and\ \citenamefont
  {Loss}(2004)}]{CoishLoss2004}%
  \BibitemOpen
  \bibfield  {author} {\bibinfo {author} {\bibfnamefont {W.~A.}\ \bibnamefont
  {Coish}}\ and\ \bibinfo {author} {\bibfnamefont {D.}~\bibnamefont {Loss}},\
  }\bibfield  {title} {\bibinfo {title} {Hyperfine interaction in a quantum
  dot: Non-markovian electron spin dynamics},\ }\href
  {https://doi.org/10.1103/PhysRevB.70.195340} {\bibfield  {journal} {\bibinfo
  {journal} {Phys. Rev. B}\ }\textbf {\bibinfo {volume} {70}},\ \bibinfo
  {pages} {195340} (\bibinfo {year} {2004})}\BibitemShut {NoStop}%
\bibitem [{\citenamefont {Deng}\ and\ \citenamefont {Hu}(2005)}]{DengHu2005}%
  \BibitemOpen
  \bibfield  {author} {\bibinfo {author} {\bibfnamefont {C.}~\bibnamefont
  {Deng}}\ and\ \bibinfo {author} {\bibfnamefont {X.}~\bibnamefont {Hu}},\
  }\bibfield  {title} {\bibinfo {title} {Selective dynamic nuclear spin
  polarization in a spin-blocked double dot},\ }\href
  {https://doi.org/10.1103/PhysRevB.71.033307} {\bibfield  {journal} {\bibinfo
  {journal} {Phys. Rev. B}\ }\textbf {\bibinfo {volume} {71}},\ \bibinfo
  {pages} {033307} (\bibinfo {year} {2005})}\BibitemShut {NoStop}%
\bibitem [{\citenamefont {Barnes}\ \emph {et~al.}(2011)\citenamefont {Barnes},
  \citenamefont {Cywi\ifmmode~\acute{n}\else \'{n}\fi{}ski},\ and\
  \citenamefont {Das~Sarma}}]{BarnesDaSarma2011}%
  \BibitemOpen
  \bibfield  {author} {\bibinfo {author} {\bibfnamefont {E.}~\bibnamefont
  {Barnes}}, \bibinfo {author} {\bibfnamefont {L.}~\bibnamefont
  {Cywi\ifmmode~\acute{n}\else \'{n}\fi{}ski}},\ and\ \bibinfo {author}
  {\bibfnamefont {S.}~\bibnamefont {Das~Sarma}},\ }\bibfield  {title} {\bibinfo
  {title} {Master equation approach to the central spin decoherence problem:
  Uniform coupling model and role of projection operators},\ }\href
  {https://doi.org/10.1103/PhysRevB.84.155315} {\bibfield  {journal} {\bibinfo
  {journal} {Phys. Rev. B}\ }\textbf {\bibinfo {volume} {84}},\ \bibinfo
  {pages} {155315} (\bibinfo {year} {2011})}\BibitemShut {NoStop}%
\bibitem [{\citenamefont {Fischer}\ \emph {et~al.}(2020)\citenamefont
  {Fischer}, \citenamefont {Kleinjohann}, \citenamefont {Anders},\ and\
  \citenamefont {Glazov}}]{fischer2020}%
  \BibitemOpen
  \bibfield  {author} {\bibinfo {author} {\bibfnamefont {A.}~\bibnamefont
  {Fischer}}, \bibinfo {author} {\bibfnamefont {I.}~\bibnamefont
  {Kleinjohann}}, \bibinfo {author} {\bibfnamefont {F.~B.}\ \bibnamefont
  {Anders}},\ and\ \bibinfo {author} {\bibfnamefont {M.~M.}\ \bibnamefont
  {Glazov}},\ }\bibfield  {title} {\bibinfo {title} {Kinetic approach to
  nuclear-spin polaron formation},\ }\href@noop {} {\bibfield  {journal}
  {\bibinfo  {journal} {Physical Review B}\ }\textbf {\bibinfo {volume}
  {102}},\ \bibinfo {pages} {165309} (\bibinfo {year} {2020})}\BibitemShut
  {NoStop}%
\bibitem [{\citenamefont {Smirnov}\ and\ \citenamefont
  {Shumilin}(2021)}]{smirnov2021}%
  \BibitemOpen
  \bibfield  {author} {\bibinfo {author} {\bibfnamefont {D.}~\bibnamefont
  {Smirnov}}\ and\ \bibinfo {author} {\bibfnamefont {A.}~\bibnamefont
  {Shumilin}},\ }\bibfield  {title} {\bibinfo {title} {Electric current noise
  in mesoscopic organic semiconductors induced by nuclear spin fluctuations},\
  }\href@noop {} {\bibfield  {journal} {\bibinfo  {journal} {Physical Review
  B}\ }\textbf {\bibinfo {volume} {103}},\ \bibinfo {pages} {195440} (\bibinfo
  {year} {2021})}\BibitemShut {NoStop}%
\bibitem [{\citenamefont {Roy}\ \emph {et~al.}(2015)\citenamefont {Roy},
  \citenamefont {Yang}, \citenamefont {Crooker},\ and\ \citenamefont
  {Sinitsyn}}]{Roy2015}%
  \BibitemOpen
  \bibfield  {author} {\bibinfo {author} {\bibfnamefont {D.}~\bibnamefont
  {Roy}}, \bibinfo {author} {\bibfnamefont {L.}~\bibnamefont {Yang}}, \bibinfo
  {author} {\bibfnamefont {S.~A.}\ \bibnamefont {Crooker}},\ and\ \bibinfo
  {author} {\bibfnamefont {N.~A.}\ \bibnamefont {Sinitsyn}},\ }\bibfield
  {title} {\bibinfo {title} {Cross-correlation spin noise spectroscopy of
  heterogeneous interacting spin systems},\ }\href
  {https://doi.org/10.1038/srep09573} {\bibfield  {journal} {\bibinfo
  {journal} {Scientific Reports}\ }\textbf {\bibinfo {volume} {5}},\ \bibinfo
  {pages} {9573} (\bibinfo {year} {2015})}\BibitemShut {NoStop}%
\bibitem [{\citenamefont {Al-Hassanieh}\ \emph {et~al.}(2006)\citenamefont
  {Al-Hassanieh}, \citenamefont {Dobrovitski}, \citenamefont {Dagotto},\ and\
  \citenamefont {Harmon}}]{Al-Hassanieh2006}%
  \BibitemOpen
  \bibfield  {author} {\bibinfo {author} {\bibfnamefont {K.~A.}\ \bibnamefont
  {Al-Hassanieh}}, \bibinfo {author} {\bibfnamefont {V.~V.}\ \bibnamefont
  {Dobrovitski}}, \bibinfo {author} {\bibfnamefont {E.}~\bibnamefont
  {Dagotto}},\ and\ \bibinfo {author} {\bibfnamefont {B.~N.}\ \bibnamefont
  {Harmon}},\ }\bibfield  {title} {\bibinfo {title} {Numerical modeling of the
  central spin problem using the spin-coherent-state $p$ representation},\
  }\href {https://doi.org/10.1103/PhysRevLett.97.037204} {\bibfield  {journal}
  {\bibinfo  {journal} {Phys. Rev. Lett.}\ }\textbf {\bibinfo {volume} {97}},\
  \bibinfo {pages} {037204} (\bibinfo {year} {2006})}\BibitemShut {NoStop}%
\bibitem [{\citenamefont {Chen}\ \emph {et~al.}(2007)\citenamefont {Chen},
  \citenamefont {Bergman},\ and\ \citenamefont {Balents}}]{ChenBalents2007}%
  \BibitemOpen
  \bibfield  {author} {\bibinfo {author} {\bibfnamefont {G.}~\bibnamefont
  {Chen}}, \bibinfo {author} {\bibfnamefont {D.~L.}\ \bibnamefont {Bergman}},\
  and\ \bibinfo {author} {\bibfnamefont {L.}~\bibnamefont {Balents}},\
  }\bibfield  {title} {\bibinfo {title} {Semiclassical dynamics and long-time
  asymptotics of the central-spin problem in a quantum dot},\ }\href
  {https://doi.org/10.1103/PhysRevB.76.045312} {\bibfield  {journal} {\bibinfo
  {journal} {Phys. Rev. B}\ }\textbf {\bibinfo {volume} {76}},\ \bibinfo
  {pages} {045312} (\bibinfo {year} {2007})}\BibitemShut {NoStop}%
\bibitem [{\citenamefont {Hackmann}\ \emph {et~al.}(2014)\citenamefont
  {Hackmann}, \citenamefont {Smirnov}, \citenamefont {Glazov},\ and\
  \citenamefont {Anders}}]{HackmannGlazov2014}%
  \BibitemOpen
  \bibfield  {author} {\bibinfo {author} {\bibfnamefont {J.}~\bibnamefont
  {Hackmann}}, \bibinfo {author} {\bibfnamefont {D.~S.}\ \bibnamefont
  {Smirnov}}, \bibinfo {author} {\bibfnamefont {M.~M.}\ \bibnamefont
  {Glazov}},\ and\ \bibinfo {author} {\bibfnamefont {F.~B.}\ \bibnamefont
  {Anders}},\ }\bibfield  {title} {\bibinfo {title} {Spin noise in a quantum
  dot ensemble: From a quantum mechanical to a semi-classical description},\
  }\href@noop {} {\bibfield  {journal} {\bibinfo  {journal} {physica status
  solidi (b)}\ }\textbf {\bibinfo {volume} {251}},\ \bibinfo {pages} {1270}
  (\bibinfo {year} {2014})}\BibitemShut {NoStop}%
\bibitem [{\citenamefont {J\"aschke}\ \emph {et~al.}(2017)\citenamefont
  {J\"aschke}, \citenamefont {Fischer}, \citenamefont {Evers}, \citenamefont
  {Belykh}, \citenamefont {Greilich}, \citenamefont {Bayer},\ and\
  \citenamefont {Anders}}]{Jaeschke2017}%
  \BibitemOpen
  \bibfield  {author} {\bibinfo {author} {\bibfnamefont {N.}~\bibnamefont
  {J\"aschke}}, \bibinfo {author} {\bibfnamefont {A.}~\bibnamefont {Fischer}},
  \bibinfo {author} {\bibfnamefont {E.}~\bibnamefont {Evers}}, \bibinfo
  {author} {\bibfnamefont {V.~V.}\ \bibnamefont {Belykh}}, \bibinfo {author}
  {\bibfnamefont {A.}~\bibnamefont {Greilich}}, \bibinfo {author}
  {\bibfnamefont {M.}~\bibnamefont {Bayer}},\ and\ \bibinfo {author}
  {\bibfnamefont {F.~B.}\ \bibnamefont {Anders}},\ }\bibfield  {title}
  {\bibinfo {title} {Nonequilibrium nuclear spin distribution function in
  quantum dots subject to periodic pulses},\ }\href
  {https://doi.org/10.1103/PhysRevB.96.205419} {\bibfield  {journal} {\bibinfo
  {journal} {Phys. Rev. B}\ }\textbf {\bibinfo {volume} {96}},\ \bibinfo
  {pages} {205419} (\bibinfo {year} {2017})}\BibitemShut {NoStop}%
\bibitem [{\citenamefont {Girard}(1984)}]{Girard_1984}%
  \BibitemOpen
  \bibfield  {author} {\bibinfo {author} {\bibfnamefont {P.~R.}\ \bibnamefont
  {Girard}},\ }\bibfield  {title} {\bibinfo {title} {The quaternion group and
  modern physics},\ }\href {https://doi.org/10.1088/0143-0807/5/1/007}
  {\bibfield  {journal} {\bibinfo  {journal} {European Journal of Physics}\
  }\textbf {\bibinfo {volume} {5}},\ \bibinfo {pages} {25} (\bibinfo {year}
  {1984})}\BibitemShut {NoStop}%
\bibitem [{\citenamefont {Bechtold}\ \emph {et~al.}(2016)\citenamefont
  {Bechtold}, \citenamefont {Li}, \citenamefont {M\"uller}, \citenamefont
  {Simmet}, \citenamefont {Ardelt}, \citenamefont {Finley},\ and\ \citenamefont
  {Sinitsyn}}]{Bechtold2016}%
  \BibitemOpen
  \bibfield  {author} {\bibinfo {author} {\bibfnamefont {A.}~\bibnamefont
  {Bechtold}}, \bibinfo {author} {\bibfnamefont {F.}~\bibnamefont {Li}},
  \bibinfo {author} {\bibfnamefont {K.}~\bibnamefont {M\"uller}}, \bibinfo
  {author} {\bibfnamefont {T.}~\bibnamefont {Simmet}}, \bibinfo {author}
  {\bibfnamefont {P.-L.}\ \bibnamefont {Ardelt}}, \bibinfo {author}
  {\bibfnamefont {J.~J.}\ \bibnamefont {Finley}},\ and\ \bibinfo {author}
  {\bibfnamefont {N.~A.}\ \bibnamefont {Sinitsyn}},\ }\bibfield  {title}
  {\bibinfo {title} {{Quantum Effects in Higher-Order Correlators of a
  Quantum-Dot Spin Qubit}},\ }\href
  {https://doi.org/10.1103/PhysRevLett.117.027402} {\bibfield  {journal}
  {\bibinfo  {journal} {Phys. Rev. Lett.}\ }\textbf {\bibinfo {volume} {117}},\
  \bibinfo {pages} {027402} (\bibinfo {year} {2016})}\BibitemShut {NoStop}%
\bibitem [{\citenamefont {Evers}\ \emph {et~al.}(2018)\citenamefont {Evers},
  \citenamefont {Belykh}, \citenamefont {Kopteva}, \citenamefont {Yugova},
  \citenamefont {Greilich}, \citenamefont {Yakovlev}, \citenamefont {Reuter},
  \citenamefont {Wieck},\ and\ \citenamefont {Bayer}}]{Evers2018}%
  \BibitemOpen
  \bibfield  {author} {\bibinfo {author} {\bibfnamefont {E.}~\bibnamefont
  {Evers}}, \bibinfo {author} {\bibfnamefont {V.~V.}\ \bibnamefont {Belykh}},
  \bibinfo {author} {\bibfnamefont {N.~E.}\ \bibnamefont {Kopteva}}, \bibinfo
  {author} {\bibfnamefont {I.~A.}\ \bibnamefont {Yugova}}, \bibinfo {author}
  {\bibfnamefont {A.}~\bibnamefont {Greilich}}, \bibinfo {author}
  {\bibfnamefont {D.~R.}\ \bibnamefont {Yakovlev}}, \bibinfo {author}
  {\bibfnamefont {D.}~\bibnamefont {Reuter}}, \bibinfo {author} {\bibfnamefont
  {A.~D.}\ \bibnamefont {Wieck}},\ and\ \bibinfo {author} {\bibfnamefont
  {M.}~\bibnamefont {Bayer}},\ }\bibfield  {title} {\bibinfo {title} {Decay and
  revival of electron spin polarization in an ensemble of (in,ga)as quantum
  dots},\ }\href {https://doi.org/10.1103/PhysRevB.98.075309} {\bibfield
  {journal} {\bibinfo  {journal} {Phys. Rev. B}\ }\textbf {\bibinfo {volume}
  {98}},\ \bibinfo {pages} {075309} (\bibinfo {year} {2018})}\BibitemShut
  {NoStop}%
\bibitem [{\citenamefont {Fauseweh}\ \emph {et~al.}(2017)\citenamefont
  {Fauseweh}, \citenamefont {Schering}, \citenamefont {H\"udepohl},\ and\
  \citenamefont {Uhrig}}]{FausewehSchering2017}%
  \BibitemOpen
  \bibfield  {author} {\bibinfo {author} {\bibfnamefont {B.}~\bibnamefont
  {Fauseweh}}, \bibinfo {author} {\bibfnamefont {P.}~\bibnamefont {Schering}},
  \bibinfo {author} {\bibfnamefont {J.}~\bibnamefont {H\"udepohl}},\ and\
  \bibinfo {author} {\bibfnamefont {G.~S.}\ \bibnamefont {Uhrig}},\ }\bibfield
  {title} {\bibinfo {title} {Efficient algorithms for the dynamics of large and
  infinite classical central spin models},\ }\href
  {https://doi.org/10.1103/PhysRevB.96.054415} {\bibfield  {journal} {\bibinfo
  {journal} {Phys. Rev. B}\ }\textbf {\bibinfo {volume} {96}},\ \bibinfo
  {pages} {054415} (\bibinfo {year} {2017})}\BibitemShut {NoStop}%
\bibitem [{\citenamefont {Uhrig}\ \emph {et~al.}(2014)\citenamefont {Uhrig},
  \citenamefont {Hackmann}, \citenamefont {Stanek}, \citenamefont {Stolze},\
  and\ \citenamefont {Anders}}]{Uhrig2014}%
  \BibitemOpen
  \bibfield  {author} {\bibinfo {author} {\bibfnamefont {G.~S.}\ \bibnamefont
  {Uhrig}}, \bibinfo {author} {\bibfnamefont {J.}~\bibnamefont {Hackmann}},
  \bibinfo {author} {\bibfnamefont {D.}~\bibnamefont {Stanek}}, \bibinfo
  {author} {\bibfnamefont {J.}~\bibnamefont {Stolze}},\ and\ \bibinfo {author}
  {\bibfnamefont {F.~B.}\ \bibnamefont {Anders}},\ }\bibfield  {title}
  {\bibinfo {title} {Conservation laws protect dynamic spin correlations from
  decay: Limited role of integrability in the central spin model},\ }\href
  {https://doi.org/10.1103/PhysRevB.90.060301} {\bibfield  {journal} {\bibinfo
  {journal} {Phys. Rev. B}\ }\textbf {\bibinfo {volume} {90}},\ \bibinfo
  {pages} {060301} (\bibinfo {year} {2014})}\BibitemShut {NoStop}%
\bibitem [{\citenamefont {Fischer}\ \emph {et~al.}(2018)\citenamefont
  {Fischer}, \citenamefont {Evers}, \citenamefont {Varwig}, \citenamefont
  {Greilich}, \citenamefont {Bayer},\ and\ \citenamefont
  {Anders}}]{Fischer2018}%
  \BibitemOpen
  \bibfield  {author} {\bibinfo {author} {\bibfnamefont {A.}~\bibnamefont
  {Fischer}}, \bibinfo {author} {\bibfnamefont {E.}~\bibnamefont {Evers}},
  \bibinfo {author} {\bibfnamefont {S.}~\bibnamefont {Varwig}}, \bibinfo
  {author} {\bibfnamefont {A.}~\bibnamefont {Greilich}}, \bibinfo {author}
  {\bibfnamefont {M.}~\bibnamefont {Bayer}},\ and\ \bibinfo {author}
  {\bibfnamefont {F.~B.}\ \bibnamefont {Anders}},\ }\bibfield  {title}
  {\bibinfo {title} {{Signatures of long-range spin-spin interactions in an
  (In,Ga)As quantum dot ensemble}},\ }\href
  {https://doi.org/10.1103/PhysRevB.98.205308} {\bibfield  {journal} {\bibinfo
  {journal} {Phys. Rev. B}\ }\textbf {\bibinfo {volume} {98}},\ \bibinfo
  {pages} {205308} (\bibinfo {year} {2018})}\BibitemShut {NoStop}%
\bibitem [{\citenamefont {Spatzek}\ \emph {et~al.}(2011)\citenamefont
  {Spatzek}, \citenamefont {Greilich}, \citenamefont {Economou}, \citenamefont
  {Varwig}, \citenamefont {Schwan}, \citenamefont {Yakovlev}, \citenamefont
  {Reuter}, \citenamefont {Wieck}, \citenamefont {Reinecke},\ and\
  \citenamefont {Bayer}}]{Spatzek2011}%
  \BibitemOpen
  \bibfield  {author} {\bibinfo {author} {\bibfnamefont {S.}~\bibnamefont
  {Spatzek}}, \bibinfo {author} {\bibfnamefont {A.}~\bibnamefont {Greilich}},
  \bibinfo {author} {\bibfnamefont {S.~E.}\ \bibnamefont {Economou}}, \bibinfo
  {author} {\bibfnamefont {S.}~\bibnamefont {Varwig}}, \bibinfo {author}
  {\bibfnamefont {A.}~\bibnamefont {Schwan}}, \bibinfo {author} {\bibfnamefont
  {D.~R.}\ \bibnamefont {Yakovlev}}, \bibinfo {author} {\bibfnamefont
  {D.}~\bibnamefont {Reuter}}, \bibinfo {author} {\bibfnamefont {A.~D.}\
  \bibnamefont {Wieck}}, \bibinfo {author} {\bibfnamefont {T.~L.}\ \bibnamefont
  {Reinecke}},\ and\ \bibinfo {author} {\bibfnamefont {M.}~\bibnamefont
  {Bayer}},\ }\bibfield  {title} {\bibinfo {title} {Optical control of coherent
  interactions between electron spins in ingaas quantum dots},\ }\href@noop {}
  {\bibfield  {journal} {\bibinfo  {journal} {Phys. Rev. Lett.}\ }\textbf
  {\bibinfo {volume} {107}},\ \bibinfo {pages} {137402} (\bibinfo {year}
  {2011})}\BibitemShut {NoStop}%
\bibitem [{\citenamefont {Mizel}\ and\ \citenamefont {Lidar}(2004)}]{Mizel04}%
  \BibitemOpen
  \bibfield  {author} {\bibinfo {author} {\bibfnamefont {A.}~\bibnamefont
  {Mizel}}\ and\ \bibinfo {author} {\bibfnamefont {D.~A.}\ \bibnamefont
  {Lidar}},\ }\bibfield  {title} {\bibinfo {title} {Exchange interaction
  between three and four coupled quantum dots: Theory and applications to
  quantum computing},\ }\href {https://doi.org/10.1103/PhysRevB.70.115310}
  {\bibfield  {journal} {\bibinfo  {journal} {Phys. Rev. B}\ }\textbf {\bibinfo
  {volume} {70}},\ \bibinfo {pages} {115310} (\bibinfo {year}
  {2004})}\BibitemShut {NoStop}%
\bibitem [{\citenamefont {Patanasemakul}\ \emph {et~al.}(2012)\citenamefont
  {Patanasemakul}, \citenamefont {Panyakeow},\ and\ \citenamefont
  {Kanjanachuchai}}]{Patanasemakul2012}%
  \BibitemOpen
  \bibfield  {author} {\bibinfo {author} {\bibfnamefont {N.}~\bibnamefont
  {Patanasemakul}}, \bibinfo {author} {\bibfnamefont {S.}~\bibnamefont
  {Panyakeow}},\ and\ \bibinfo {author} {\bibfnamefont {S.}~\bibnamefont
  {Kanjanachuchai}},\ }\bibfield  {title} {\bibinfo {title} {Chirped ingaas
  quantum dot molecules for broadband applications},\ }\href
  {https://doi.org/10.1186/1556-276X-7-207} {\bibfield  {journal} {\bibinfo
  {journal} {Nanoscale Research Letters}\ }\textbf {\bibinfo {volume} {7}},\
  \bibinfo {pages} {207} (\bibinfo {year} {2012})}\BibitemShut {NoStop}%
\bibitem [{\citenamefont {Wang}\ \emph {et~al.}(2006)\citenamefont {Wang},
  \citenamefont {Mazur}, \citenamefont {Shultz}, \citenamefont {Salamo},
  \citenamefont {Mishima},\ and\ \citenamefont {Johnson}}]{Wang2006}%
  \BibitemOpen
  \bibfield  {author} {\bibinfo {author} {\bibfnamefont {Z.~M.}\ \bibnamefont
  {Wang}}, \bibinfo {author} {\bibfnamefont {Y.~I.}\ \bibnamefont {Mazur}},
  \bibinfo {author} {\bibfnamefont {J.~L.}\ \bibnamefont {Shultz}}, \bibinfo
  {author} {\bibfnamefont {G.~J.}\ \bibnamefont {Salamo}}, \bibinfo {author}
  {\bibfnamefont {T.~D.}\ \bibnamefont {Mishima}},\ and\ \bibinfo {author}
  {\bibfnamefont {M.~B.}\ \bibnamefont {Johnson}},\ }\bibfield  {title}
  {\bibinfo {title} {{One-dimensional postwetting layer in InGaAs GaAs(100)
  quantum-dot chains}},\ }\href {https://doi.org/10.1063/1.2169868} {\bibfield
  {journal} {\bibinfo  {journal} {Journal of Applied Physics}\ }\textbf
  {\bibinfo {volume} {99}},\ \bibinfo {pages} {033705} (\bibinfo {year}
  {2006})}\BibitemShut {NoStop}%
\bibitem [{\citenamefont {Sugaya}\ \emph {et~al.}(2013)\citenamefont {Sugaya},
  \citenamefont {Oshima}, \citenamefont {Matsubara},\ and\ \citenamefont
  {Niki}}]{Sugaya2013}%
  \BibitemOpen
  \bibfield  {author} {\bibinfo {author} {\bibfnamefont {T.}~\bibnamefont
  {Sugaya}}, \bibinfo {author} {\bibfnamefont {R.}~\bibnamefont {Oshima}},
  \bibinfo {author} {\bibfnamefont {K.}~\bibnamefont {Matsubara}},\ and\
  \bibinfo {author} {\bibfnamefont {S.}~\bibnamefont {Niki}},\ }\bibfield
  {title} {\bibinfo {title} {Ingaas quantum dot superlattice with vertically
  coupled states in ingap matrix},\ }\href {https://doi.org/10.1063/1.4812567}
  {\bibfield  {journal} {\bibinfo  {journal} {Journal of Applied Physics}\
  }\textbf {\bibinfo {volume} {114}},\ \bibinfo {pages} {014303} (\bibinfo
  {year} {2013})}\BibitemShut {NoStop}%
\bibitem [{\citenamefont {Smirnov}\ \emph {et~al.}(2014)\citenamefont
  {Smirnov}, \citenamefont {Glazov},\ and\ \citenamefont
  {Ivchenko}}]{Smirnov2014}%
  \BibitemOpen
  \bibfield  {author} {\bibinfo {author} {\bibfnamefont {D.}~\bibnamefont
  {Smirnov}}, \bibinfo {author} {\bibfnamefont {M.}~\bibnamefont {Glazov}},\
  and\ \bibinfo {author} {\bibfnamefont {E.}~\bibnamefont {Ivchenko}},\
  }\bibfield  {title} {\bibinfo {title} {Effect of exchange interaction on the
  spin fluctuations of localized electrons},\ }\href@noop {} {\bibfield
  {journal} {\bibinfo  {journal} {Physics of the Solid State}\ }\textbf
  {\bibinfo {volume} {56}},\ \bibinfo {pages} {254} (\bibinfo {year}
  {2014})}\BibitemShut {NoStop}%
\bibitem [{\citenamefont {Schwan}\ \emph {et~al.}(2011)\citenamefont {Schwan},
  \citenamefont {Meiners}, \citenamefont {Henriques}, \citenamefont {Maia},
  \citenamefont {Quivy}, \citenamefont {Spatzek}, \citenamefont {Varwig},
  \citenamefont {Yakovlev},\ and\ \citenamefont {Bayer}}]{Schwan2011}%
  \BibitemOpen
  \bibfield  {author} {\bibinfo {author} {\bibfnamefont {A.}~\bibnamefont
  {Schwan}}, \bibinfo {author} {\bibfnamefont {B.-M.}\ \bibnamefont {Meiners}},
  \bibinfo {author} {\bibfnamefont {A.~B.}\ \bibnamefont {Henriques}}, \bibinfo
  {author} {\bibfnamefont {A.~D.~B.}\ \bibnamefont {Maia}}, \bibinfo {author}
  {\bibfnamefont {A.~A.}\ \bibnamefont {Quivy}}, \bibinfo {author}
  {\bibfnamefont {S.}~\bibnamefont {Spatzek}}, \bibinfo {author} {\bibfnamefont
  {S.}~\bibnamefont {Varwig}}, \bibinfo {author} {\bibfnamefont {D.~R.}\
  \bibnamefont {Yakovlev}},\ and\ \bibinfo {author} {\bibfnamefont
  {M.}~\bibnamefont {Bayer}},\ }\bibfield  {title} {\bibinfo {title}
  {{Dispersion of electron g-factor with optical transition energy in
  (In,Ga)As/GaAs self-assembled quantum dots}},\ }\href
  {https://doi.org/10.1063/1.3588413} {\bibfield  {journal} {\bibinfo
  {journal} {Applied Physics Letters}\ }\textbf {\bibinfo {volume} {98}},\
  \bibinfo {pages} {233102} (\bibinfo {year} {2011})}\BibitemShut {NoStop}%
\bibitem [{\citenamefont {Beugeling}\ \emph {et~al.}(2016)\citenamefont
  {Beugeling}, \citenamefont {Uhrig},\ and\ \citenamefont
  {Anders}}]{Beugeling2016}%
  \BibitemOpen
  \bibfield  {author} {\bibinfo {author} {\bibfnamefont {W.}~\bibnamefont
  {Beugeling}}, \bibinfo {author} {\bibfnamefont {G.~S.}\ \bibnamefont
  {Uhrig}},\ and\ \bibinfo {author} {\bibfnamefont {F.~B.}\ \bibnamefont
  {Anders}},\ }\bibfield  {title} {\bibinfo {title} {Quantum model for mode
  locking in pulsed semiconductor quantum dots},\ }\href@noop {} {\bibfield
  {journal} {\bibinfo  {journal} {Phys. Rev. B}\ }\textbf {\bibinfo {volume}
  {94}},\ \bibinfo {pages} {245308} (\bibinfo {year} {2016})}\BibitemShut
  {NoStop}%
\bibitem [{\citenamefont {Beugeling}\ \emph {et~al.}(2017)\citenamefont
  {Beugeling}, \citenamefont {Uhrig},\ and\ \citenamefont
  {Anders}}]{Beugeling2017}%
  \BibitemOpen
  \bibfield  {author} {\bibinfo {author} {\bibfnamefont {W.}~\bibnamefont
  {Beugeling}}, \bibinfo {author} {\bibfnamefont {G.~S.}\ \bibnamefont
  {Uhrig}},\ and\ \bibinfo {author} {\bibfnamefont {F.~B.}\ \bibnamefont
  {Anders}},\ }\bibfield  {title} {\bibinfo {title} {{Influence of the nuclear
  Zeeman effect on mode locking in pulsed semiconductor quantum dots}},\
  }\href@noop {} {\bibfield  {journal} {\bibinfo  {journal} {Phys. Rev. B}\
  }\textbf {\bibinfo {volume} {96}},\ \bibinfo {pages} {115303} (\bibinfo
  {year} {2017})}\BibitemShut {NoStop}%
\bibitem [{\citenamefont {Coish}\ and\ \citenamefont
  {Baugh}(2009)}]{Coish2009}%
  \BibitemOpen
  \bibfield  {author} {\bibinfo {author} {\bibfnamefont {W.~A.}\ \bibnamefont
  {Coish}}\ and\ \bibinfo {author} {\bibfnamefont {J.}~\bibnamefont {Baugh}},\
  }\bibfield  {title} {\bibinfo {title} {Nuclear spins in nanostructures},\
  }\href {https://doi.org/10.1002/pssb.200945229} {\bibfield  {journal}
  {\bibinfo  {journal} {physica status solidi (b)}\ }\textbf {\bibinfo {volume}
  {246}},\ \bibinfo {pages} {2203} (\bibinfo {year} {2009})}\BibitemShut
  {NoStop}%
\bibitem [{\citenamefont {Bechtold}\ \emph {et~al.}(2015)\citenamefont
  {Bechtold}, \citenamefont {Rauch}, \citenamefont {Simmet}, \citenamefont
  {Ardelt}, \citenamefont {Regler}, \citenamefont {M\"uller}, \citenamefont
  {Sinitsyn},\ and\ \citenamefont {Finley}}]{Bechtold2015}%
  \BibitemOpen
  \bibfield  {author} {\bibinfo {author} {\bibfnamefont {A.}~\bibnamefont
  {Bechtold}}, \bibinfo {author} {\bibfnamefont {A.}~\bibnamefont {Rauch}},
  \bibinfo {author} {\bibfnamefont {T.}~\bibnamefont {Simmet}}, \bibinfo
  {author} {\bibfnamefont {P.-L.}\ \bibnamefont {Ardelt}}, \bibinfo {author}
  {\bibfnamefont {A.}~\bibnamefont {Regler}}, \bibinfo {author} {\bibfnamefont
  {K.}~\bibnamefont {M\"uller}}, \bibinfo {author} {\bibfnamefont {N.~A.}\
  \bibnamefont {Sinitsyn}},\ and\ \bibinfo {author} {\bibfnamefont {J.~J.}\
  \bibnamefont {Finley}},\ }\bibfield  {title} {\bibinfo {title} {{Three-stage
  decoherence dynamics of an electron spin qubit in an optically active quantum
  dot}},\ }\href@noop {} {\bibfield  {journal} {\bibinfo  {journal} {Nature
  Phys.}\ }\textbf {\bibinfo {volume} {11}},\ \bibinfo {pages} {1005} (\bibinfo
  {year} {2015})}\BibitemShut {NoStop}%
\bibitem [{\citenamefont {Dyakonov}(2008)}]{Dyakonov2008}%
  \BibitemOpen
  \bibfield  {author} {\bibinfo {author} {\bibfnamefont {M.~I.}\ \bibnamefont
  {Dyakonov}},\ }\href@noop {} {\emph {\bibinfo {title} {Spin Physics in
  Semiconductors}}}\ (\bibinfo  {publisher} {Springer-Verlag},\ \bibinfo
  {address} {Berlin Heidelberg},\ \bibinfo {year} {2008})\BibitemShut {NoStop}%
\bibitem [{\citenamefont {Slichter}(1996)}]{Slichter1996}%
  \BibitemOpen
  \bibfield  {author} {\bibinfo {author} {\bibfnamefont {C.~P.}\ \bibnamefont
  {Slichter}},\ }\href@noop {} {\emph {\bibinfo {title} {Principles of Magnetic
  Resonance}}}\ (\bibinfo  {publisher} {Springer Science \& Business Media,,
  Berlin},\ \bibinfo {year} {1996})\BibitemShut {NoStop}%
\bibitem [{\citenamefont {Urbaszek}\ \emph {et~al.}(2013)\citenamefont
  {Urbaszek}, \citenamefont {Marie}, \citenamefont {Amand}, \citenamefont
  {Krebs}, \citenamefont {Voisin}, \citenamefont {Maletinsky}, \citenamefont
  {H\"ogele},\ and\ \citenamefont {Imamoglu}}]{Urbaszek2013}%
  \BibitemOpen
  \bibfield  {author} {\bibinfo {author} {\bibfnamefont {B.}~\bibnamefont
  {Urbaszek}}, \bibinfo {author} {\bibfnamefont {X.}~\bibnamefont {Marie}},
  \bibinfo {author} {\bibfnamefont {T.}~\bibnamefont {Amand}}, \bibinfo
  {author} {\bibfnamefont {O.}~\bibnamefont {Krebs}}, \bibinfo {author}
  {\bibfnamefont {P.}~\bibnamefont {Voisin}}, \bibinfo {author} {\bibfnamefont
  {P.}~\bibnamefont {Maletinsky}}, \bibinfo {author} {\bibfnamefont
  {A.}~\bibnamefont {H\"ogele}},\ and\ \bibinfo {author} {\bibfnamefont
  {A.}~\bibnamefont {Imamoglu}},\ }\bibfield  {title} {\bibinfo {title}
  {Nuclear spin physics in quantum dots: An optical investigation},\ }\href
  {https://doi.org/10.1103/RevModPhys.85.79} {\bibfield  {journal} {\bibinfo
  {journal} {Rev. Mod. Phys.}\ }\textbf {\bibinfo {volume} {85}},\ \bibinfo
  {pages} {79} (\bibinfo {year} {2013})}\BibitemShut {NoStop}%
\bibitem [{\citenamefont {Kleinjohann}\ \emph {et~al.}(2018)\citenamefont
  {Kleinjohann}, \citenamefont {Evers}, \citenamefont {Schering}, \citenamefont
  {Greilich}, \citenamefont {Uhrig}, \citenamefont {Bayer},\ and\ \citenamefont
  {Anders}}]{Kleinjohann2018}%
  \BibitemOpen
  \bibfield  {author} {\bibinfo {author} {\bibfnamefont {I.}~\bibnamefont
  {Kleinjohann}}, \bibinfo {author} {\bibfnamefont {E.}~\bibnamefont {Evers}},
  \bibinfo {author} {\bibfnamefont {P.}~\bibnamefont {Schering}}, \bibinfo
  {author} {\bibfnamefont {A.}~\bibnamefont {Greilich}}, \bibinfo {author}
  {\bibfnamefont {G.~S.}\ \bibnamefont {Uhrig}}, \bibinfo {author}
  {\bibfnamefont {M.}~\bibnamefont {Bayer}},\ and\ \bibinfo {author}
  {\bibfnamefont {F.~B.}\ \bibnamefont {Anders}},\ }\bibfield  {title}
  {\bibinfo {title} {Magnetic field dependence of the electron spin revival
  amplitude in periodically pulsed quantum dots},\ }\href
  {https://doi.org/10.1103/PhysRevB.98.155318} {\bibfield  {journal} {\bibinfo
  {journal} {Phys. Rev. B}\ }\textbf {\bibinfo {volume} {98}},\ \bibinfo
  {pages} {155318} (\bibinfo {year} {2018})}\BibitemShut {NoStop}%
\bibitem [{\citenamefont {Fradkin}\ and\ \citenamefont
  {Stone}(1988)}]{Fradkin1988}%
  \BibitemOpen
  \bibfield  {author} {\bibinfo {author} {\bibfnamefont {E.}~\bibnamefont
  {Fradkin}}\ and\ \bibinfo {author} {\bibfnamefont {M.}~\bibnamefont
  {Stone}},\ }\bibfield  {title} {\bibinfo {title} {Topological terms in one-
  and two-dimensional quantum heisenberg antiferromagnets},\ }\href
  {https://doi.org/10.1103/PhysRevB.38.7215} {\bibfield  {journal} {\bibinfo
  {journal} {Phys. Rev. B}\ }\textbf {\bibinfo {volume} {38}},\ \bibinfo
  {pages} {7215} (\bibinfo {year} {1988})}\BibitemShut {NoStop}%
\bibitem [{\citenamefont {Fradkin}(2013)}]{Fradkin2013}%
  \BibitemOpen
  \bibfield  {author} {\bibinfo {author} {\bibfnamefont {E.}~\bibnamefont
  {Fradkin}},\ }\href {https://doi.org/10.1017/CBO9781139015509} {\emph
  {\bibinfo {title} {Field Theories of Condensed Matter Physics}}},\ \bibinfo
  {edition} {2nd}\ ed.\ (\bibinfo  {publisher} {Cambridge University Press},\
  \bibinfo {year} {2013})\BibitemShut {NoStop}%
\bibitem [{\citenamefont {Schering}\ \emph {et~al.}(2018)\citenamefont
  {Schering}, \citenamefont {H\"udepohl}, \citenamefont {Uhrig},\ and\
  \citenamefont {Fauseweh}}]{ScheringSD-SCA2018}%
  \BibitemOpen
  \bibfield  {author} {\bibinfo {author} {\bibfnamefont {P.}~\bibnamefont
  {Schering}}, \bibinfo {author} {\bibfnamefont {J.}~\bibnamefont
  {H\"udepohl}}, \bibinfo {author} {\bibfnamefont {G.~S.}\ \bibnamefont
  {Uhrig}},\ and\ \bibinfo {author} {\bibfnamefont {B.}~\bibnamefont
  {Fauseweh}},\ }\bibfield  {title} {\bibinfo {title} {Nuclear frequency
  focusing in periodically pulsed semiconductor quantum dots described by
  infinite classical central spin models},\ }\href
  {https://doi.org/10.1103/PhysRevB.98.024305} {\bibfield  {journal} {\bibinfo
  {journal} {Phys. Rev. B}\ }\textbf {\bibinfo {volume} {98}},\ \bibinfo
  {pages} {024305} (\bibinfo {year} {2018})}\BibitemShut {NoStop}%
\bibitem [{\citenamefont {Khaetskii}\ \emph {et~al.}(2003)\citenamefont
  {Khaetskii}, \citenamefont {Loss},\ and\ \citenamefont
  {Glazman}}]{Khaetskii2003}%
  \BibitemOpen
  \bibfield  {author} {\bibinfo {author} {\bibfnamefont {A.}~\bibnamefont
  {Khaetskii}}, \bibinfo {author} {\bibfnamefont {D.}~\bibnamefont {Loss}},\
  and\ \bibinfo {author} {\bibfnamefont {L.}~\bibnamefont {Glazman}},\
  }\bibfield  {title} {\bibinfo {title} {Electron spin evolution induced by
  interaction with nuclei in a quantum dot},\ }\href@noop {} {\bibfield
  {journal} {\bibinfo  {journal} {Phys. Rev. B}\ }\textbf {\bibinfo {volume}
  {67}},\ \bibinfo {pages} {195329} (\bibinfo {year} {2003})}\BibitemShut
  {NoStop}%
\bibitem [{\citenamefont {Yang}\ \emph {et~al.}(2014)\citenamefont {Yang},
  \citenamefont {Glasenapp}, \citenamefont {Greilich}, \citenamefont {Reuter},
  \citenamefont {Wieck}, \citenamefont {Yakovlev}, \citenamefont {Bayer},\ and\
  \citenamefont {Crooker}}]{Yang2014}%
  \BibitemOpen
  \bibfield  {author} {\bibinfo {author} {\bibfnamefont {L.}~\bibnamefont
  {Yang}}, \bibinfo {author} {\bibfnamefont {P.}~\bibnamefont {Glasenapp}},
  \bibinfo {author} {\bibfnamefont {A.}~\bibnamefont {Greilich}}, \bibinfo
  {author} {\bibfnamefont {D.}~\bibnamefont {Reuter}}, \bibinfo {author}
  {\bibfnamefont {A.~D.}\ \bibnamefont {Wieck}}, \bibinfo {author}
  {\bibfnamefont {D.~R.}\ \bibnamefont {Yakovlev}}, \bibinfo {author}
  {\bibfnamefont {M.}~\bibnamefont {Bayer}},\ and\ \bibinfo {author}
  {\bibfnamefont {S.~A.}\ \bibnamefont {Crooker}},\ }\bibfield  {title}
  {\bibinfo {title} {Two-colour spin noise spectroscopy and fluctuation
  correlations reveal homogeneous linewidths within quantum-dot ensembles},\
  }\href {https://doi.org/10.1038/ncomms5949} {\bibfield  {journal} {\bibinfo
  {journal} {Nature Communications}\ }\textbf {\bibinfo {volume} {5}},\
  \bibinfo {pages} {4949} (\bibinfo {year} {2014})}\BibitemShut {NoStop}%
\bibitem [{\citenamefont {Fr\"ohling}\ \emph {et~al.}(2018)\citenamefont
  {Fr\"ohling}, \citenamefont {Anders},\ and\ \citenamefont
  {Glazov}}]{FroehlingGlazov2018}%
  \BibitemOpen
  \bibfield  {author} {\bibinfo {author} {\bibfnamefont {N.}~\bibnamefont
  {Fr\"ohling}}, \bibinfo {author} {\bibfnamefont {F.~B.}\ \bibnamefont
  {Anders}},\ and\ \bibinfo {author} {\bibfnamefont {M.}~\bibnamefont
  {Glazov}},\ }\bibfield  {title} {\bibinfo {title} {Nuclear spin noise in the
  central spin model},\ }\href {https://doi.org/10.1103/PhysRevB.97.195311}
  {\bibfield  {journal} {\bibinfo  {journal} {Phys. Rev. B}\ }\textbf {\bibinfo
  {volume} {97}},\ \bibinfo {pages} {195311} (\bibinfo {year}
  {2018})}\BibitemShut {NoStop}%
\bibitem [{\citenamefont {Kozlov}(2007)}]{Kozlov2007}%
  \BibitemOpen
  \bibfield  {author} {\bibinfo {author} {\bibfnamefont {G.~G.}\ \bibnamefont
  {Kozlov}},\ }\bibfield  {title} {\bibinfo {title} {Exactly solvable spin
  dynamics of an electron coupled to a large number of nuclei; the
  electron-nuclear spin echo in a quantum dot},\ }\href@noop {} {\bibfield
  {journal} {\bibinfo  {journal} {Journal of Experimental and Theoretical
  Physics}\ }\textbf {\bibinfo {volume} {105}},\ \bibinfo {pages} {803}
  (\bibinfo {year} {2007})}\BibitemShut {NoStop}%
\bibitem [{\citenamefont {Press}\ \emph {et~al.}(2010)\citenamefont {Press},
  \citenamefont {De~Greve}, \citenamefont {McMahon}, \citenamefont {Ladd},
  \citenamefont {Friess}, \citenamefont {Schneider}, \citenamefont {Kamp},
  \citenamefont {H\"ofling}, \citenamefont {Forchel},\ and\ \citenamefont
  {Yamamoto}}]{Press2010}%
  \BibitemOpen
  \bibfield  {author} {\bibinfo {author} {\bibfnamefont {D.}~\bibnamefont
  {Press}}, \bibinfo {author} {\bibfnamefont {K.}~\bibnamefont {De~Greve}},
  \bibinfo {author} {\bibfnamefont {P.~L.}\ \bibnamefont {McMahon}}, \bibinfo
  {author} {\bibfnamefont {T.~D.}\ \bibnamefont {Ladd}}, \bibinfo {author}
  {\bibfnamefont {B.}~\bibnamefont {Friess}}, \bibinfo {author} {\bibfnamefont
  {C.}~\bibnamefont {Schneider}}, \bibinfo {author} {\bibfnamefont
  {M.}~\bibnamefont {Kamp}}, \bibinfo {author} {\bibfnamefont {S.}~\bibnamefont
  {H\"ofling}}, \bibinfo {author} {\bibfnamefont {A.}~\bibnamefont {Forchel}},\
  and\ \bibinfo {author} {\bibfnamefont {Y.}~\bibnamefont {Yamamoto}},\
  }\bibfield  {title} {\bibinfo {title} {Ultrafast optical spin echo in a
  single quantum dot},\ }\href {https://doi.org/10.1038/nphoton.2010.83}
  {\bibfield  {journal} {\bibinfo  {journal} {Nat Photon}\ }\textbf {\bibinfo
  {volume} {4}},\ \bibinfo {pages} {367 } (\bibinfo {year} {2010})}\BibitemShut
  {NoStop}%
\bibitem [{\citenamefont {Yugova}\ \emph {et~al.}(2009)\citenamefont {Yugova},
  \citenamefont {Glazov}, \citenamefont {Ivchenko},\ and\ \citenamefont
  {Efros}}]{Yugova2009}%
  \BibitemOpen
  \bibfield  {author} {\bibinfo {author} {\bibfnamefont {I.~A.}\ \bibnamefont
  {Yugova}}, \bibinfo {author} {\bibfnamefont {M.~M.}\ \bibnamefont {Glazov}},
  \bibinfo {author} {\bibfnamefont {E.~L.}\ \bibnamefont {Ivchenko}},\ and\
  \bibinfo {author} {\bibfnamefont {{\relax Al}.~L.}\ \bibnamefont {Efros}},\
  }\bibfield  {title} {\bibinfo {title} {Pump-probe faraday rotation and
  ellipticity in an ensemble of singly charged quantum dots},\ }\href@noop {}
  {\bibfield  {journal} {\bibinfo  {journal} {Phys. Rev. B}\ }\textbf {\bibinfo
  {volume} {80}},\ \bibinfo {pages} {104436} (\bibinfo {year}
  {2009})}\BibitemShut {NoStop}%
\end{thebibliography}%

\end{document}